\documentclass[12pt]{amsart}

\usepackage{amssymb,amsfonts,latexsym,amscd}
\usepackage{amsmath}

\usepackage{hyperref}

\hypersetup{
unicode=true, 
pdftoolbar=true, 
pdfmenubar=true, 
pdffitwindow=true, 
pdftitle={SPECTRAL THEORY OF A MATHEMATICAL MODEL IN QUANTUM FIELD THEORY FOR ANY SPIN}, 
pdfauthor={JEAN-CLAUDE GUILLOT}, 
pdfsubject={article}, 
pdfnewwindow=true, 
pdfkeywords={polonium, P6}, 
colorlinks=true, 
linkcolor=red, 
citecolor=blue, 
filecolor=blue, 
urlcolor=blue 
}
\hypersetup{
linktocpage=true
}

\newcommand{\gF}{{\mathfrak F}}  

\newcommand{\cD}{{\mathcal{D}}} 

\newcommand{{\adat}}{\mathrm{ad}_{A_t}}
\def\C{{\mathbb{C}}} 
\def\N{{\mathbb{N}}} 
\def\Z{{\mathbb{Z}}} 
\def\R{{\mathbb{R}}} 

\def\vp{\varphi}

\def\1{{\mathbf{1}}}

\renewcommand\d{\mathrm{d}}
\renewcommand\Re{\mathrm{Re}}

\renewcommand\Im{\mathrm{Im}} 

\newtheorem{theorem}{Theorem}[section]

\newtheorem{lemma}[theorem]{Lemma}

\newtheorem{proposition}[theorem]{Proposition}
\newtheorem{remark}[theorem]{Remark}

\newtheorem{hypothesis}[theorem]{Hypothesis}

\newcommand{\arxivlink}[1]{\href{http://arxiv.org/abs/#1}{arXiv:#1}}

\begin{document}

\title[A Model in Quantum Field Theory for Any Spin]{Spectral Theory of a Mathematical Model in Quantum Field Theory for Any Spin}

\author[J.-C. Guillot]{Jean-Claude Guillot}

\address[J.-C. Guillot]{
Centre de Math\'ematiques Appliqu\'ees, UMR 7641, \'Ecole
Polytechnique - C.N.R.S, 91128 Palaiseau Cedex, France}

\email{Jean-Claude.Guillot@polytechnique.edu}

\maketitle
\begin{center}
\textit{To my friend James Ralston.}
\end{center}

\begin{abstract}
In this paper we use the formalism of S.Weinberg in order to construct a mathematical model based on the weak decay of hadrons and nuclei. In particular we consider a model which generalizes the weak decay of the nucleus of the cobalt. We associate with this model a Hamiltonian with cutoffs in a Fock space. The Hamiltonian is self-adjoint and has a ground state. By using the commutator theory we get a limiting absorption principle from which we deduce that the spectrum of the Hamiltonian is absolutely continuous above the energy of a ground state and below the first threshold.
The coupling constant is supposed sufficiently small and no infrared regularization is assumed.

\end{abstract}

\maketitle
\tableofcontents
\section{Introduction}

\setcounter{equation}{0}

This article initiates the study of mathematical models based on Quantum Field Theory without any restriction concerning the spins of the involved particles.

Precisely, in this paper, we study a mathematical model which generalizes the weak decay of the nucleus $^{60}_{27}Co$ of spin 5 into the nucleus $^{60}_{28}Ni^{*}$ of spin 4, one electron and the antineutrino associated to the electron. This experiment by C.S.Wu and her collaborators showed that parity conservation is violated in the $\beta$ decay of atomic nuclei. See \cite{GreinerMüller1996}. The same approach can be applied to many examples of weak decays of hadrons and nuclei with both Fermi and Gamow-Teller transitions. See \cite{Perkins1987} and \cite{GreinerMüller1996}.

The mathematical model is based on the construction of free causal fields associated with two massive bosons of spins $j_{1}$ and $j_{2}$ respectively, a massive fermion of spin $j_{3}$ and a massless fermion of helicity $-j_{4}$ which is the antiparticle of a massless fermion of helicity $j_{4}$. These free causal fields are constructed according to the formalism described by S.Weinberg in \cite{WeinbergI1964,WeinbergII1964,WeinbergIII1964,Weinberg1965,Weinberg1969,Weinberg1995}( see also \cite{StreaterWightman1964},\cite{Grigore1995}).

This construction depends on the unitary irreducible representations of the Poincar\'{e} group for massive and massless particles and on the finite dimensional representations of $SL(2,\,\C)$  which are both well known (see \cite{Wigner1939}, \cite{Wightman1960}, \cite{Guillot1968}, \cite{Varadarajan1970}, \cite{Thaller1992}, \cite{Naimark1964} and \cite{vanderWaerden1974}). Relativistic covariance laws and microscopic causality conditions determine unique free causal fields up to over-all scales. Note that in this paper we only consider fields associated with irreducible finite dimensional representations of $SL(2,\,\C)$ because we are only concerned with a weak decay for which parity is not conserved.

As it is well known, the construction of the unitary irreducible representations of the Poincar\'{e} group for massive particles of any spin and for massless ones with any finite helicity is based on the theory of E.P.Wigner and G.W.Mackey. We choose the realizations of the unitary irreducible representations of the Poincar\'{e} group given  by E.P.Wigner because they are important from the physical point of view and because they allow a clear distinction between the canonical and helicity formalisms.

The interaction between particles is the one given by S.Weinberg in \cite[chap 5]{Weinberg1995}. As for the weak interactions we do not suppose that the interaction commutes with space inversion.

After introducing convenient cutoffs for the associated Hamiltonian the mathematical method used to study the spectral properties of the Hamiltonian  is based on the one applied to a mathematical model associated with the weak decay of the intermediate vector bosons $W^{\pm}$ into the family of leptons which has been recently developed by \cite{BarbarouxGuillot2009,AschbacherGuillot2011}. The existence of a ground state and the proof that the spectrum of the Hamiltonian is absolutely continuous above the energy of the ground state and below the first threshold for a sufficiently small coupling constant are our main results. Our methods are largely taken from \cite{Bachetal2006,Frohlichetal2008,Chenetal2009} and are based on \cite{Mourre1981,Amreinetal1996,Sahbani1997,GeorgescuGerard1999,GoleniaJecko2007,Gerard2008}. No infrared regularization is assumed.

In the framework of non-relativistic QED similar results have been successfully obtained for the massless Pauli-Fierz models (see \cite{BachSigal1998,BachSigal1999,BachSoffer1999,Georgescuetal2004,Frohlichetal2008,Frohlichetal2011,BonyFaupin2012} and references therein ).

For other mathematical models in Quantum Field Theory see, for example, \cite{Amouretal2007,BarbarouxDGuillot2004, GerardPanati2008} and for string-localized quantum fields see \cite{PY2012} and references therein.

The paper is organized as follows. In the next section we recall the realizations of the unitary irreducible representations of the Poincar\'{e} group obtained by E.P.Wigner. In section 3 we first introduce the Fock spaces and the creation and annihilation operators with their usual commutation or anticommutation relations for massive particles. We then recall the construction of the finite dimensional irreducible representations of $SL(2,\,\C)$ and we give a very detailed review of the construction of free causal fields for a massive particle of any spin following the formalism of S.Weinberg and associated with a finite dimensional irreducible representation of $SL(2,\,\C)$. Similarly in section 4 we recall the construction of free causal fields for massless particles of any finite helicity according to the same formalism as for the massive particles. In section 5 we describe the model for the weak decay of a massive boson into a massive boson, a massive fermion and a massless fermion which can be an antineutrino generalizing the model for the weak decay of the nucleus $^{60}_{27}Co$ . In section 6 we associate a self-adjoint Hamiltonian in a Fock space with this model and in section 7 we finally give our main results concerning the spectrum of the self-adjoint Hamiltonian.

\section{The Poincar\'{e} Group}\label{section2}

\setcounter{equation}{0}

Let us recall that the Minkowski space is $\R^4$ equipped with the Lorentz inner product which is the bilinear form L defined by
\begin{equation}\label{2.1}
    L(x,y)=x^{0}y^{0}-x^{1}y^{1}-x^{2}y^{2}-x^{3}y^{3}\ .
\end{equation}
$x^{0}= ct$, where $t$ is the time coordinate and $c$ the speed of light. $(x^{1},x^{2},x^{3})$ is a set of cartesian coordinates on $\R^3$.

From now on we choose units such that $c=\hbar=1$.

The Lorentz form L is associated with the metric 
\begin{equation}\label{2.2}
\begin{split}
  & ds^{2}=g_{\mu\nu}dx^{\mu}dx^{\mu}=g^{\mu\nu}dx_{\mu}dx_{\mu} \\
  & g_{\mu\nu}=g^{\mu\nu}=\mathrm{diag}(1,-1,-1,-1)\ .
\end{split}
\end{equation}
 with $\mu, \nu=0,1,2,3.$ and where  we denote by $x^{\mu}$ ( resp.$x_{\mu}$) the vector $(x^{0},x^{1},x^{2},x^{3})$ ( resp.$(x_{0},x_{1},x_{2},x_{3})$). We use the Einstein summation convention.

A point $x\in \R^4$ may be written as $(x^{0},\textbf{x})$ where $\textbf{x}=(x^{1},x^{2},x^{3})$.

Note that $x^{\mu}=g^{\mu\nu}x_{\nu}$ and $x_{\mu}=g_{\mu\nu}x^{\nu}$.

 The restricted Lorentz group or proper Lorentz group, denoted by $\mathcal{L}$, is the group of all linear real transformations $\Lambda=(\Lambda^{\mu}_{\ \nu})$ such that

 \begin{equation}\label{2.3}
   L(\Lambda x,\Lambda y)=L(x,y)\ .
 \end{equation}
 \begin{equation}\label{2.4}
   \det\Lambda=1
 \end{equation}
 \begin{equation}\label{2.5}
   \Lambda^{0}_{0}\geq1
 \end{equation}

 The rotation group $SO(3)$ is the orthogonal subgroup of $\mathcal{L}$ that fixes the point $(1,0,0,0)$.

 The inhomogeneous Lorentz group is the group of transformations of $\R^4$ generated by $\mathcal{L}$ and the group of translations isomorphic to $\R^4$ itself. The inhomogeneous group is the semi-direct product of $\mathcal{L}$ and $\R^4$ , denoted by $\mathcal{L}\ltimes\R^4$, with group law given by
\begin{equation}\label{2.6}
  (\Lambda_{1},a_{1})(\Lambda_{2},a_{2})=(\Lambda_{1}\Lambda_{2},a_{1}+ \Lambda_{1}a_{2})\ .
\end{equation}
where $\Lambda_{j}\in\mathcal{L}$ and $a_{j}\in\R^4$, $j=1,2$.

The action of $(\Lambda,a)$ on $\R^4$ is
\begin{equation}\label{2.7}
    (\Lambda,a)x=\Lambda x + a\ .
\end{equation}

According to E.P.Wigner and V.Bargmann ( see \cite{Bargmann1954},\cite{Thaller1992}, \cite{Wightman1960} and \cite{Wigner1939} ), in relativistic quantum mechanics, every projective representation of the inhomogeneous Lorentz group has a lift to an unitary representation of the universal covering group of the inhomogeneous Lorentz group. It is well known that the universal covering group of the inhomogeneous Lorentz group is the semi-direct product of $SL(2,\C)$ and of $\R^4$ with the following group law
\begin{equation}\label{2.8}
    (A,a)(B,b)=(AB,a + \Lambda(A)b)\ .
\end{equation}

 Recall that $SL(2,\C)$ is the group of the $2\times2$ complex matrices A such that $\det(A)=1$ . $\Lambda(A)$ is the image of $A$ in the Lorentz group by the double covering of $\mathcal{L}$
by $SL(2,\C)$ and is defined below.

The usual three Pauli matrices $(\sigma_{1},\sigma_{2},\sigma_{3})$ together with $\sigma_{0}$, the unit $2\times2$ matrix on $\C^2$, generate the $2\times2$ hermitian matrices. We set $\boldsymbol{\sigma}=(\sigma_{1},\sigma_{2},\sigma_{3})$. We identify $\R^4$ with a hermitian matrix by the map
\begin{equation}\label{2.9}
    p=(p^0,\textbf{p})\longmapsto p^{\mu}\sigma_{\mu}=\left(
                                                         \begin{array}{cc}
                                                           p^0+p^3 & p^1-ip^2 \\
                                                           p^1+ip^2 & p^0-p^3 \\
                                                         \end{array}
                                                       \right)
\end{equation}
where $\mu=0,1,2,3.$ and $ p^{\mu}\sigma_{\mu}= p^{0}\sigma_{0}+ p^{1}\sigma_{1}+ p^{2}\sigma_{2}+ p^{3}\sigma_{3}$.

Every $A\in SL(2,\C)$ acts on  $p^{\mu}\sigma_{\mu}$ by
\begin{equation}\label{2.10}
   p^{\mu}\sigma_{\mu}\longmapsto A(p^{\mu}\sigma_{\mu})A^\ast
\end{equation}
and there exists $\Lambda(A)\in \mathcal{L}$ such that
\begin{equation}\label{2.11}
    (\Lambda(A)p)^\mu\sigma_{\mu}=A(p^{\mu}\sigma_{\mu})A^\ast
\end{equation}
with
\begin{equation}\label{2.12}
    \Lambda(A)^\mu_{\ \nu}= \frac{1}{2} Tr(\sigma_{\mu}A\sigma_{\nu}A^\ast)
\end{equation}
The map $A\longmapsto \Lambda(A)$ is a double covering of $\mathcal{L}$ by $SL(2,\C)$ such that  $\Lambda(A)=\Lambda(-A)$.

From now on we call Poincar\'{e} group the universal covering group of the inhomogeneous Lorentz group with the group law defined by \eqref{2.8}. The Poincar\'{e} group is denoted by $\mathcal{P}$.

The subgroup $SU(2)$ of $2\times2$ unitary matrices of $SL(2,\C)$ is the universal covering group of $SO(3)$. The covering map is the restriction of the one of $SL(2,\C)$ to $SU(2)$.

Let $R(\textbf{n},\theta)$ be the rotation of axis $\textbf{n}$ and  angle $\theta\in [0,2\pi)$ in $\R^4$ . We have
\begin{equation}\label{2.13}
    \begin{split}
      &\textbf{x}'= (\cos\theta) \textbf{x} + (1-\cos\theta) (\textbf{x}.\textbf{n})\textbf{n} + \sin\theta ( \textbf{n} \wedge \textbf{x})  \\
      &x'\,^{0}= x^0
    \end{split}
\end{equation}
where $\textbf{x}.\textbf{n}=x^1n^1 + x^2n^2 + x^3n^3$ and $|\textbf{n}|=1$ with $|\textbf{n}|=\sqrt{\textbf{n}.\textbf{n}}$.

The following $2\times2$ matrix
\begin{equation}\label{2.14}
    A(\textbf{n},\theta)=\cos\frac{\theta}{2}\sigma_0 - i\sin\frac{\theta}{2}(\textbf{n}.\boldsymbol{\sigma})=\mathrm{e}^{-i\theta\textbf{n}.\frac{\boldsymbol{\sigma}}{2}}
\end{equation}
is associated with $R(\textbf{n},\theta)$ by the double covering of $\mathcal{L}$ by $SL(2,\C)$. Here $\textbf{n}.\boldsymbol{\sigma}=n^1\sigma_1 + n^2\sigma_2 + n^3\sigma_3$.

Thus $R(\textbf{n},\theta)=\Lambda(A(\textbf{n},\theta)$.

Let $L(\chi,\textbf{m})$ be the pure Lorentz transformation in $\mathcal{L}$ in the direction $\textbf{m}=(m^1,m^2,m^3)$ and with rapidity $v=\tanh\chi$ in the Minkowski space. We have
\begin{equation}\label{2.15}
\begin{split}
  &\textbf{x}'= \textbf{x} + (\cosh\chi-1)(\textbf{x}.\textbf{m})\textbf{m} + x^0(\sinh\chi)\textbf{m}  \\
  &x'\,^{0}=(\cosh\chi)x^0 + (\textbf{x}.\textbf{m})\sinh\chi
\end{split}
\end{equation}
where $\textbf{x}.\textbf{m}=x^1m^1 + x^2m^2 + x^3m^3$ and $|\textbf{m}|=1$.

In $SL(2,\C)$ the following $2\times2$ matrix
\begin{equation}\label{2.16}
    A(\chi,\textbf{m})=\cosh\frac{\chi}{2}\sigma_0 + \sinh\frac{\chi}{2}(\textbf{m}. \boldsymbol{\sigma})=\mathrm{e}^{\chi\textbf{m}.\frac{\boldsymbol{\sigma}}{2}}
\end{equation}
is associated with $L(\chi,\textbf{m})$ by the double covering of $\mathcal{L}$ by $SL(2,\C)$. Here $\textbf{m}.\boldsymbol{\sigma}=m^1\sigma_1 + m^2\sigma_2 + m^3\sigma_3$.

Thus $L(\chi,\textbf{m})=\Lambda(A(\chi,\textbf{m})$.

For $R(\textbf{n},\theta)$  and for $L(\chi,\textbf{m})$ we follow the active point of view of
transformations. See, for example, \cite{Werle1966}.

Note that

\begin{equation}\label{2.17}
    \begin{split}
       A(\textbf{n},\theta)A(\chi,\textbf{m}) &= A(\chi,R(\textbf{n},\theta)\textbf{m}) A(\textbf{n},\theta) \\
       A(\chi,\textbf{m}) A(\textbf{n},\theta) &=  A(\textbf{n},\theta)A(\chi,R(\textbf{n},\theta)^{-1}\textbf{m})
    \end{split}
\end{equation}

In relativistic quantum mechanics elementary systems are associated with unitary irreducible representations of $\mathcal{P}$. From this point of view elementary particles are elementary systems (see \cite{Wightman1960}). It can be also necessary to introduce the extended  Poincar\'{e} group by considering discrete transformations such as space-inversion and time-reversal.

The description of irreducible unitary representations of $\mathcal{P}$ has been first accomplished by E.P.Wigner (see \cite{Wigner1939}). It is now treated as an application of the work of G.W. Mackey using induced representations. Many articles and books have been devoted to this theory. We only mention some of them. See \cite{Varadarajan1970}, \cite{BarutRaczka1986}, \cite{Thaller1992}, \cite{Folland2008} and references therein.

We still keep the realization of the physical irreducible unitary representations of the $\mathcal{P}$ obtained by E.P.Wigner because they are associated with spectral representations of maximal sets of commuting observables as the momenta, the spins or the helicities which are fundamental in dealing with kinematical problems for elementary particles.

We have to consider two cases in physics. First, the case of a positive mass $m >0$ and a spin $j$, with j integer or half-integer, i.e.$j\in\N$ or $j\in\N +\frac{1}{2}$. Second, the case of a mass $m=0$ and a helicity $j\in\Z$ or $j\in\Z +\frac{1}{2}$ for which the spin is $\mid j \mid$. In both cases the energy has to be positive.

\subsection{Positive mass and spin j}\mbox{}

Let $\Omega_m$ be the orbit corresponding to the mass $m>0$. We have
\begin{equation}\label{2.18}
    \Omega_m=\{p\in\R^4 ;\, p_\mu p^\mu=m^2,\, p^0>0\}
\end{equation}
Observe that $p\in\Omega_m$ if and only if $p=(\omega_\textbf{p},\textbf{p})$ where $\omega_\textbf{p}=\sqrt{\mid\textbf{p}\mid^2 + m^2}$.
Here $\mid\textbf{p}\mid=\sqrt{p_1^2 + p_2^2 + p_3^2}$.

The Lorentz invariant measure on $\Omega_m$ is $\frac{\d^3\textbf{p}}{2\omega_\textbf{p}}$.

Set
\begin{equation}\label{2.19}
    k_m=(m,0,0,0)
\end{equation}
The little group of $k_m$ is $SU(2)$ which determines the spin of the particle.

The unitary irreducible representations of $SU(2)$ are finite dimensional ones and they are well known . See, for example, \cite{Edmonds1957}, \cite{Rose1957}, \cite{Liubarski1960}, \cite{Naimark1964} and \cite{vanderWaerden1974}.

 Let $D^j(.)$ be the unitary irreducible representation of $SU(2)$ defined on a Hilbert space of dimension $2j + 1$ that, for simplicity, we suppose to be $\C^{2j + 1}$. The irreducible unitary representation of mass $m>0$ and spin $j$ is defined on the Hilbert space $L^2(\Omega_m,\frac{\d^3\textbf{p}}{2\omega_\textbf{p}},\C^{2j + 1})$ with the scalar product
\begin{equation}\label{2.20}
    (F,G)=\int_{\Omega_m} F(p).G(p)_{2j + 1}\,\frac{\d^3\textbf{p}}{2\omega_\textbf{p}}
\end{equation}
where $\{F,G\}\longmapsto F(p).G(p)_{2j + 1}$ is the scalar product in $\C^{2j + 1}$ which is linear with respect to $G$ and anti-linear with respect to $F$.

The unitary irreducible representation of $\mathcal{P}$ of mass $m>0$ and spin $j$ depends on a field of transformations of the restricted Lorentz group $p \longmapsto \Lambda(A_p)$ such that, for every $p$,
\begin{equation}\label{2.21}
    \Lambda(A_p) k_m=p
\end{equation}
Given the field $p\longmapsto \Lambda(A_p)$, the unitary irreducible representation of the $\mathcal{P}$ of mass $m>0$ and spin $j$, denoted by $U^{[m,j]}( A,a)$, is then
\begin{equation}\label{2.22}
   ( U^{[m,j]}( A,a)F)(p)=\mathrm{e}^{ia.p}D^j(A_p^{-1} A A_{\Lambda(A)^{-1}p}) F(\Lambda(A)^{-1}p)\ .
\end{equation}
where $a.p= a_\mu p^\mu$ and $F\in L^2(\Omega_m,\frac{\d^3\textbf{p}}{2\omega_\textbf{p}},\C^{2j + 1})$ .

In physics one considers two interesting examples of the field $p\longmapsto \Lambda(A_p)$.

\subsubsection{\textbf{The canonical formalism}}\mbox{}

In that case $\Lambda(A_p)$ is the pure Lorentz transformation in the direction $\frac{\textbf{p}}{|\textbf{p}|}$. We then have
\begin{equation}\label{2.23}
\begin{split}
  &\textbf{p}= (\sinh\chi)m \frac{\textbf{p}}{|\textbf{p}|} \\
  & p^0=  (\cosh\chi)m
\end{split}
\end{equation}
This pure Lorentz transformation is associated with the following element of $SL(2,\C)$, denoted by $A^C_p$, by the double covering of $\mathcal{L}$ by $SL(2,\C)$:
\begin{multline}\label{2.24}
   A^C_p= \frac{1}{2}( \sqrt{\frac{p_0 + |\textbf{p}|}{m}} + \sqrt{\frac{p_0 - |\textbf{p}|}{m}})  \\
   +\frac{1}{2}( \sqrt{\frac{p_{0} + |\textbf{p}|}{m}} - \sqrt{\frac{p_{0} - |\textbf{p}|}{m}})(\frac{\textbf{p}.\boldsymbol{\sigma}}{|\textbf{p}|})\ .
\end{multline}

An easy computation shows that
\begin{equation}\label{2.25}
    A^C_p=\frac{(m + p_0) \sigma_0 + \textbf{p}.\boldsymbol{\sigma}}{\sqrt{2m(m + p_0)}}\ .
\end{equation}
For the choice of  $A^C_p$ the corresponding formalism is called canonical. See \cite{Wightman1960}.
\subsubsection{\textbf{The helicity formalism}}\mbox{}

In that case $\Lambda( A_p)$ is the product of a pure Lorentz transformation $\Lambda_{\textbf{p}}^H$ such that
\begin{equation}\label{2.26}
    \Lambda_{\textbf{p}}^Hk_m=(p^0,0,0,\mid\textbf{p}\mid)\ .
\end{equation}
and of a rotation $R_{\textbf{p}}^H$ which carries the third axis into the direction of $\textbf{p}$.

There are several choices for $R_{\textbf{p}}^H$.

Let $\textbf{k}$ is the unit vector of the third axis. For $\frac{\textbf{p}}{|\textbf{p}|}\neq \pm\textbf{k}$  $R_{\textbf{p}}^H$ will be the rotation of axis $\frac{\textbf{k}\wedge\frac{\textbf{p}}{|\textbf{p}|}}{|\textbf{k}\wedge\frac{\textbf{p}}{|\textbf{p}|}|}$ and angle $\theta=(\textbf{k},\frac{\textbf{p}}{|\textbf{p}|})$ with $0\leq\theta\leq\pi$.

Let $B_{\frac{\textbf{p}}{|\textbf{p}|}}$ denote the lift \eqref{2.14} in $SL(2,\C)$ of the rotation $R_{\textbf{p}}^H$ for $\frac{\textbf{p}}{|\textbf{p}|}\neq \pm\textbf{k}$.We have  
 \begin{equation}\label{2.27}
    B_{\frac{\textbf{p}}{|\textbf{p}|}}=\mathrm{e}^{-i\arccos\frac{p^{3}}{|\textbf{p}|}(-\frac{p^{2}}{|\textbf{p}|}\frac{\sigma_{1}}{2}+ \frac{p^{1}}{|\textbf{p}|}\frac{\sigma_{2}}{2})} .
\end{equation}

For  $\frac{\textbf{p}}{|\textbf{p}|}= \pm\textbf{k}$ we set $ B_{\pm \textbf{k}}= I$.

We consider the lift in $SL(2,\C)$ of $\Lambda_{\textbf{p}}^H$ given by \eqref{2.25}.

From now on we consider the lift of $R_{\textbf{p}}^H\Lambda_{\textbf{p}}^H$ in $SL(2,\C)$ denoted by $A^H_p$ and given by

 \begin{equation}\label{2.28}
 \begin{split}
   A^{H}_p=&\mathrm{e}^{-i\arccos\frac{p^{3}}{|\textbf{p}|}(-\frac{p^{2}}{|\textbf{p}|}\frac{\sigma_{1}}{2}+ \frac{p^{1}}{|\textbf{p}|}\frac{\sigma_{2}}{2})} A^{C}_{(p^0,0,0,\mid\textbf{p}\mid)} \\
       &= \frac{1}{(\sqrt{2m(m + p_0)}}\mathrm{e}^{-i\arccos\frac{p^{3}}{|\textbf{p}|}(-\frac{p^{2}}{|\textbf{p}|}\frac{\sigma_{1}}{2}+ \frac{p^{1}}{|\textbf{p}|}\frac{\sigma_{2}}{2})}\left(
                                                                    \begin{array}{cc}
                                                                      \alpha(p) & 0 \\
                                                                      0 & \beta(p) \\
                                                                    \end{array}
                                                                  \right)
   \end{split}
 \end{equation}
Here $\alpha(p)=(m+p^0+|\textbf{p}|)$ and $\beta(p)=(m+p^0-|\textbf{p}|)$.

Note that for a lift of $R_{\textbf{p}}^H\Lambda_{\textbf{p}}^H$ in $SL(2,\C)$ we can also choose
 \begin{equation*}
\begin{split}
 \widetilde{A}^{H}_p=&\mathrm{e}^{-i\varphi\frac{\sigma_{3}}{2}}\mathrm{e}^{-i\theta\frac{\sigma_{2}}{2}} A^{C}_{(p^0,0,0,\mid\textbf{p}\mid)}\\                                                                                                                                   =& \frac{1}{(\sqrt{2m(m + p_0)}} \left(
                                   \begin{array}{cc}
                                     \alpha(p)\mathrm{e}^{-i\frac{\varphi}{2}}\cos\frac{\theta}{2} & -\beta(p)\mathrm{e}^{-i\frac{\varphi}{2}}\sin\frac{\theta}{2} \\
                                     \alpha(p)\mathrm{e}^{i\frac{\varphi}{2}}\sin\frac{\theta}{2} & \beta(p)\mathrm{e}^{i\frac{\varphi}{2}}\cos\frac{\theta}{2} \\
                                   \end{array}
                                 \right).
 \end{split}
\end{equation*}
Here $\theta$ ( resp.$\varphi$) is the polar( resp.azimutal) angle of $\textbf{p}$ with $0\leq\theta\leq\pi$ ( resp. $0\leq\varphi<2\pi$).

The corresponding formalism is called the helicity one. See \cite{JacobWick1959}, \cite{Weinberg1969}, \cite{GuillotPetitI1966} and \cite{Guillot1968}.

Note that $\widetilde{A}^H_p$ is defined up to a rotation of axis $\textbf{k}$. For example we can also choose $\widetilde{A}^{H}_p= \mathrm{e}^{-i\varphi\frac{\sigma_{3}}{2}}\mathrm{e}^{-i\theta\frac{\sigma_{2}}{2}}\mathrm{e}^{i\varphi\frac{\sigma_{3}}{2}} A^{C}_{(p^0,0,0,\mid\textbf{p}\mid)}$.

See \cite{Guillot1968}.

\subsection{Mass m=0 and helicity j}\mbox{}

Let $\Omega$ be the light cone:
\begin{equation}\label{2.29}
    \Omega=\{ p^{\mu}p_{\mu}=0\, ;\, p^0>0 \}\ .
\end{equation}
Set
\begin{equation}\label{2.30}
    k_0=(1,0,0,1)
\end{equation}
The little group of $k_0$ is the spinorial group of the euclidean group in $\R^2$ ,i.e., the group of rigid motions in $\R^2$ denoted by $E_2$.
$E_2$ is the set of motions $(R(\varphi), a)$ in $\R^2$ such that, for $u$ , $v$ and $a$ $\in\R^2$,
\begin{equation}\label{2.31}
    u=(R(\varphi))v + a\ .
\end{equation}
Here $R(\varphi)$ is a rotation of angle $\varphi$ in $\R^2$ whose center is the origin ${0}$.

The group law of $E_2$ is
\begin{equation}\label{2.32}
    (R(\varphi_1) , a_1)(R(\varphi_2) , a_2)=(R(\varphi_1 + \varphi_2) , a_1 + (R(\varphi_1)a_2)\ .
\end{equation}

The spinorial group of $E_2$ is the following set of $2\times2$ matrices:
\begin{equation}\label{2.33}
    \{z,\varphi\}=\left(
                  \begin{array}{cc}
                    \mathrm{e}^{-i\frac{\varphi}{2}} & z \\
                    0 &\mathrm{e}^{i\frac{\varphi}{2}} \\
                  \end{array}
                \right)
\end{equation}
where $z\in\C$ and $\varphi\in\R$ with the group law
\begin{equation}\label{2.34}
    \{z_1,\varphi_1\}\{z_2,\varphi_2\}=\{z_1\mathrm{e}^{i\frac{\varphi_2}{2}} + z_2\mathrm{e}^{-i\frac{\varphi_1}{2}} , \varphi_1 + \varphi_2\}\ .
\end{equation}

The spinorial group of $E_2$ is a double covering of $E_2$. The $2\longmapsto1$ homomorphism of the spinorial group over $E_2$ is given by
\begin{equation}\label{2.35}
    \{z,\varphi\}\longmapsto( R(\varphi), a(z\mathrm{e}^{i\frac{\varphi}{2}}))\ .
\end{equation}
where
\begin{equation}\label{2.36}
    a(z\mathrm{e}^{i\frac{\varphi}{2}})=(\Re(z\mathrm{e}^{i\frac{\varphi}{2}}),\Im(z\mathrm{e}^{\frac{i\varphi}{2}}))\in\R^2 .
\end{equation}
Note that $\{z,\varphi\}$ and $\{-z,\varphi + 2\pi\}$ correspond to the same element in $E_{2}$

The unitary irreducible representations of the spinorial group associated to a finite helicity are of dimension one. They are indexed by $j\in\Z$ or $j\in\Z$ + $\frac{1}{2}$. They are given by
\begin{equation}\label{2.37}
    L^j(\{z,\varphi\})=\mathrm{e}^{-ij\varphi}
\end{equation}
 $j$ is the helicity and $|j|$ is the spin.

\begin{remark}\label{Rem2.1}
The spinorial group of $E_2$ is isomorphic to the group generated by the following set of $2\times2$ matrices:
\begin{equation}\label{2.38}
    \big[z,\varphi\big]=\left(
                  \begin{array}{cc}
                    \mathrm{e}^{-i\frac{\varphi}{2}} & z\mathrm{e}^{i\frac{\varphi}{2}} \\
                    0 &\mathrm{e}^{i\frac{\varphi}{2}} \\
                  \end{array}
                  \right)=
                  \left(
                    \begin{array}{cc}
                      1 & z \\
                      0 & 1 \\
                    \end{array}
                  \right)
                  \left(
                    \begin{array}{cc}
                      \mathrm{e}^{-i\frac{\varphi}{2}} & 0 \\
                      0 & \mathrm{e}^{i\frac{\varphi}{2}} \\
                    \end{array}
                  \right)
\end{equation}

with the group law

\begin{equation}\label{2.39}
    \big[z_1,\varphi_1\big]\big[z_2,\varphi_2\big]=\big[z_1+ z_2\mathrm{e}^{-i\varphi_1} , \varphi_1 + \varphi_2\big]\ .
\end{equation}
\end{remark}

The unitary irreducible representation of $\mathcal{P}$ of m=$0$ and helicity $j$ depends on a field of transformations of the restricted Lorentz group $p\longmapsto\Lambda(A_p)$ ($A_p\in SL(2,\C)$) such that, for every $p\in\Omega$, we have
\begin{equation}\label{2.40}
    \Lambda(A_p)k_0=p\ .
\end{equation}
Given the field $p\rightarrow\Lambda(A_p)$ the unitary irreducible representation of $\mathcal{P}$ of m=$0$ and helicity $j$,
denoted by $U^{[j]}( A, a)$, is given by
\begin{equation}\label{2.41}
    (U^{[j]}( A,a)G)(p)=\mathrm{e}^{ia.p} L^j(A_p^{-1} A A_{\Lambda(A)^{-1}p}) G(\Lambda(A)^{-1}p)\ .
\end{equation}
where $G(.)\in L^2(\Omega,\frac{\d^3\textbf{p}}{2|\textbf{p}|})$\ . Recall that $p=( |\textbf{p}|,\textbf{p})$.

Two important choices of $A_p$ are made in physics.

\subsubsection{\textbf{The canonical formalism}}\mbox{}
This formalism corresponds to the choice made by A.S.Wightman ( see \cite{Wightman1960} and \cite{GuillotPetitI1966}):
\begin{equation}\label{2.42}
    A^1_p=\left(
            \begin{array}{cc}
              \sqrt{\frac{|\textbf{p}| + p^3}{2}} & 0 \\
              \frac{p^1+ip^2}{\sqrt{2(|\textbf{p}| + p^3)}} & \sqrt{\frac{2}{|\textbf{p}| + p^3}} \\
            \end{array}
          \right)
\end{equation}

\subsubsection{\textbf{The helicity formalism}}\mbox{}
In that case $A^2_p$ is the lift in $SL(2,\C)$, associated with \eqref{2.14} and \eqref{2.16}, of the product of a pure Lorentz transformation $\Lambda_{|\textbf{p}|}^H$ such that
\begin{equation}\label{2.43}
    \Lambda_{|\textbf{p}|}^Hk_0=(|\textbf{p}|,0,0,|\textbf{p}|)\ .
\end{equation}
and of the same rotation $R_{p}^H$ as the one defined for a positive mass. By \eqref{2.27} we obtain
\begin{equation}\label{2.44}
    A^2_p= B_{\frac{\textbf{p}}{|\textbf{p}|}}\left(
            \begin{array}{cc}
              |\textbf{p}|^{\frac{1}{2}} & 0
                \\
              0 & |\textbf{p}|^{-\frac{1}{2}} \\
            \end{array}
          \right)
\end{equation}

See \cite{WeinbergII1964}, \cite{WeinbergIII1964}, \cite{GuillotPetitI1966}, \cite{GuillotPetitII1966} and \cite{Guillot1968}\ .

  \begin{remark}\label{Rem2.2}
  The helicity $j$ is Lorentz invariant. Nevertheless note that photons and gravitons have helicity $\pm1$ and $\pm2$ respectively because of the symmetry of space inversion of the electromagnetic and gravitational interactions. On the other hand it is well known that the parity conservation is violated in weak interactions. Thus one has to distinguish the neutrinos with helicity $-\frac{1}{2}$ from the antineutrinos with helicity $\frac{1}{2}$ in the Standard Model. It is conventional to call a particle with helicity $j>0$ right-handed and a particle with helicity $-j$ left-handed.
 \end{remark}

\subsection{The representations of the Poincar\'{e} group in physical spaces}\mbox{}


For most applications to Quantum Field Theory it is more convenient to use the spaces $L^{2}( \R^{3},\C^{2j+1})$ for $m>0$ and $L^{2}(\R^{3})$ for $m=0$ instead of the spaces $L^2(\Omega_m,\frac{\d^3\textbf{p}}{2\omega_\textbf{p}},\C^{2j + 1})$ and $L^2(\Omega,\frac{\d^3\textbf{p}}{2|\textbf{p}|})$ respectively.

The following map
\begin{equation}\label{2.45}
    (V_{m}f)(\textbf{p})= (2\omega_\textbf{p})^{-\frac{1}{2}}f(\omega_\textbf{p},\textbf{p})\ .
\end{equation}
is a unitary map from $L^2(\Omega_m,\frac{\d^3\textbf{p}}{2\omega_\textbf{p}},\C^{2j + 1})$
  onto $L^{2}(\R^{3},\C^{2j+1})$
and
\begin{equation}\label{2.46}
    (V_{0}f)(\textbf{p})= (2\omega_\textbf{p})^{-\frac{1}{2}}f(|\textbf{p}|,\textbf{p})\ .
\end{equation}
 is a unitary map from $L^2(\Omega,\frac{\d^3\textbf{p}}{2|\textbf{p}|})$  onto $L^{2}(\R^{3})$.

We have for both cases
\begin{equation}\label{2.47}
    (V_{\textbf{.}}^{-1}g)(p^{0},\textbf{p})=\sqrt{2p^{0}}g(\textbf{p})\ .
\end{equation}
 where $g(\textbf{p})\in L^{2}( \R^{3},\C^{2j+1})$ when $m>0$ with $p^{0}=\omega_\textbf{p}$ and where $g(\textbf{p})\in L^{2}(\R^{3})$ when $m=0$ with $p^{0}=|\textbf{p}|$.

For any field $p\longmapsto\Lambda(A_p)$ of Lorentz transformations such that, for $m>0$ and $p\in \Omega_{m}$,
\begin{equation}\label{2.48}
    \Lambda(A_p)k_{m}= p
\end{equation}
one easily gets the form of the unitary irreducible representation of $\mathcal{P}$ corresponding the mass $m>0$ and spin $j$ in the space $L^{2}(\R^{3},\C^{2j+1})$.

We obtain, for $f(\textbf{p})$ belonging to $L^{2}(\R^{3},\C^{2j+1})$,
\begin{equation}\label{2.49}
    \begin{split}
       &\{(V_{m}U^{[m,j]}(A,a)V_{m}^{-1})f\}(\textbf{p})  \\
        &=\Big(\frac{\omega_{\textbf{p}_{\Lambda(A)^{-1}p}}}{\omega_{\textbf{p}}}\Big)^{\frac{1}{2}}\mathrm{e}^{ia.p}                       D^j(A_p^{-1} A A_{\Lambda(A)^{-1}p})f(\textbf{p}_{\Lambda(A)^{-1}p})\ .
    \end{split}
\end{equation}
Here $p^{0}=\omega_{\textbf{p}}$, i.e., $p=( \omega_{\textbf{p}}, \textbf{p})$ and $\textbf{p}_{\Lambda(A)^{-1}p}$ is the three-vector part of $\Lambda(A)^{-1}p$ such that
\begin{equation}\label{2.50}
    \Lambda(A)^{-1}p=(\omega_{\textbf{p}_{\Lambda(A)^{-1}p}}\, \textbf{p}_{\Lambda(A)^{-1}p})
\end{equation}
For any field of Lorentz transformations $p\longmapsto\Lambda(A_{p})$ such that
\begin{equation}\label{2.51}
    \Lambda(A_{p})k_{0}=p\ , p\in\Omega.
\end{equation}
 we easily get in a similar way the unitary irreducible representations of $\mathcal{P}$ in the massless case for helicity $j$ in the space
$L^{2}(\R^{3})$.

Thus we obtain, for $g(\textbf{p})$ belonging to $L^{2}(\R^{3})$,
\begin{equation}\label{2.52}
    \begin{split}
       &\{(V_{0}U^{[j]}(A,a)V_{0}^{-1})g\}(\textbf{p})  \\
        &=\Big(\frac{|\textbf{p}_{\Lambda(A)^{-1}p}|}{|\textbf{p}|}\Big)^{\frac{1}{2}}\mathrm{e}^{ia.p} L^j(A_p^{-1} A A_{\Lambda(A)^{-1}p})   g(\textbf{p}_{\Lambda(A)^{-1}p})\ .
    \end{split}
\end{equation}
where $p=(|\textbf{p}|, \textbf{p})$ and $\Lambda(A)^{-1}p=(|\textbf{p}_{\Lambda(A)^{-1}p}|,\textbf{p}_{\Lambda(A)^{-1}p})$.

We now set

\begin{equation}\label{2.53}
  \widetilde{U}^{[m,j]}(A,a)= V_{m}U^{[m,j]}(A,a)V_{m}^{-1}
\end{equation}
\begin{equation}\label{2.54}
  \widetilde{U}^{[j]}(A,a)=V_{0}U^{[j]}(A,a)V_{0}^{-1}
\end{equation}


\begin{remark}\label{Rem2.3}
In \cite{WeinbergI1964}, \cite{WeinbergII1964}, \cite{WeinbergIII1964}, \cite{Weinberg1965}, \cite{Weinberg1969} and \cite{Weinberg1995} the irreducible representations of $\mathcal{P}$ are written down in the space of generalized eigenfunctions of momenta, spins and helicities denoted by $\Psi_{p,\sigma}$ and $\Psi_{p,j}$ respectively. From the mathematical point of view the corresponding space is a subspace of the space of distributions $\mathfrak{D'}(\R^{3},\C^{2j+1})$ for $m>0$ and spin $j$ and of $\mathfrak{D'}(\R^{3})$ for $m=0$ and helicity $j$. From the knowledge of the representations $\widetilde{U}^{[m,j]}$ and $\widetilde{U}^{[j]}$ in the spaces $L^{2}(\R^{3},\C^{2j+1})$ and $L^{2}(\R^{3})$ respectively it is not difficult to get the corresponding representations in the spaces of distributions by duality. For simplicity we keep the same notations $\widetilde{U}^{[m,j]}$ and $\widetilde{U}^{[j]}$ for the representations in the spaces of distributions.

In the massive case we get
\begin{equation}\label{2.55}
    \begin{split}
       &(\widetilde{U}^{[m,j]}(A,a))\Psi_{p,\sigma}=  \\
       &\Big(\frac{\omega_{\textbf{p}_{\Lambda(A)p}}}{\omega_{\textbf{p}}}\Big)^{\frac{1}{2}}\mathrm{e}^{ia.\Lambda(A)p}\sum_{-{j}\leq\sigma'\leq{j}} D^j_{\sigma'\sigma}(A^{-1}_{\Lambda(A)p} A A_{p})\Psi_{\Lambda(A)p,\sigma'}\ .
    \end{split}
\end{equation}
In the massless case, we obtain
\begin{equation}\label{2.56}
    \begin{split}
       &(\widetilde{U}^{[j]}(A,a))\Psi_{p,j}=  \\
       &\Big(\frac{|\textbf{p}_{\Lambda(A)p}|}{|\textbf{p}|}\Big)^{\frac{1}{2}}\mathrm{e}^{ia.\Lambda(A)p} L^j(A^{-1}_{\Lambda(A)p} A A_{p})\Psi_{\Lambda(A)p,j}\ .
    \end{split}
\end{equation}
\eqref{2.55} and \eqref{2.56} are the representations of $\mathcal{P}$ given in \cite[(2.5.23) and (2.5.42)]{Weinberg1995}.
\end{remark}
\begin{remark}\label{Rem2.4}
 Let $\mathrm{\mathbf{P}}=(\mathrm{P}^{1},\mathrm{P}^{2},\mathrm{P}^{3})$ be the momentum three-vector
  and let $\mathrm{\mathbf{J}}=(\mathrm{J}^{1},\mathrm{J}^{2},\mathrm{J}^{3})$ be the angular momentum three-vector. Let us consider the massive case for a given spin $j$. In the canonical formalism we obtain a spectral representation of the maximal set of commuting self-adjoint operators generated by $(
  \mathrm{\mathbf{P}},\mathrm{S}^{3})$ where $\mathrm{S}^{3}$ is $\mathrm{J}^{3}$ in the rest frame of the particle generated by $\Lambda((A^{c}_{p})^{-1})$. In the helicity formalism one gets a spectral representation of the maximal set of commuting self-adjoint operators generated by
 $(\mathrm{\mathbf{P}},\mathrm{H}^{3})$ where $\mathrm{H}^{3}$ is the helicity operator $\big(\sqrt{\sum_{i=1}^{3}(\mathrm{P}^{i})^{2}}\big)^{-1}(\sum_{l=1}^{3}\mathrm{P}^{l}\mathrm{J}^{l})$. $\mathrm{S}^{3}$ and $\mathrm{H}^{3}$ have the same spectrum $(-j,-j+1,\dots,j-1,j)$.
 \end{remark}

\section{Free causal fields for a massive particle of any spin}\label{section3}

\setcounter{equation}{0}

In this chapter we now introduce the construction of unique free causal fields for particles with $m>0$ and spin $j$. For that we follow the formalism of S.Weinberg as described in \cite{WeinbergI1964}, \cite{WeinbergII1964},  \cite{WeinbergIII1964}, \cite{Weinberg1965}, \cite{Weinberg1969} and \cite[chapter 5]{Weinberg1995}. See also \cite{StreaterWightman1964}.

\subsection{Fock spaces for massive particles of any spin}\mbox{}

Consider a particle with mass $m>0$ and spin $j$.

Set
\begin{equation}\label{3.1}
    \Z_{j}=(-j, -j+1,\ldots ,j-1,j)
\end{equation}
and
\begin{equation}\label{3.2}
    \Sigma_{j}=\R^{3}\times\Z_{j}
\end{equation}
In the following $(\textbf{p},s)$ will be the quantum variables for a massive particle of spin $j$ and for both the canonical and helicity formalisms. Here $\textbf{p}\in\R^{3}$ and $s\in\Z_{j}$. In the sequel,we will identify $L^{2}(\R^{3},\C^{2j+1})$ with $L^{2}(\Sigma_{j})$. For simplicity we keep the same notations $\widetilde{U}^{[m,j]}$  for the representations of $\mathcal{P}$ in these two Hilbert spaces.

We shall sometimes use the notations $\xi=(\textbf{p},s)$ and $\int_{\Sigma_j} \d \xi =\sum_{s\in\Sigma_{j}} \int \d^{3}\textbf{p}$.

Let $\gF^{[m,j]}_{s}$ (resp.$\gF^{[m,j]}_{a}$) be the bosonic (resp. fermionic)Fock space for bosons (resp.fermions) of mass $m>0$ and spin $j$.
 We have
 \begin{equation}\label{3.3}
    \gF^{[m,j]}_{s}
 =  \bigoplus_{n=0}^\infty \otimes_s^n
 L^2(\Sigma_j)\ .
\end{equation}
 where $\otimes_s^n$ denotes the symmetric $n$-th tensor product
and $\otimes_s^0 L^2(\Sigma_j)$ =$\C$.

and

\begin{equation}\label{3.4}
   \gF^{[m,j]}_{a}
 = \bigoplus_{n=0}^\infty \otimes_a^n
 L^2(\Sigma_j)\ .
\end{equation}
where $\otimes_a^n$ denotes the antisymmetric $n$-th tensor
product and $\otimes_a^0 L^2(\Sigma_j)$ =$ \C$.

In the case where a massive particle is not its own antiparticle we introduce a Fock space for both the particles and antiparticles denoted by $\widetilde{\mathfrak{F}}^{[m,j]}_{s}$ and $\widetilde{\mathfrak{F}}^{[m,j]}_{a}$ respectively and defined by
\begin{equation}\label{3.5}
    \begin{split}
      & \widetilde{\mathfrak{F}}^{[m,j]}_{s}= \gF^{[m,j]}_{s}\otimes\gF^{[m,j]}_{s}\ ,  \\
      & \widetilde{\mathfrak{F}}^{[m,j]}_{a}= \gF^{[m,j]}_{s}\otimes\gF^{[m,j]}_{a}\ .
    \end{split}
\end{equation}



The unitary irreducible representations  $\widetilde{U}^{[m,j]}$ of $\mathcal{P}$ induce two unitary representations of $\mathcal{P}$ in $\gF^{[m,j]}_{s}$ and $\gF^{[m,j]}_{a}$ which are denoted by $\Gamma(\widetilde{U}^{[m,j]})$ where $\Gamma(.)$ is defined, for example, in \cite[section X.7]{ReedSimon1975}, \cite[4.53]{Folland2008} and \cite[5.48]{Dimock2011}.

The unitary representation of $\mathcal{P}$ in $\widetilde{\mathfrak{F}}^{[m,j]}_{s}$
and $\widetilde{\mathfrak{F}}^{[m,j]}_{a}$ respectively is then $\Gamma(\widetilde{U}^{[m,j]}\otimes\widetilde{U}^{[m,j]})$ .

We now introduce the creation and annihilation operators for bosons and fermions.

$a_{\epsilon}(\xi;m,j)$  (resp. $a^{*}_{\epsilon}(\xi;m,j)$ ) is the annihilation  (resp.creation)  operator for a massive boson of mass $m>0$ and spin $j$ if $\epsilon=+$ and  for the corresponding massive antiparticle if $\epsilon=-$.

In the case where a particle is its own antiparticle $a(\xi;m,j)$ (resp. $a^{*}(\xi;m,j)$ ) is the annihilation (resp. creation ) operator for the corresponding particle.

Similarly, $b_{\epsilon}(\xi;m,j)$ (resp. $b^{*}_{\epsilon}(\xi;m,j)$ ) is the annihilation\\ (resp.creation)  operator for a massive fermion of mass $m>0$ and spin $j$ if $\epsilon=+$ and for the corresponding massive antiparticle if $\epsilon=-$.

In the case where a particle is its own antiparticle $b(\xi;m,j)$ (resp. $b^{*}(\xi;m,j)$ ) is the annihilation (resp. creation ) operator for the corresponding particle.

See \cite[section X.7]{ReedSimon1975}, \cite[section 4.5]{Folland2008}, \cite{BarbarouxGuillot2009} and \cite[section 5.4]{Dimock2011} for the definition of annihilation and creation operators.

The operators $a_{\epsilon}(\xi;m,j)$ and $a^{*}_{\epsilon}(\xi;m,j)$ fulfill the usual commutation relations (CCR), whereas $b_{\epsilon}(\xi;m,j)$ and $b^{*}_{\epsilon}(\xi;m,j)$ fulfill the canonical anticommutation relation (CAR). See \cite{Weinberg1995}. Futhermore, the $a's$ commute with the $b's$.

In addition, in the case where several fermions are involved, we follow the convention described in \cite[sections 4.1 and 4.2]{Weinberg1995}. This means that we will assume that fermionic annihilation and creation operators of different species of particles anticommute ( see \cite[arXiv]{BarbarouxGuillot2009} for explicit definitions ).

Therefore, the following canonical anticommutation and commutation
relations hold for a couple of massive particles with $m>0$ and $m'>0$ and spins $j$ and $j'$,
\begin{equation}\label{3.6}
\begin{split}
 &\{ b_{\epsilon}(\xi;m,j), b^*_{\epsilon'}(\xi';m',j')\} =
 \delta_{\epsilon \epsilon'}\delta_{jj'}\delta_{mm'} \delta(\xi - \xi') \ ,\\
 &[ a_{\epsilon}(\xi;m,j), a^*_{\epsilon'}(\xi';m',j')] =
 \delta_{\epsilon \epsilon'}\delta_{jj'}\delta_{mm'} \delta(\xi - \xi').
 \end{split}
\end{equation}
\begin{equation}\label{3.7}
\begin{split}
&\{ b_{\epsilon}(\xi;m,j), b_{\epsilon'}(\xi';m',j')\} =0 \\
&[ a_{\epsilon}(\xi;m,j), a^*_{\epsilon'}(\xi';m',j')] =0.
\end{split}
\end{equation}
\begin{equation}\label{3.8}
[ b_{ \epsilon}(\xi;m,j), a_{\epsilon'}(\xi';m',j')]
 = [ b_{ \epsilon}(\xi;m,j), a^*_{\epsilon'}(\xi';m',j')]=0
\end{equation}
where $\{b, b'\} = bb' + b'b$ and $[a,a'] = aa' - a'a$.

We now introduce
\begin{equation}\label{3.9}
 a_{ \epsilon}(m,j)(\vp) = \int_{\Sigma_j} a_{
  \epsilon}(\xi;m,j) \overline{\vp(\xi)} \d \xi\ ,
   a^*_{ \epsilon}(m,j)(\vp) = \int_{\Sigma_j} a^*_{
  \epsilon}(\xi;m,j) {\vp(\xi)} \d \xi\ .
 \end{equation}
\begin{equation}\label{3.10}
b_{ \epsilon}(m,j)(\vp) = \int_{\Sigma_j} b_{
  \epsilon}(\xi;m,j) \overline{\vp(\xi)} \d \xi\ ,
    b^*_{ \epsilon}(m,j)(\vp) = \int_{\Sigma_j} b^*_{
  \epsilon}(\xi;m,j) {\vp(\xi)} \d \xi\ .
\end{equation}

We recall that,  for $\vp\in L^2(\Sigma_j)$, the operators $b_{ \epsilon}(m,j)$ and $b^*_{ \epsilon}(m,j)$ are bounded operators on $\gF^{[m,j]}_{a}$ satisfying
\begin{equation}\label{3.11}
  \| b^\sharp_{\epsilon}(m,j)(\vp)\| = \|\vp\|_{L^2}\ .
\end{equation}
where $b^\sharp$ is b or $b^*$.

We now study the transformation rules of the annihilation and creation operators by $\Gamma(\widetilde{U}^{[m,j]})$.

By\cite[Lemma 2.7]{DerezinskiGerard1999}(see also \cite[thm 18]{Derezinski2006} and \cite[4.54]{Folland2008}) we obtain for  $f \in L^{2}(\Sigma_{j})$
\begin{equation}\label{3.12}
   \Gamma(\widetilde{U}^{[m,j]}(A,a)) a^*_{ \epsilon}(m,j)(f)\Gamma(\widetilde{U}^{[m,j]}(A,a))^{-1}=  a^*_{ \epsilon}(m,j)(\widetilde{U}^{[m,j]}(A,a)f)
\end{equation}

We now use the explicit notation $(\textbf{p},s)$ for $\xi$.

Note that, for $B \in SU(2)$ ,
\begin{equation}\label{3.13}
    D^{j}( B )= (D^{j}( B^{-1)}))^{*}
\end{equation}
where $T^{*}$ is the adjoint of the operator $T$.

By \eqref{3.9} and \eqref{3.12} we get
\begin{equation}\label{3.14}
\begin{split}
   &\Gamma(\widetilde{U}^{[m,j]}(A,a)) a^*_{ \epsilon}(m,j)(f)\Gamma(\widetilde{U}^{[m,j]}(A,a))^{-1} \\
    &=\sum_{s}\int \Big(\Gamma(\widetilde{U}^{[m,j]}(A,a)) a^{*}_{\epsilon}(\textbf{p},s; m,j)  \Gamma(\widetilde{U}^{[m,j]}(A,a))^{-1}\Big)f_{s}(\textbf{p})\d^{3}\textbf{p} \\
    &=\sum_{s}\int  a^{*}_{\epsilon}(\textbf{p},s; m,j)\Big(\widetilde{U}^{[m,j]}(A,a)f\Big)_{s}(\textbf{p}) \d^{3}\textbf{p}\ .
\end{split}
\end{equation}

By \eqref{2.49}, \eqref{2.50}, \eqref{3.13}and \eqref{3.14} we easily obtain
\begin{equation}\label{3.15}
    \begin{split}
       &\Gamma(\widetilde{U}^{[m,j]}(A,a)) a^*_{\epsilon}(\textbf{p},s; m,j) \Gamma(\widetilde{U}^{[m,j]}(A,a))^{-1}  \\
        &=\sum_{s'}\Big(\frac{\omega_{\textbf{p}_{\Lambda(A)p}}}{\omega_{\textbf{p}}}\big)^{\frac{1}{2}}\mathrm{e}^{ia.\Lambda(A)p}\overline{ D_{ss'}^j(A_p^{-1} A A_{\Lambda(A)p})}a^*_{\epsilon}(\textbf{p}_{\Lambda(A)p} ,s'; m,j)\ .
    \end{split}
\end{equation}
$\overline{z}$ is the complex conjugate of any comlex number z .

By taking the adjoint of \eqref{3.15} we get
\begin{equation}\label{3.16}
    \begin{split}
       &\Gamma(\widetilde{U}^{[m,j]}(A,a)) a_{\epsilon}(\textbf{p},s; m,j) \Gamma(\widetilde{U}^{[m,j]}(A,a))^{-1}  \\
        &=\sum_{s'}\Big(\frac{\omega_{\textbf{p}_{\Lambda(A)p}}}{\omega_{\textbf{p}}}\Big)^{\frac{1}{2}}\mathrm{e}^{-ia.\Lambda(A)p} D_{ss'}^j(A_p^{-1} A A_{\Lambda(A)p})a_{\epsilon}(\textbf{p}_{\Lambda(A)p} ,s'; m,j)\ .
    \end{split}
\end{equation}

\eqref{3.15} and \eqref{3.16} are the equations \cite[(5.1.12) and (5.1.11)]{Weinberg1995} written down with our choice of the space-time metric \eqref{2.2} instead of the one used by S.Weinberg in \cite{Weinberg1995}.

By \cite[thm 18]{Derezinski2006} we also have
\begin{equation}\label{3.17}
    \begin{split}
       &\Gamma(\widetilde{U}^{[m,j]}(A,a)) b^*_{\epsilon}(\textbf{p},s; m,j) \Gamma(\widetilde{U}^{[m,j]}(A,a))^{-1}  \\
        &=\sum_{s'}\Big(\frac{\omega_{\textbf{p}_{\Lambda(A)p}}}{\omega_{\textbf{p}}}\Big)^{\frac{1}{2}}\mathrm{e}^{ia.\Lambda(A)p}\overline{ D_{ss'}^j(A_p^{-1} A A_{\Lambda(A)p})}b^*_{\epsilon}(\textbf{p}_{\Lambda(A)p} ,s'; m,j)\ .
    \end{split}
\end{equation}
and
\begin{equation}\label{3.18}
    \begin{split}
       &\Gamma(\widetilde{U}^{[m,j]}(A,a)) b_{\epsilon}(\textbf{p},s; m,j) \Gamma(\widetilde{U}^{[m,j]}(A,a))^{-1}  \\
        &=\sum_{s'}\Big(\frac{\omega_{\textbf{p}_{\Lambda(A)p}}}{\omega_{\textbf{p}}}\Big)^{\frac{1}{2}}\mathrm{e}^{-ia.\Lambda(A)p} D_{ss'}^j(A_p^{-1} A A_{\Lambda(A)p})b_{\epsilon}(\textbf{p}_{\Lambda(A)p} ,s'; m,j)\ .
    \end{split}
\end{equation}

Note that, in \eqref{3.12},\eqref{3.14},\eqref{3.15}, \eqref{3.16}, \eqref{3.17} and \eqref{3.18}, $A_{p}$ is $A^{C}_{p}$ or $A^{H}_{p}$ depending on the formalism we consider.
It is important to remark that the operators of creation and annihilation both in the canonical and helicity formalism depend on the formalism we consider.

We further note that
\begin{equation}\label{3.19}
    ^{C}\widetilde{U}^{[m,j]}(A,a)= D^{j}(A^{C-1}_{\textbf{.}}A^{H}_{\textbf{.}})^{H}\widetilde{U}^{[m,j]}(A,a)D^{j}(A^{C-1}_{\textbf{.}}A^{H}_{\textbf{.}})^{-1}
\end{equation}

In view of \eqref{2.17} and \eqref{2.27} we get for $\frac{\textbf{p}}{|\textbf{p}|}\neq \pm\textbf{k}$
\begin{equation}\label{3.20}
    A(\chi,\frac{\textbf{p}}{|\textbf{p}|})=B_{\frac{\textbf{p}}{|\textbf{p}|}}A(\chi,\textbf{k})B_{\frac{\textbf{p}}{|\textbf{p}|}}^{-1}
\end{equation}


Combining \eqref{3.20} with \eqref{3.19} we obtain
\begin{equation}\label{3.21}
    \begin{split}
     &a^{C*}_{\epsilon}(\textbf{p},s; m,j)=\sum_{s'}D_{s's}^{j}\big((B_{\frac{\textbf{p}}{|\textbf{p}|}})^{-1}\big)a^{H*}_{\epsilon}(\textbf{p},s'; m,j) \\
     &a^{C}_{\epsilon}(\textbf{p},s; m,j)=\sum_{s'}D_{s's}^{j}(B_{\frac{\textbf{p}}{|\textbf{p}|}})a^{H}_{\epsilon}(\textbf{p},s'; m,j)\ .
     \end{split}
\end{equation}
\begin{equation}\label{3.22}
    \begin{split}
     &a^{H*}_{\epsilon}(\textbf{p},s; m,j)=\sum_{s'}D_{s's}^{j}(B_{\frac{\textbf{p}}{|\textbf{p}|}})a^{C*}_{\epsilon}(\textbf{p},s'; m,j) \\
     &a^{H}_{\epsilon}(\textbf{p},s; m,j)=\sum_{s'}D_{s's}^{j}\big((B_{\frac{\textbf{p}}{|\textbf{p}|}})^{-1}\big)a^{C}_{\epsilon}(\textbf{p},s'; m,j)\ .
    \end{split}
\end{equation}
and likewise for $ b^{*}_{\epsilon}(\textbf{p},s; m,j)$ and  $ b_{\epsilon}(\textbf{p},s; m,j)$.

In the following we will omit the superscripts $C$ and $H$ for $a^{\sharp}$ and $b^{\sharp}$  when the formalism that we are using is well determined.

The construction of free causal fields associated with a massive particle of spin $j$ depends on the knowledge of the irreducible finite dimensional representations of $SL(2,\C)$ that we now study.

\subsection{The irreducible finite dimensional representations of $SL(2,\C)$}\mbox{}

These representations are well known. See, for example, \cite{Naimark1964} and \cite{vanderWaerden1974}. Once again we shall follow the method used by S.Weinberg ( see \cite[subsection 5.6]{Weinberg1995} ) in order to construct such representations.

Let us recall the Lie algebra of $SL(2,\C)$.

Let $\mathrm{J}_{i}$, $i=1,2,3$, be the generators of the rotations as defined as follows. For the lift \eqref{2.14} in $SL(2,\C)$ of the rotation of axis $\textbf{n}=(n^{1},n^{2},n^{3})$ and angle $\theta$ we have
\begin{equation}\label{3.23}
 A(\textbf{n},\theta)= \mathrm{e}^{-i\theta(\sum_{l=1}^{3}n^{l}\mathrm{J}_{l})}\ \mbox{with}\  \mathrm{J}_{l}= \frac{\sigma_{l}}{2}.
\end{equation}

Let $\mathrm{K}_{i}$, $i=1,2,3$,  be the generators of the pure Lorentz transformations as defined as follows. For the lift \eqref{2.16} in  $SL(2,\C)$ of the  pure Lorentz transformation in the direction $\textbf{m}=(m^{1},m^{2},m^{3})$ and with rapidity $v=th\chi$ we have
\begin{equation}\label{3.24}
 A(\chi,\textbf{m})= \mathrm{e}^{-i\chi(\sum_{l=1}^{3}m^{l}\mathrm{K}_{l})}\ \mbox{with}\ \mathrm{K}_{l}= i\frac{\sigma_{l}}{2}.
\end{equation}

 We have
 \begin{equation}\label{3.25}
    \begin{split}
       & [\mathrm{J}_{i} , \mathrm{J}_{j}]= i\epsilon_{ijk}\mathrm{J}_{k}\ . \\
       & [\mathrm{J}_{i} , \mathrm{K}_{j}]= i\epsilon_{ijk}\mathrm{K}_{k}\ . \\
       & [\mathrm{K}_{i} , \mathrm{K}_{j}]= -i\epsilon_{ijk}\mathrm{J}_{k}\ .
    \end{split}
 \end{equation}
 where $\epsilon_{ijk}$ is totally antisymmetric with $\epsilon_{123}=+1$ .

 $\mathrm{J}_{i}$, $i=1,2,3$ and $\mathrm{K}_{i}$, $i=1,2,3$ generate the Lie algebra of $SL(2,\C)$.

In any linear finite dimensional representation of the Lie algebra of $SL(2,\C)$ we denote $\widetilde{\mathrm{J}}_{i}$ and $\widetilde{\mathrm{K}}_{j}$, $(i,j=1,2,3)$, the representation of $\mathrm{J}_{i}$ and $\mathrm{K}_{j}$, $(i,j=1,2,3)$ . $\widetilde{\mathrm{J}}_{i}$ and $\widetilde{\mathrm{K}}_{j}$, $(i,j=1,2,3)$ satisfy \eqref{3.25} and generate the representation of the Lie algebra of $SL(2,\C)$. $\mathrm{e}^{-i\theta(\sum_{l=1}^{3}n^{l}\widetilde{\mathrm{J}}_{l})}$ is then the representation of a rotation of axis $\textbf{n}=(n^{1},n^{2},n^{3})$ and angle $\theta$ and $\mathrm{e}^{-i\chi(\sum_{l=1}^{3}m^{l}\widetilde{\mathrm{K}}_{l})}$ is the representation of the  pure Lorentz transformation in the direction $\textbf{m}=(m^{1},m^{2},m^{3})$ and with rapidity $v=th\chi$.


 Set

\begin{equation}\label{3.26}
 \begin{split}
     &\mathrm{M}_{ij}=-\mathrm{M}_{ji}=\epsilon_{ijk} \mathrm{J}_{k} \\
     &\mathrm{M}_{i0}=-\mathrm{M}_{0i}= \mathrm{K}_{i} \\
     &\mathrm{M}_{00}=\mathrm{M}_{ii}=0 .
     \end{split}
 \end{equation}

Equations \eqref{3.25} and \eqref{3.26} then read
 \begin{equation}\label{3.27}
    \big[\mathrm{M}_{\mu\nu},\mathrm{M}_{\rho\sigma}\big]= i\big(g_{\mu\sigma}\mathrm{M}_{\nu\varrho}+g_{\nu\rho}\mathrm{M}_{\mu\sigma}-g_{\nu\sigma}\mathrm{M}_{\mu\rho}-g_{\mu\rho}\mathrm{M}_{\nu\sigma}\big).
\end{equation}
where $\mu$,$\nu$,$\rho$ and $\sigma$ run over the values 0,1,2,3.


 The generators $\mathrm{M}_{\mu\nu}$, $(\mu,\nu =0,1,2,3 )$, satisfying \eqref{3.27}  generate also the Lie algebra of $SL(2,\C)$ .


  Any $A\in SL(2,\C)$ can be written down in the following form
 \begin{equation}\label{3.28}
   A= \mathrm{e}^{-\frac{i}{2}\omega^{\mu\nu}\mathrm{M}_{\mu\nu}}
 \end{equation}
where
\begin{equation}\label{3.29}
    \begin{split}
     &\omega^{\mu\nu}=-\omega^{\nu\mu}\in\R \\
     &\mathrm{M}_{ij}=-\mathrm{M}_{ji}=\epsilon_{ijk}\frac{\sigma_{k}}{2} \\
     &\mathrm{M}_{i0}=-\mathrm{M}_{0i}=i\frac{\sigma_{i}}{2} .
     \end{split}
 \end{equation}
 In the case of the lift \eqref{2.14} in $SL(2,\C)$ of a rotation of axis $\textbf{n}$ and angle $\theta$, we have
 \begin{equation}\label{3.30}
    \omega^{ij}=\epsilon_{ijk}n^{k}\theta\ ,   \omega^{i0}=0 .
 \end{equation}
and,in the case of the lift \eqref{2.16} in $SL(2,\C)$ of a pure Lorentz transformation in the direction $\textbf{m}$ and with rapidity $v=\tanh\chi$, we have
 \begin{equation}\tag{\ref{3.30}$a$}
\omega^{i0 }= m^{i}\chi\ ,  \omega^{ij}=0 .
 \end{equation}

 By \eqref{3.29} we have
 \begin{equation}\label{3.31}
  A= \mathrm{e}^{-i\sum_{k=1}^{3}(\chi_{k}\mathrm{K}_{k}+ \theta_{k}\mathrm{J}_{k}}=\mathrm{e}^{i\sum_{k=1}^{3}(\chi_{k}-i\theta_{k})\frac{\sigma_{k}}{2}}
 \end{equation}
 with
\begin{equation}\tag{\ref{3.31}$a$}
\omega^{k0 }=\chi_{k}\ ,  \omega^{ij}\epsilon_{ijk}=\theta_{k} .                                                                                                                                                                                                                                                                               \end{equation}

 For any $A\in SL(2,\C)$ we also get an another representation by using the polar decomposition. We have the following unique decomposition $A=|A^*|U$ where $U$ is unitary and $|A|=\sqrt{AA^*}$ is self-adjoint (see\cite{Kato1966}). Furthermore $\Lambda(A)$ is a rotation if and only if $A$ is unitary and $\Lambda(A)$ is a Lorentz transformation if and only if $A$ is self-adjoint (see\cite{Wightman1960}). For every $A$ we have $\Lambda(A)=\Lambda(A_1)\Lambda(A_2)$ where $A_1$ is self-adjoint and $A_2$ is unitary.
Therefore it follows from \eqref{2.14} and \eqref{2.16} that we have, for any $A\in SL(2,\C)$,

\begin{equation}\label{3.32}
    A=\mathrm{e}^{\tilde{\chi}(\sum_{l=1}^{3}\tilde{m}^{l}\frac{\sigma_{l}}{2})}\mathrm{e}^{-i\tilde{\theta}(\sum_{l=1}^{3}\tilde{n}^{l}\frac{\sigma_{l}}{2})}
\end{equation}
for some $\tilde{\chi}$,$\tilde{m}^{i}$,$\tilde{\theta}$ and $\tilde{n}^{k}$ depending on $A$.

We now introduce for $j=(1,2,3)$ ,
 \begin{equation}\label{3.33}
    \begin{split}
       & \mathcal{A}_{j}=\frac{1}{2}(\mathrm{J}_{j} + i\mathrm{K}_{j})\ . \\
       & \mathcal{B}_{j}=\frac{1}{2}(\mathrm{J}_{j} - i\mathrm{K}_{j})\ .
    \end{split}
 \end{equation}

 We have
 \begin{equation}\label{3.34}
    \begin{split}
       & [\mathcal{A}_{i} , \mathcal{A}_{j}]= i\epsilon_{ijk}\mathcal{A}_{k}\ . \\
       & [\mathcal{B}_{i} , \mathcal{B}_{j}]= i\epsilon_{ijk}\mathcal{B}_{k}\ . \\
       & [\mathcal{A}_{i} , \mathcal{B}_{j}]=0\ .
    \end{split}
 \end{equation}

 By \eqref{3.34} the irreducible finite dimensional representations of $SL(2,\C)$ are characterized by a couple of two positive integers and/or half-integers $(J_{1},J_{2})$ representing the spins of two uncoupled particles. The generators of the spin $J_{1}$ are denoted by $\mathcal{J}_{1}^{(1)}$,
 $\mathcal{J}_{2}^{(1)}$, $\mathcal{J}_{3}^{(1)}$ and likewise for the spin $J_{2}$. The associated representation of $SL(2,\C)$ will be denoted $D^{[J_{1},J_{2}]}(.)$ where $D^{[J_{1},J_{2}]}(A)$ is a matrix defined on $\C^{(2J_{1} + 1)(2J_{2} + 1)}$. See \cite{Edmonds1957}.

 $\mathcal{J}_{\textbf{.}}^{(1)}$ are represented by the standard spin matrices for spin $J^{1}$. We have
 \begin{equation}\label{3.35}
    \begin{split}
       & (\mathcal{J}_{3}^{(1)})_{M_{1},M'_{1}}= M_{1}\delta_{M_{1},M'_{1}} \\
       & (\mathcal{J}_{1}^{(1)}\pm i\mathcal{J}_{2}^{(1)})_{M_{1},M'_{1}}= \delta_{M_{1},M'_{1}\pm1}\sqrt{J_{1}(J_{1} + 1)-M'_{1}(M'_{1} \pm 1)}\ .
    \end{split}
 \end{equation}
where $M_{1},M'_{1}\in (-J_{1},-J_{1} + 1,\dots,J_{1}- 1,J_{1})$ and likewise for $\mathcal{J}_{\textbf{.}}^{(2)}$ .

The matrices of $\mathcal{A}_{\textbf{.}}$ and $\mathcal{B}_{\textbf{.}}$ with respect to the tensor product of the canonical basis for the spins $J_{1}$ and $J_{2}$ are now given by
\begin{equation}\label{3.36}
    \begin{split}
       & (\mathcal{A}_{\textbf{.}})_{M_{1}M_{2},M'_{1}M'_{2}}= \delta_{M_{2},M'_{2}}(\mathcal{J}^{(1)}_{\textbf{.}})_{M_{1},M'_{1}}\ , \\
       & (\mathcal{B}_{\textbf{.}})_{M_{1}M_{2},M'_{1}M'_{2}}= \delta_{M_{1},M'_{1}}(\mathcal{J}^{(2)}_{\textbf{.}})_{M_{2},M'_{2}}\ .
    \end{split}
\end{equation}

We have then get the following representation of the generators of the Lie algebra of $SL(2,\C)$
\begin{equation}\label{3.37}
    \begin{split}
       & (\widetilde{\mathrm{J}}_{\textbf{.}})_{M_{1}M_{2},M'_{1}M'_{2}}= \delta_{M_{2},M'_{2}}(\mathcal{J}^{(1)}_{\textbf{.}})_{M_{1},M'_{1}} + \delta_{M_{1},M'_{1}}(\mathcal{J}^{(2)}_{\textbf{.}})_{M_{2},M'_{2}}\ . \\
       & (\widetilde{\mathrm{K}}_{\textbf{.}})_{M_{1}M_{2},M'_{1}M'_{2}}=-i\delta_{M_{2},M'_{2}}(\mathcal{J}^{(1)}_{\textbf{.}})_{M_{1},M'_{1}} + i\delta_{M_{1},M'_{1}}(\mathcal{J}^{(2)}_{\textbf{.}})_{M_{2},M'_{2}}\ .
    \end{split}
\end{equation}

Note that, for $B$ $\in$ $SU(2)$, we have
\begin{equation}\label{3.38}
    D^{[J_{1},J_{2}]}_{M_{1}M_{2},M'_{1}M'_{2}}(B)= D^{J_{1}}_{M_{1},M'_{1}}(B) D^{J_{2}}_{M_{2},M'_{2}}(B)\ .
\end{equation}

Note that
\begin{equation}\label{3.39}
  \begin{split}
     D^{[\frac{1}{2},0]}(A) &= =(A^*)^{-1}=\mathrm{e}^{-\chi(\sum_{l=1}^{3}m^{l}\frac{\sigma_{l}}{2})}\mathrm{e}^{-i\theta(\sum_{l=1}^{3}n^{l}\frac{\sigma_{l}}{2})}  \\
     D^{[0,\frac{1}{2}]} (A)&= A=\mathrm{e}^{\chi(\sum_{l=1}^{3}m^{l}\frac{\sigma_{l}}{2})}\mathrm{e}^{-i\theta(\sum_{l=1}^{3}n^{l}\frac{\sigma_{l}}{2})}
  \end{split}
\end{equation}

Recall that $A\longmapsto(A^*)^{-1}$ is an automorphism of $SL(2,\C)$.

\subsubsection{\textbf{Computation of $D^{[J_{1},J_{2}]}(A^{C}_{p})$}}\mbox{}

It follows from \eqref{2.23} that
\begin{equation}\label{3.40}
 D^{[J_{1},J_{2}]}(A^{C}_{p})=D^{[J_{1},J_{2}]}(\mathrm{e}^{-i\chi\sum_{l=1}^{3}\frac{p^{l}}{|\textbf{p}|}\widetilde{\mathrm{K}}_{l}}).
 \end{equation}
where, by \eqref{2.23}, we have $\mathrm{e}^{\chi}= \frac{|\textbf{p}|+\omega_{\textbf{p}}}{m}$ .

By \eqref{3.35} and \eqref{3.37} we now get
\begin{equation}\label{3.41}
\begin{split}
   & D^{[J_{1},J_{2}]}_{M_{1}M_{2},M'_{1}M'_{2}}(A^{C}_{p})= \\
&(\mathrm{e}^{-(\mathrm{Ln}\frac{|\textbf{p}|+\omega_{\textbf{p}}}{m})\sum_{l=1}^{3}\frac{p^{l}}{|\textbf{p}|}\mathcal{J}^{(1)}_{l}})_{M_{1}M'_{1}}  (\mathrm{e}^{(\mathrm{Ln}\frac{|\textbf{p}|+\omega_{\textbf{p}}}{m})\sum_{l=1}^{3}\frac{p^{l}}{|\textbf{p}|}\mathcal{J}^{(2)}_{l}})_{M_{2}M'_{2}}\ .                                                   \end{split}
\end{equation}

\subsubsection{\textbf{Computation of $D^{[J_{1},J_{2}]}(A^{H}_{p})$}}\mbox{}

  For $B_{\frac{\textbf{p}}{|\textbf{p}|}}$ we choose the lift \eqref{2.14} in $SL(2,\C)$ of the rotation  of axis $\textbf{k}\wedge\frac{\textbf{p}}{|\textbf{p}|}$ and angle $\theta$=$(\textbf{k},\frac{\textbf{p}}{|\textbf{p}|})$ where $\textbf{k}$ is the unit vector of the third axis.

   By \eqref{3.27},  \eqref{3.28}, \eqref{3.29} and \eqref{3.30} we get
 \begin{equation}\label{3.42}
    B_{\frac{\textbf{p}}{|\textbf{p}|}}=\mathrm{e}^{-i\arccos\frac{p^{3}}{|\textbf{p}|}(-\frac{p^{2}}{|\textbf{p}|}\frac{\sigma_{1}}{2}+ \frac{p^{1}}{|\textbf{p}|}\frac{\sigma_{2}}{2})} .
\end{equation}
According to the helicity formalism we have
\begin{equation}\label{3.43}
    A^{H}_{p}= B_{\frac{\textbf{p}}{|\textbf{p}|}}A^{C}_{(p^{0},0,0,|\textbf{p}|)}
\end{equation}
 By \eqref{3.41} we obtain
\begin{equation}\label{3.44}
   D^{[J_{1},J_{2}]}_{M_{1}M_{2}M'_{1}M'_{2}}(A^{C}_{(p^{0},0,0,|\textbf{p}|)})=
       (\frac{|\textbf{p}|+\omega_{\textbf{p}}}{m})^{M'_{2}-M'_{1}}\delta_{M_{1}M'_{1}}\delta_{M_{2}M'_{2}}\ .
\end{equation}
By \eqref{3.38}, \eqref{3.42} and \eqref{3.44} we finally get
\begin{equation}\label{3.45}
    \begin{split}
       &D^{[J_{1},J_{2}]}_{M_{1}M_{2},M'_{1}M'_{2}}(A^{H}_{p})= \\
       & D^{J_{1}}_{M_{1},M'_{1}}(B_{\frac{\textbf{p}}{|\textbf{p}|}}) D^{J_{2}}_{M_{2},M'_{2}}(B_{\frac{\textbf{p}}{|\textbf{p}|}})(\frac{|\textbf{p}|+\omega_{\textbf{p}}}{m})^{M'_{2}-M'_{1}}\ .
    \end{split}
\end{equation}
where
\begin{equation}\label{3.46}
    D^{J_{l}}(B_{\frac{\textbf{p}}{|\textbf{p}|}})=\mathrm{e}^{-i\arccos\frac{p^{3}}{|\textbf{p}|}(-\frac{p^{2}}{|\textbf{p}|}\mathcal{J}^{(l)}_{1}+ \frac{p^{1}}{|\textbf{p}|}\mathcal{J}^{(l)}_{2})}\ ,l=1,2 .
\end{equation}

\subsection{ Free causal fields for a massive particle of spin $j$}\mbox{}

Consider a particle of mass $m>0$ and spin $j$. Let $(J_{1},J_{2})$ be two spins such that
\begin{equation}\label{3.47}
    |J_{1}-J_{2}| \leq j \leq J_{1}+J_{2} \ .
\end{equation}

One can prove the existence of unique causal free fields denoted by
\begin{equation*}
   \Big(^{\sharp}\Psi^{[J_{1},J_{2}]\epsilon}_{M_{1}M_{2}}(x)\Big)_{M_{1}M_{2}}
\end{equation*}
where $M_{1}\in (-J_{1},-J_{1} + 1,\dots,J_{1}- 1,J_{1})$ and $M_{2}\in (-J_{2},-J_{2} + 1,\dots,J_{2}- 1,J_{2})$  and where $\sharp$=$C$ or $H$ and $\epsilon=\pm$, involving particles and antiparticles. $C$ is for the canonical formalism and $H$ is for the helicity formalism.

Set
\begin{equation}\label{3.48}
   ^{\sharp} \widetilde{V}^{[m,j]}(A,a)=^{\sharp}\widetilde{U}^{[m,j]}(A,a)\oplus^{\sharp}\widetilde{U}^{[m,j]}(A,a)
\end{equation}

The causal free fields have to satisfy the two fundamental conditions:

(a)The relativistic covariance law:
\begin{equation}\label{3.49}
    \begin{split}
       &\big(\Gamma(^{\sharp}\widetilde{V}^{[m,j]}(A,a))\big)(^{\sharp}\Psi^{[J_{1},J_{2}]\epsilon}_{M_{1}M_{2}})(x)\big( \Gamma(^{\sharp}\widetilde{V}^{[m,j]}(A,a))\big)^{-1} \\
       & =\sum_{M'_{1}M'_{2}}D^{[J_{1},J_{2}]}_{M_{1}M_{2}M'_{1}M'_{2}}(A^{-1})(^{\sharp}\Psi^{[J_{1},J_{2}]\epsilon}_{M'_{1}M'_{2}})(\Lambda(A)x+ a)\ .
    \end{split}
\end{equation}
where $x\in \R^4$.

(b)The microscopic causality in the bosonic case:
\begin{equation}\label{3.50}
    \begin{split}
       & [^{\sharp}\Psi^{[J_{1},J_{2}]\epsilon}_{M_{1}M_{2}}(x),^{\sharp}\Psi^{[J_{1},J_{2}]\epsilon}_{M'_{1}M'_{2}}(y)]=  \\
       & =[^{\sharp}\Psi^{[J_{1},J_{2}]\epsilon}_{M_{1}M_{2}}(x),^{\sharp}\Psi^{[J_{1},J_{2}]\epsilon\dag}_{M'_{1}M'_{2}}(y)]=0 \ ,
    \end{split}
\end{equation}

and

(c)The microscopic causality in the fermionic case:
\begin{equation}\label{3.51}
    \begin{split}
       &\{^{\sharp}\Psi^{[J_{1},J_{2}]\epsilon}_{M_{1}M_{2}}(x),^{\sharp}\Psi^{[J_{1},J_{2}]\epsilon}_{M'_{1}M'_{2}}(y)\}=  \\
       & =\{^{\sharp}\Psi^{[J_{1},J_{2}]\epsilon}_{M_{1}M_{2}}(x),^{\sharp}\Psi^{[J_{1},J_{2}]\epsilon\dag}_{M'_{1}M'_{2}}(y)\}=0 \ .
    \end{split}
\end{equation}
for x-y space-like.

From now on we restrict ourselves to the case of a massive boson of spin $j$. We suppose that the massive boson is not its own antiparticle. The case of a massive fermion is strictly similar and we shall omit the details. Moreover when a particle is its own antiparticle the results are an easy consequence of what it follows.

Mimicking \cite[chapter5]{Weinberg1995} we set
\begin{equation}\label{3.52}
    \begin{split}
       & (^{\sharp}_{1}\Upsilon^{[J_{1},J_{2}]\epsilon}_{ M_{1}M_{2}})(x) \\
       & =\sum_{s}\int \d^{3}\textbf{p}(^{\sharp}u_{ M_{1}M_{2}}^{[J_{1},J_{2}]})(x;\textbf{p},s;m,j)a_{\epsilon}(\textbf{p},s;m,j)\ .
    \end{split}
\end{equation}

and
\begin{equation}\label{3.53}
    \begin{split}
       & (^{\sharp}_{2}\Upsilon^{[J_{1},J_{2}]\epsilon'}_{ M_{1}M_{2}})(x) \\
       & =\sum_{s}\int \d^{3}\textbf{p}(^{\sharp}v_{ M_{1}M_{2}}^{[J_{1},J_{2}]})(x;\textbf{p},s;m,j)a_{\epsilon'}^{*}(\textbf{p},s;m,j)\ .
    \end{split}
\end{equation}
Here $\epsilon\neq\epsilon'$. For simplicity we have omit the superscript $\sharp$ for the creation and annihilation operators which depend on the formalism we consider.

$(^{\sharp}_{1}\Upsilon^{[J_{1},J_{2}]\epsilon}_{ M_{1}M_{2}})(x)$ and $(^{\sharp}_{2}\Upsilon^{[J_{1},J_{2}]\epsilon'}_{ M_{1}M_{2}})(x)$ are supposed to satisfy \eqref{3.49}.

For simplicity we also omit the superscripts $[J_{1},J_{2}]\epsilon$, $[J_{1},J_{2}]\epsilon'$ and $[J_{1},J_{2}]$. We will finally give the complete formulae later.

Combining \eqref{3.15} and \eqref{3.16} with \eqref{3.42}, \eqref{3.46}, \eqref{3.47} and \eqref{3.50} we obtain
\begin{equation}\label{3.54}
    \begin{split}
    &\Big(\frac{p^{0}}{(\Lambda(A)p)^{0}}\Big)^{\frac{1}{2}}\sum_{M'_{1}M'_{2}}D_{M_{1}M_{2}M'_{1}M'_{2}}(A)(\mathrm{e}^{-ia.\Lambda(A)p})  (^{\sharp}u_{M'_{1}M'_{2}})(x;\textbf{p},s;m,j)\\ &=\sum_{s'}D^{j}_{s's}(A^{\sharp-1}_{\Lambda(A)p)}A A^{\sharp}_{p})                                    (^{\sharp}u_{ M_{1}M_{2}})(\Lambda(A)x+ a; \textbf{p}_{\Lambda(A)p},s';m,j)\ .
    \end{split}
\end{equation}

and

\begin{equation}\label{3.55}
    \begin{split}
    &\Big(\frac{p^{0}}{(\Lambda(A)p)^{0}}\Big)^{\frac{1}{2}}\sum_{M'_{1}M'_{2}}D_{M_{1}M_{2}M'_{1}M'_{2}}(A)(\mathrm{e}^{ia.\Lambda(A)p})   (^{\sharp}v_{M'_{1}M'_{2}})(x;\textbf{p},s;m,j)\\ &=\sum_{s'}(\overline{D^{j}_{s's}(A^{\sharp-1}_{\Lambda(A)p)}A A^{\sharp}_{p})})                       (^{\sharp}v_{ M_{1}M_{2}})(\Lambda(A)x+ a; \textbf{p}_{\Lambda(A)p},s';m,j)\ .
    \end{split}
\end{equation}

By \eqref{3.51} and \eqref{3.52} with $A=1$ and for any $a\in\R^{3}$,  $^{\sharp}u_{M_{1}M_{2}}(x;\textbf{p},s;m,j)$ \\ and $^{\sharp}v_{M_{1}M_{2}}(x;\textbf{p},s;m,j)$
have the form $e^{-ia.x}(^{\sharp}u_{M_{1}M_{2}})(\textbf{p},s;m,j)$\\ and $e^{ia.x}(^{\sharp}v_{M_{1}M_{2}})(\textbf{p},s;m,j)$ respectively .

Following the convention in Physics we set (see \cite[chapter5]{Weinberg1995})

\begin{equation}\label{3.56}
  (^{\sharp}u_{ M_{1}M_{2}})(x;\textbf{p},s;m,j)=\big(2\pi\big)^{-{3}/{2}}e^{-ip.x}(^{\sharp}u_{ M_{1}M_{2}})(\textbf{p},s;m,j)
\end{equation}
\begin{equation}\label{3.57}
  (^{\sharp}v_{ M_{1}M_{2}})(x;\textbf{p},s;m,j)=\big(2\pi\big)^{-{3}/{2}}e^{ip.x}(^{\sharp}v_{ M_{1}M_{2}})(\textbf{p},s;m,j)
\end{equation}

This, together with \eqref{3.54} and \eqref{3.55}, yields

\begin{equation}\label{3.58}
    \begin{split}
    &\Big(\frac{p^{0}}{(\Lambda(A)p)^{0}}\Big)^{\frac{1}{2}}\sum_{M'_{1}M'_{2}}D_{M_{1}M_{2}M'_{1}M'_{2}}(A)(^{\sharp}u_{M'_{1}M'_{2}})(\textbf{p},s;m,j)\\ &=\sum_{s'}D^{j}_{s's}(A^{\sharp-1}_{\Lambda(A)p)}A A^{\sharp}_{p})(^{\sharp}u_{ M_{1}M_{2}})(\textbf{p}_{\Lambda(A)p},s';m,j)\ .
    \end{split}
\end{equation}

and

\begin{equation}\label{3.59}
    \begin{split}
    &\Big(\frac{p^{0}}{(\Lambda(A)p)^{0}}\Big)^{\frac{1}{2}}\sum_{M'_{1}M'_{2}}D_{M_{1}M_{2}M'_{1}M'_{2}}(A)(^{\sharp}v_{M'_{1}M'_{2}}) (\textbf{p},s;m,j)\\ &=\sum_{s'}(\overline{D^{j}_{s's}(A^{\sharp-1}_{\Lambda(A)p)}A A^{\sharp}_{p})})(^{\sharp}v_{ M_{1}M_{2}}) (\textbf{p}_{\Lambda(A)p},s';m,j)\ .
    \end{split}
\end{equation}

Letting $p=k_{m}$, where $k_{m}$ is defined in \eqref{2.19}, and $A=A^{\sharp}_{p}$ with $p\in\Omega_{m}$\\ in \eqref{3.58} and \eqref{3.59} one easily shows that
\begin{equation}\label{3.60}
    \begin{split}
       & (^{\sharp}u^{[J_{1},J_{2}]}_{M_{1}M_{2}})(\textbf{p},s ;m,j) \\ &=\Big(\frac{m}{p^{0}}\Big)^{\frac{1}{2}}\sum_{M'_{1}M'_{2}}D^{[J_{1},J_{2}]}_{M_{1}M_{2}M'_{1}M'_{2}}(A^{\sharp}_{p})   (^{\sharp}u^{[J_{1},J_{2}]}_{M'_{1}M'_{2}})(0,s;m,j)\ .
    \end{split}
\end{equation}

and

\begin{equation}\label{3.61}
    \begin{split}
       & (^{\sharp}v^{[J_{1},J_{2}]}_{M_{1}M_{2}})(\textbf{p},s ;m,j) \\ &=\Big(\frac{m}{p^{0}}\Big)^{\frac{1}{2}}\sum_{M'_{1}M'_{2}}D^{[J_{1},J_{2}]}_{M_{1}M_{2}M'_{1}M'_{2}}(A^{\sharp}_{p})   (^{\sharp}v^{[J_{1},J_{2}]}_{M'_{1}M'_{2}})(0,s;m,j)\ .
    \end{split}
\end{equation}
where we have introduced the superscript $[J_{1},J_{2}]$ again and where $\sharp$=$C$ or $H$ .

By using \eqref{3.7} and \eqref{3.58} with $\textbf{p}=0$ and $A \in SU(2)$ S.Weinberg shows that \\ ( see \cite [section 5.7]{Weinberg1995})

\begin{equation}\label{3.62}
  (^{\sharp}u^{[J_{1},J_{2}]}_{M_{1}M_{2}})(0,s;m,j)=\Big(\frac{1}{2m}\Big)^{\frac{1}{2}}(J_{1}J_{2}js|J_{1}M_{1}J_{2}M_{2})
\end{equation}
\begin{equation}\label{3.63}
  (^{\sharp}v^{[J_{1},J_{2}]}_{M_{1}M_{2}})(0,s;m,j)=(-1)^{j+s}(^{\sharp}u^{[J_{1},J_{2}]}_{M_{1}M_{2}})(0,-s;m,j)
\end{equation}
where $(J_{1}J_{2}js|J_{1}M_{1}J_{2}M_{2})$ is the Clebsch-Gordan coefficient in the notation of A.R.Edmonds ( see \cite{Edmonds1957}) .The  Clebsch-Gordan coefficient vanishes unless $s=M_{1}+M_{2}$ so that we have
\begin{equation}\label{3.64}
    (J_{1}J_{2}js|J_{1}M_{1}J_{2}M_{2})=(J_{1}J_{2}js|J_{1}M_{1}J_{2}M_{2})\delta_{s,M_{1}+M_{2}} .
\end{equation}
$j$ is of the same type, integer or half-integer, as $J_{1}+ J_{2}$ and $ |J_{1}-J_{2}|$.

It follows from \eqref{3.41}  that, for the canonical formalism,

\begin{equation}\label{3.65}
    \begin{split}
       & (^{C}u^{[J_{1},J_{2}]}_{M_{1}M_{2}})(\textbf{p},s;m,j)\\
       &=\frac{1}{\sqrt{2\omega_{\textbf{p}}}}\sum_{M'_{1}M'_{2}}\big((\mathrm{e}^{-(\ln\frac{|\textbf{p}|+\omega_{\textbf{p}}}{m})\sum_{l=1}^{3}\frac{p^{l}}{|\textbf{p}|} \mathcal{J}^{(1)}_{l}})_{M_{1}M'_{1}}(\mathrm{e}^{\ln\frac{|\textbf{p}|+\omega_{\textbf{p}}}{m})\sum_{l=1}^{3}\frac{p^{l}}{|\textbf{p}|}\mathcal{J}^{(2)}_{l}})_{M_{2}M'_{2}}\\
       &\times(J_{1}J_{2}js|J_{1}M'_{1}J_{2}M'_{2})\big).
    \end{split}
\end{equation}

and

\begin{equation}\label{3.66}
    (^{C}v^{[J_{1},J_{2}]}_{M_{1}M_{2}})(\textbf{p},s;m,j)=(-1)^{j+s}(^{C}u^{[J_{1},J_{2}]}_{M_{1}M_{2}})(\textbf{p},-s;m,j)\ .
\end{equation}

By \eqref{3.45} we now get for the helicity formalism
\begin{equation}\label{3.67}
    \begin{split}
       & (^{H}u^{[J_{1},J_{2}]}_{M_{1}M_{2}})(\textbf{p},s;m,j)\\
       &=\frac{1}{\sqrt{2\omega_{\textbf{p}}}}\sum_{M'_{1}M'_{2}}\big(D^{J_{1}}_{M_{1}M'_{1}}(B_{\frac{\textbf{p}}{|\textbf{p}|}}) D^{J_{2}}_{M_{2}M'_{2}}(B_{\frac{\textbf{p}}{|\textbf{p}|}})\\ &\times (\frac{|\textbf{p}|+\omega_{\textbf{p}}}{m})^{M'_{2}-M'_{1}}(J_{1}J_{2}js|J_{1}M'_{1}J_{2}M'_{2})\big) .
    \end{split}
\end{equation}

and

\begin{equation}\label{3.68}
    (^{H}v^{[J_{1},J_{2}]}_{M_{1}M_{2}})(\textbf{p},s;m,j)=(-1)^{j+s}(^{H}u^{[J_{1},J_{2}]}_{M_{1}M_{2}})(\textbf{p},-s;m,j)\ .
\end{equation}

 We now set
\begin{equation}\label{3.69}
    ^{\sharp}\Psi^{[J_{1},J_{2}]\epsilon}_{M_{1}M_{2}}(\textbf{.})=\alpha (^{\sharp}_{1}\Upsilon^{[J_{1},J_{2}]\epsilon}_{M_{1}M_{2}}(\textbf{.}))+ \beta (^{\sharp}_{2}\Upsilon^{[J_{1},J_{2}]\epsilon}_{M_{1}M_{2}}(\textbf{.}))
\end{equation}

$^{\sharp}\Psi^{[J_{1},J_{2}]\epsilon}_{M_{1}M_{2}}(x)$ satisfies the relativistic covariance law given by \eqref{3.46}. In order to verify the microscopic causality condition given by \eqref{3.47} S.Weinberg has carefully shown that one must have $|\alpha|=|\beta|$ with
\begin{equation}\label{3.70}
\beta=(-1)^{2J_{2}}\gamma\alpha \quad,  |\gamma|=1
\end{equation}
$\gamma$ is the same for every field for a given particle.

$\alpha$ and $\gamma$ can be eliminated so that we finally obtain  in the bosonic case when $j\in\N$ and in the case of the canonical formalism

\begin{equation}\label{3.71}
    \begin{split}
     &^{C}\Psi^{[J_{1},J_{2}]\epsilon}_{M_{1}M_{2}}(x)\\                                                                &=(2\pi)^{-\frac{3}{2}}\sum_{s}\int \d^{3}\textbf{p}\bigg(\frac{1}{\sqrt{2\omega_{\textbf{p}}}}\Big(\sum_{M'_{1}M'_{2}}(\mathrm{e}^{-(\ln\frac{|\textbf{p}|+\omega_{\textbf{p}}}{m})\sum_{l=1}^{3}\frac{p^{l}}{|\textbf{p}|}    \mathcal{J}^{(1)}_{l}})_{M_{1},M'_{1}}\\ &(\mathrm{e}^{(\ln\frac{|\textbf{p}|+\omega_{\textbf{p}}}{m})\sum_{l=1}^{3}\frac{p^{l}}{|\textbf{p}|}\mathcal{J}^{(2)}_{l}})_{M_{2},M'_{2}}(J_{1}J_{2}js|J_{1}M'_{1}J_{2}M'_{2})\Big)\times\\   &\mathrm{e}^{-ip.x}a_{\epsilon}(\textbf{p},s;m,j)\\ &+(-1)^{2J_{2}+j+s}\frac{1}{\sqrt{2\omega_{\textbf{p}}}}\Big(\sum_{M'_{1}M'_{2}}(\mathrm{e}^{-(\ln\frac{|\textbf{p}|+\omega_{\textbf{p}}}{m})\sum_{l=1}^{3}\frac{p^{l}}{|\textbf{p}|} \mathcal{J}^{(1)}_{l}})_{M_{1},M'_{1}}\\   &(\mathrm{e}^{(\ln\frac{|\textbf{p}|+\omega_{\textbf{p}}}{m})\sum_{l=1}^{3}\frac{p^{l}}{|\textbf{p}|}\mathcal{J}^{(2)}_{l}})_{M_{2},M'_{2}}
    \times(J_{1}J_{2}j(-s)|J_{1}M'_{1}J_{2}M'_{2})\Big)\times\\                                   &\mathrm{e}^{ip.x}a_{\epsilon'}^{*}(\textbf{p},s;m,j)\bigg)\ .
    \end{split}
\end{equation}

We also have
\begin{equation}\label{3.72}
    \begin{split}
    &^{C}\Psi^{[J_{1},J_{2}]\epsilon}_{M_{1}M_{2}}(x)\\
    &=(2\pi)^{-\frac{3}{2}}\sum_{s}\int \d^{3}\textbf{p}\frac{1}{\sqrt{2\omega_{\textbf{p}}}}\Big(\sum_{M'_{1}M'_{2}}(\mathrm{e}^{-(\ln\frac{|\textbf{p}|+\omega_{\textbf{p}}}{m})\sum_{l=1}^{3}\frac{p^{l}}{|\textbf{p}|}    \mathcal{J}^{(1)}_{l}})_{M_{1}M'_{1}}\\ &\times(\mathrm{e}^{(\ln\frac{|\textbf{p}|+\omega_{\textbf{p}}}{m})\sum_{l=1}^{3}\frac{p^{l}}{|\textbf{p}|}\mathcal{J}^{(2)}_{l}})_{M_{2}M'_{2}}(J_{1}J_{2}js|J_{1}M'_{1}J_{2}M'_{2})\Big)\\   &(\mathrm{e}^{-ip.x}a_{\epsilon}(\textbf{p},s;m,j)+(-1)^{2J_{2}+j-s}\mathrm{e}^{ip.x}a_{\epsilon'}^{*}(\textbf{p},-s;m,j)) .
    \end{split}
\end{equation}

On the other hand we obtain in the case of the helicity formalism

\begin{equation}\label{3.73}
    \begin{split}
    &^{H}\Psi^{[J_{1},J_{2}]\epsilon}_{M_{1}M_{2}}(x)\\                                                                &=(2\pi)^{-\frac{3}{2}}\sum_{s}\int
     \d^{3}\textbf{p}\bigg(\frac{1}{\sqrt{2\omega_{\textbf{p}}}}\Big(\sum_{M'_{1}M'_{2}}D^{J_{1}}_{M_{1},M'_{1}}(B_{\frac{\textbf{p}}{|\textbf{p}|}})   D^{J_{2}}_{M_{2},M'_{2}}(B_{\frac{\textbf{p}}{|\textbf{p}|}})\\
     &(\frac{|\textbf{p}|+\omega_{\textbf{p}}}{m})^{M'_{2}-M'_{1}}                                             (J_{1}J_{2}js|J_{1}M'_{1}J_{2}M'_{2})\Big)  \mathrm{e}^{-ip.x}a_{\epsilon}(\textbf{p},s;m,j)\\
    &+(-1)^{2J_{2}+j+s}\frac{1}{\sqrt{2\omega_{\textbf{p}}}}\Big(\sum_{M'_{1}M'_{2}}D^{J_{1}}_{M_{1},M'_{1}}(B_{\frac{\textbf{p}}{|\textbf{p}|}})    D^{J_{2}}_{M_{2},M'_{2}}(B_{\frac{\textbf{p}}{|\textbf{p}|}})\\
     &(\frac{|\textbf{p}|+\omega_{\textbf{p}}}{m})^{M'_{2}-M'_{1}}                                         (J_{1}J_{2}j(-s)|J_{1}M'_{1}J_{2}M'_{2})\Big)\mathrm{e}^{ip.x}a_{\epsilon'}^{*}(\textbf{p},s;m,j)\bigg)\ .
    \end{split}
\end{equation}

 We also obtain
\begin{equation}
\begin{split}
    &^{H}\Psi^{[J_{1},J_{2}]\epsilon}_{M_{1}M_{2}}(x)\\
    &=(2\pi)^{-\frac{3}{2}}\sum_{s}\int\d^{3}\textbf{p}\frac{1}{\sqrt{2\omega_{\textbf{p}}}}\Big(\sum_{M'_{1}M'_{2}}\big(\mathrm{e}^{-i\arccos\frac{p^{3}}{|\textbf{p}|}(-\frac{p^{2}}{|\textbf{p}|}\mathcal{J}^{(1)}_{1}+ \frac{p^{1}}{|\textbf{p}|}\mathcal{J}^{(1)}_{2})}\big)_{M_{1}M'_{1}}\\   &\big(\mathrm{e}^{-i\arccos\frac{p^{3}}{|\textbf{p}|}(-\frac{p^{2}}{|\textbf{p}|}\mathcal{J}^{(2)}_{1}+ \frac{p^{1}}{|\textbf{p}|}\mathcal{J}^{(2)}_{2})}\big)_{M_{2}M'_{2}}
     (\frac{|\textbf{p}|+\omega_{\textbf{p}}}{m})^{M'_{2}-M'_{1}}(J_{1}J_{2}js|J_{1}M'_{1}J_{2}M'_{2})\Big)\\  &(\mathrm{e}^{-ip.x}a_{\epsilon}(\textbf{p},s;m,j)
     +(-1)^{2J_{2}+j-s}\mathrm{e}^{ip.x}a_{\epsilon'}^{*}(\textbf{p},-s;m,j)).
    \end{split}
\end{equation}

Similarly, in the fermionic case when $j\in \N+1/2$, we obtain in the case of the canonical formalism
\begin{equation}
\begin{split}
  &^{C}\widetilde\Psi^{[J_{1},J_{2}]\epsilon}_{M_{1}M_{2}}(x)\\                                                         &=(2\pi)^{-\frac{3}{2}}\sum_{s}\int \d^{3}\textbf{p}\bigg(\frac{1}{\sqrt{2\omega_{\textbf{p}}}}\Big(\sum_{M'_{1}M'_{2}}(\mathrm{e}^{-(\ln\frac{|\textbf{p}|+\omega_{\textbf{p}}}{m})\sum_{l=1}^{3}\frac{p^{l}}{|\textbf{p}|}  \mathcal{J}^{(1)}_{l}})_{M_{1},M'_{1}}\\ &(\mathrm{e}^{(\ln\frac{|\textbf{p}|+\omega_{\textbf{p}}}{m})\sum_{l=1}^{3}\frac{p^{l}}{|\textbf{p}|}\mathcal{J}^{(2)}_{l}})_{M_{2},M'_{2}}(J_{1}J_{2}js|J_{1}M'_{1}J_{2}M'_{2})\Big)   \mathrm{e}^{-ip.x}b_{\epsilon}(\textbf{p},s;m,j)\\ &+(-1)^{2J_{2}+j+s}\frac{1}{\sqrt{2\omega_{\textbf{p}}}}\Big(\sum_{M'_{1}M'_{2}}(\mathrm{e}^{-(\ln\frac{|\textbf{p}|+\omega_{\textbf{p}}}{m})\sum_{l=1}^{3}\frac{p^{l}}{|\textbf{p}|}   \mathcal{J}^{(1)}_{l}})_{M_{1},M'_{1}}\\  &(\mathrm{e}^{(\ln\frac{|\textbf{p}|+\omega_{\textbf{p}}}{m})\sum_{l=1}^{3}\frac{p^{l}}{|\textbf{p}|}\mathcal{J}^{(2)}_{l}})_{M_{2},M'_{2}}
  (J_{1}J_{2}j(-s)|J_{1}M'_{1}J_{2}M'_{2})\Big) \mathrm{e}^{ip.x}b_{\epsilon'}^{*}(\textbf{p},s;m,j)\bigg)\ .
    \end{split}
\end{equation}

We also get
\begin{equation}
\begin{split}
    &^{C}\widetilde\Psi^{[J_{1},J_{2}]\epsilon}_{M_{1}M_{2}}(x)\\
    &=(2\pi)^{-\frac{3}{2}}\sum_{s}\int \d^{3}\textbf{p}\bigg(\frac{1}{\sqrt{2\omega_{\textbf{p}}}}\Big(\sum_{M'_{1}M'_{2}}(\mathrm{e}^{-(\ln\frac{|\textbf{p}|+\omega_{\textbf{p}}}{m})\sum_{l=1}^{3}\frac{p^{l}}{|\textbf{p}|}  \mathcal{J}^{(1)}_{l}})_{M_{1}M'_{1}}\\ &\times(\mathrm{e}^{(\ln\frac{|\textbf{p}|+\omega_{\textbf{p}}}{m})\sum_{l=1}^{3}\frac{p^{l}}{|\textbf{p}|}\mathcal{J}^{(2)}_{l}})_{M_{2}M'_{2}}(J_{1}J_{2}js|J_{1}M'_{1}J_{2}M'_{2})\Big)\\   &(\mathrm{e}^{-ip.x}b_{\epsilon}(\textbf{p},s;m,j)+(-1)^{2J_{2}+j-s}\mathrm{e}^{ip.x}b_{\epsilon'}^{*}(\textbf{p},-s;m,j))\bigg) .
    \end{split}
\end{equation}

and in the case of the helicity formalism

\begin{equation}
\begin{split}
    &^{H}\widetilde\Psi^{[J_{1},J_{2}]\epsilon}_{M_{1}M_{2}}(x)=(2\pi)^{-\frac{3}{2}}\sum_{s}\int
     \d^{3}\textbf{p}\bigg(\frac{1}{\sqrt{2\omega_{\textbf{p}}}}\Big(\sum_{M'_{1}M'_{2}}D^{J_{1}}_{M_{1},M'_{1}}(B_{\frac{\textbf{p}}{|\textbf{p}|}})   D^{J_{2}}_{M_{2},M'_{2}}(B_{\frac{\textbf{p}}{|\textbf{p}|}})\\
     &\times(\frac{|\textbf{p}|+\omega_{\textbf{p}}}{m})^{M'_{2}-M'_{1}}(J_{1}J_{2}js|J_{1}M'_{1}J_{2}M'_{2})\Big)  \mathrm{e}^{-ip.x}b_{\epsilon}(\textbf{p},s;m,j)\\
     &+(-1)^{2J_{2}+j+s}\frac{1}{\sqrt{2\omega_{\textbf{p}}}}\Big(\sum_{M'_{1}M'_{2}}D^{J_{1}}_{M_{1},M'_{1}}(B_{\frac{\textbf{p}}{|\textbf{p}|}})    D^{J_{2}}_{M_{2},M'_{2}}(B_{\frac{\textbf{p}}{|\textbf{p}|}})\\
     &\times(\frac{|\textbf{p}|+\omega_{\textbf{p}}}{m})^{M'_{2}-M'_{1}}(J_{1}J_{2}j(-s)|J_{1}M'_{1}J_{2}M'_{2})\Big)   \mathrm{e}^{ip.x}b_{\epsilon'}^{*}(\textbf{p},s;m,j)\bigg)\ .
    \end{split}
\end{equation}

We also have
\begin{equation}
\begin{split}
     &^{H}\widetilde\Psi^{[J_{1},J_{2}]\epsilon}_{M_{1}M_{2}}(x)  \\
     &=(2\pi)^{-\frac{3}{2}}\sum_{s}\int\d^{3}\textbf{p}\frac{1}{\sqrt{2\omega_{\textbf{p}}}}\Big(\sum_{M'_{1}M'_{2}}\big(\mathrm{e}^{-i\arccos\frac{p^{3}}{|\textbf{p}|}(-\frac{p^{2}}{|\textbf{p}|}\mathcal{J}^{(1)}_{1}+ \frac{p^{1}}{|\textbf{p}|}\mathcal{J}^{(1)}_{2})}\big)_{M_{1}M'_{1}}\\   &\big(\mathrm{e}^{-i\arccos\frac{p^{3}}{|\textbf{p}|}(-\frac{p^{2}}{|\textbf{p}|}\mathcal{J}^{(2)}_{1}+ \frac{p^{1}}{|\textbf{p}|}\mathcal{J}^{(2)}_{2})}\big)_{M_{2}M'_{2}}
     (\frac{|\textbf{p}|+\omega_{\textbf{p}}}{m})^{M'_{2}-M'_{1}}(J_{1}J_{2}js|J_{1}M'_{1}J_{2}M'_{2})\Big)\\  &(\mathrm{e}^{-ip.x}b_{\epsilon}(\textbf{p},s;m,j)+(-1)^{2J_{2}+j-s}\mathrm{e}^{ip.x}b_{\epsilon'}^{*}(\textbf{p},-s;m,j)).
    \end{split}
\end{equation}

\begin{remark}
Note that the construction of the fields $^{\sharp}\Psi^{[J_{1},J_{2}]\epsilon}_{M_{1}M_{2}}(x)$
 involves an irreducible representation of $SL(2,\C)$ of finite dimension.From a physical point of view, in particular in the case of an interaction invariant by space inversion, it can be more convenient to construct such fields associated to a direct sum of irreducible representations of finite dimension.For example the Dirac field for a particle of spin $1/2$ is based on the representation $[\frac{1}{2},0]\oplus[0,\frac{1}{2}]$.
\end{remark}

\subsection{Two particular cases: [j,0] and [0,j]}\mbox{}

In the bosonic case when $J_{1}=j\in\N$ and $J_{2}=0$ we have $(j0js|js'00)=\delta_{s,s'}$ for the Clebsch-Gordan coefficient  we obtain
\begin{equation}
    \begin{split}
    &^{C}\Psi^{[j,0]\epsilon}_{s}(x)=(2\pi)^{-\frac{3}{2}}\sum_{s'}\int \d^{3}\textbf{p}\bigg(\frac{1}{\sqrt{2\omega_{\textbf{p}}}}(\mathrm{e}^{-(\ln\frac{|\textbf{p}|+\omega_{\textbf{p}}}{m})\sum_{l=1}^{3}\frac{p^{l}}{|\textbf{p}|}    \mathcal{J}^{(j)}_{l}})_{ss'}\\    &\mathrm{e}^{-ip.x}a_{\epsilon}(\textbf{p},s';m,j)+\frac{1}{\sqrt{2\omega_{\textbf{p}}}}(\mathrm{e}^{-(\ln\frac{|\textbf{p}|+\omega_{\textbf{p}}}{m})\sum_{l=1}^{3}\frac{p^{l}}{|\textbf{p}|}    \mathcal{J}^{(j)}_{l}})_{s(-s')}\\                                          &(-1)^{j+s'}\mathrm{e}^{ip.x}a_{\epsilon'}^{*}(\textbf{p},s';m,j))\bigg)\ .
    \end{split}
\end{equation}
where $\big(\mathcal{J}^{(j)}_{l}\big)$, $l=1,2,3$, are the generators of the rotations in the representation $D^{j}(\textbf{.})$ of $SU(2)$.

We also have
\begin{equation}
    \begin{split}
    &^{C}\Psi^{[j,0]\epsilon}_{s}(x) \\
    &=(2\pi)^{-\frac{3}{2}}\sum_{s'}\int \d^{3}\textbf{p}\frac{1}{\sqrt{2\omega_{\textbf{p}}}}(\mathrm{e}^{-(\ln\frac{|\textbf{p}|+\omega_{\textbf{p}}}{m})\sum_{l=1}^{3}\frac{p^{l}}{|\textbf{p}|}    \mathcal{J}^{(j)}_{l}})_{ss'}\\    &\big(\mathrm{e}^{-ip.x}a_{\epsilon}(\textbf{p},s';m,j)+(-1)^{j-s'}\mathrm{e}^{ip.x}a_{\epsilon'}^{*}(\textbf{p},-s';m,j)\big)\ .
    \end{split}
\end{equation}

and
\begin{equation}
    \begin{split}
    &^{H}\Psi^{[j,0]\epsilon}_{s}(x)=(2\pi)^{-\frac{3}{2}}\sum_{s'}\int \d^{3}\textbf{p}\bigg(\frac{1}{\sqrt{2\omega_{\textbf{p}}}}\big(\mathrm{e}^{-i\arccos\frac{p^{3}}{|\textbf{p}|}(-\frac{p^{2}}{|\textbf{p}|}\mathcal{J}^{(j)}_{1}+ \frac{p^{1}}{|\textbf{p}|}\mathcal{J}^{(j)}_{2})}\big)_{ss'}\\                              &\big(\frac{|\textbf{p}|+\omega_{\textbf{p}}}{m}\big)^{-s'}    \mathrm{e}^{-ip.x}a_{\epsilon}(\textbf{p},s';m,j)
    +\frac{1}{\sqrt{2\omega_{\textbf{p}}}}\big(\mathrm{e}^{-i\arccos\frac{p^{3}}{|\textbf{p}|}(-\frac{p^{2}}{|\textbf{p}|}\mathcal{J}^{(j)}_{1}+ \frac{p^{1}}{|\textbf{p}|}\mathcal{J}^{(j)}_{2})}\big)_{s(-s')}\\                           &\big(\frac{|\textbf{p}|+\omega_{\textbf{p}}}{m}\big)^{s'}(-1)^{j+s'}\mathrm{e}^{-ip.x}a^{*}_{\epsilon}(\textbf{p},s';m,j)\bigg)\ .
    \end{split}
\end{equation}

We also get
\begin{equation}
    \begin{split}
    &^{H}\Psi^{[j,0]\epsilon}_{s}(x)  \\
    &=(2\pi)^{-\frac{3}{2}}\sum_{s'}\int\d^{3}\textbf{p}\frac{1}{\sqrt{2\omega_{\textbf{p}}}}\big(\mathrm{e}^{-i\arccos\frac{p^{3}}{|\textbf{p}|}(-\frac{p^{2}}{|\textbf{p}|}\mathcal{J}^{(j)}_{1}+ \frac{p^{1}}{|\textbf{p}|}\mathcal{J}^{(j)}_{2})}\big)_{ss'}\\
    &\big(\frac{|\textbf{p}|+\omega_{\textbf{p}}}{m}\big)^{-s'}\big(\mathrm{e}^{-ip.x}a_{\epsilon}(\textbf{p},s';m,j)
    +(-1)^{j-s'}\mathrm{e}^{ip.x}a_{\epsilon'}^{*}(\textbf{p},-s';m,j)\big).
    \end{split}
\end{equation}

In the fermionic case when $j\in\N+1/2$ we obtain

\begin{equation}
    \begin{split}
    &^{C}\widetilde\Psi^{[j,0]\epsilon}_{s}(x)=(2\pi)^{-\frac{3}{2}}\sum_{s'}\int \d^{3}\textbf{p}\bigg(\frac{1}{\sqrt{2\omega_{\textbf{p}}}}(\mathrm{e}^{-(\ln\frac{|\textbf{p}|+\omega_{\textbf{p}}}{m})\sum_{l=1}^{3}\frac{p^{l}}{|\textbf{p}|}    \mathcal{J}^{(j)}_{l}})_{ss'}\\    &\mathrm{e}^{-ip.x}b_{\epsilon}(\textbf{p},s';m,j)+\frac{1}{\sqrt{2\omega_{\textbf{p}}}}(\mathrm{e}^{-(\ln\frac{|\textbf{p}|+\omega_{\textbf{p}}}{m})\sum_{l=1}^{3}\frac{p^{l}}{|\textbf{p}|}    \mathcal{J}^{(j)}_{l}})_{s(-s')}\\                                          &(-1)^{j+s'}\mathrm{e}^{ip.x}b_{\epsilon'}^{*}(\textbf{p},s';m,j))\bigg)\ .
    \end{split}
\end{equation}

We also get
\begin{equation}
    \begin{split}
     &^{C}\widetilde\Psi^{[j,0]\epsilon}_{s}(x) \\
    &=(2\pi)^{-\frac{3}{2}}\sum_{s'}\int \d^{3}\textbf{p}\frac{1}{\sqrt{2\omega_{\textbf{p}}}}(\mathrm{e}^{-(\ln\frac{|\textbf{p}|+\omega_{\textbf{p}}}{m})\sum_{l=1}^{3}\frac{p^{l}}{|\textbf{p}|}    \mathcal{J}^{(j)}_{l}})_{ss'}\\    &(\mathrm{e}^{-ip.x}b_{\epsilon}(\textbf{p},s';m,j)+(-1)^{j-s'}\mathrm{e}^{ip.x}b_{\epsilon'}^{*}(\textbf{p},-s';m,j)) .
    \end{split}
\end{equation}

and

\begin{equation}
    \begin{split}
    &^{H}\widetilde\Psi^{[j,0]\epsilon}_{s}(x)=(2\pi)^{-\frac{3}{2}}\sum_{s'}\int \d^{3}\textbf{p}\bigg(\frac{1}{\sqrt{2\omega_{\textbf{p}}}}\big(\mathrm{e}^{-i\arccos\frac{p^{3}}{|\textbf{p}|}(-\frac{p^{2}}{|\textbf{p}|}\mathcal{J}^{(j)}_{1}+ \frac{p^{1}}{|\textbf{p}|}\mathcal{J}^{(j)}_{2})}\big)_{ss'}\\                              &\big(\frac{|\textbf{p}|+\omega_{\textbf{p}}}{m}\big)^{-s'}\mathrm{e}^{-ip.x}b_{\epsilon}(\textbf{p},s';m,j)
    +\frac{1}{\sqrt{2\omega_{\textbf{p}}}}\big(\mathrm{e}^{-i\arccos\frac{p^{3}}{|\textbf{p}|}(-\frac{p^{2}}{|\textbf{p}|}\mathcal{J}^{(j)}_{1}+ \frac{p^{1}}{|\textbf{p}|}\mathcal{J}^{(j)}_{2})}\big)_{s(-s')}\\                           &\big(\frac{|\textbf{p}|+\omega_{\textbf{p}}}{m}\big)^{s'}(-1)^{j+s'}\mathrm{e}^{-ip.x}b^{*}_{\epsilon}(\textbf{p},s';m,j)\bigg)\ .
    \end{split}
\end{equation}

Also

\begin{equation}
    \begin{split}
    &^{H}\widetilde\Psi^{[j,0]\epsilon}_{s}(x)  \\
    &=(2\pi)^{-\frac{3}{2}}\sum_{s'}\int\d^{3}\textbf{p}\frac{1}{\sqrt{2\omega_{\textbf{p}}}}\big(\mathrm{e}^{-i\arccos\frac{p^{3}}{|\textbf{p}|}(-\frac{p^{2}}{|\textbf{p}|}\mathcal{J}^{(j)}_{1}+ \frac{p^{1}}{|\textbf{p}|}\mathcal{J}^{(j)}_{2})}\big)_{ss'}\\
    &\big(\frac{|\textbf{p}|+\omega_{\textbf{p}}}{m}\big)^{-s'}\big(\mathrm{e}^{-ip.x}b_{\epsilon}(\textbf{p},s';m,j)
    +(-1)^{j-s'}\mathrm{e}^{ip.x}b_{\epsilon'}^{*}(\textbf{p},-s';m,j)\big).
    \end{split}
\end{equation}

In the bosonic case when $J_{2}=j\in\N$ and $J_{1}=0$ we have $(0jjs|00js')=\delta_{s,s'}$ for the Clebsch-Gordan coefficient and from \eqref{3.68} and \eqref{3.70} we obtain

\begin{equation}
    \begin{split}
    &^{C}\Psi^{[0,j]\epsilon}_{s}(x)=(2\pi)^{-\frac{3}{2}}\sum_{s'}\int \d^{3}\textbf{p}\bigg(\frac{1}{\sqrt{2\omega_{\textbf{p}}}}(\mathrm{e}^{(\ln\frac{|\textbf{p}|+\omega_{\textbf{p}}}{m})\sum_{l=1}^{3}\frac{p^{l}}{|\textbf{p}|}    \mathcal{J}^{(j)}_{l}})_{ss'}\\    &\mathrm{e}^{-ip.x}a_{\epsilon}(\textbf{p},s';m,j)+\frac{1}{\sqrt{2\omega_{\textbf{p}}}}(\mathrm{e}^{(\ln\frac{|\textbf{p}|+\omega_{\textbf{p}}}{m})\sum_{l=1}^{3}\frac{p^{l}}{|\textbf{p}|}    \mathcal{J}^{(j)}_{l}})_{s(-s')}\\                                          &(-1)^{3j+s'}\mathrm{e}^{ip.x}a_{\epsilon'}^{*}(\textbf{p},s';m,j))\bigg)\ .
    \end{split}
\end{equation}

We also have
\begin{equation}
    \begin{split}
    &^{C}\Psi^{[0,j]\epsilon}_{s}(x)  \\
    &=(2\pi)^{-\frac{3}{2}}\sum_{s'}\int \d^{3}\textbf{p}\frac{1}{\sqrt{2\omega_{\textbf{p}}}}(\mathrm{e}^{(\ln\frac{|\textbf{p}|+\omega_{\textbf{p}}}{m})\sum_{l=1}^{3}\frac{p^{l}}{|\textbf{p}|}    \mathcal{J}^{(j)}_{l}})_{ss'}\\    &(\mathrm{e}^{-ip.x}a_{\epsilon}(\textbf{p},s';m,j)+(-1)^{3j-s'}\mathrm{e}^{ip.x}a_{\epsilon'}^{*}(\textbf{p},-s';m,j)) .
    \end{split}
\end{equation}

and
\begin{equation}
    \begin{split}
    &^{H}\Psi^{[0,j]\epsilon}_{s}(x)=(2\pi)^{-\frac{3}{2}}\sum_{s'}\int \d^{3}\textbf{p}\bigg(\frac{1}{\sqrt{2\omega_{\textbf{p}}}}\big(\mathrm{e}^{-i\arccos\frac{p^{3}}{|\textbf{p}|}(-\frac{p^{2}}{|\textbf{p}|}\mathcal{J}^{(j)}_{1}+ \frac{p^{1}}{|\textbf{p}|}\mathcal{J}^{(j)}_{2})}\big)_{ss'}\\                              &\big(\frac{|\textbf{p}|+\omega_{\textbf{p}}}{m}\big)^{s'}\mathrm{e}^{-ip.x}a_{\epsilon}(\textbf{p},s';m,j)
    +\frac{1}{\sqrt{2\omega_{\textbf{p}}}}\big(\mathrm{e}^{-i\arccos\frac{p^{3}}{|\textbf{p}|}(-\frac{p^{2}}{|\textbf{p}|}\mathcal{J}^{(j)}_{1}+ \frac{p^{1}}{|\textbf{p}|}\mathcal{J}^{(j)}_{2})}\big)_{s(-s')}\\                           &\big(\frac{|\textbf{p}|+\omega_{\textbf{p}}}{m}\big)^{-s'}(-1)^{3j+s'}\mathrm{e}^{-ip.x}a^{*}_{\epsilon}(\textbf{p},s';m,j)\bigg)\ .
    \end{split}
\end{equation}

We also have

\begin{equation}
   \begin{split}
   &^{H}\Psi^{[0,j]\epsilon}_{s}(x)  \\
   &=(2\pi)^{-\frac{3}{2}}\sum_{s'}\int\d^{3}\textbf{p}\frac{1}{\sqrt{2\omega_{\textbf{p}}}}\big(\mathrm{e}^{-i\arccos\frac{p^{3}}{|\textbf{p}|}(-\frac{p^{2}}{|\textbf{p}|}\mathcal{J}^{(j)}_{1}+ \frac{p^{1}}{|\textbf{p}|}\mathcal{J}^{(j)}_{2})}\big)_{ss'}\\
    &\big(\frac{|\textbf{p}|+\omega_{\textbf{p}}}{m}\big)^{s'}(\mathrm{e}^{-ip.x}a_{\epsilon}(\textbf{p},s';m,j)
    +(-1)^{3j-s'}\mathrm{e}^{ip.x}a_{\epsilon'}^{*}(\textbf{p},-s';m,j)).
   \end{split}
\end{equation}

In the fermionic case for when $j\in\N+1/2$ we obtain

\begin{equation}
    \begin{split}
    &^{C}\widetilde\Psi^{[0,j]\epsilon}_{s}(x)=(2\pi)^{-\frac{3}{2}}\sum_{s'}\int \d^{3}\textbf{p}\bigg(\frac{1}{\sqrt{2\omega_{\textbf{p}}}}(\mathrm{e}^{(\ln\frac{|\textbf{p}|+\omega_{\textbf{p}}}{m})\sum_{l=1}^{3}\frac{p^{l}}{|\textbf{p}|}    \mathcal{J}^{(j)}_{l}})_{ss'}\\    &\mathrm{e}^{-ip.x}b_{\epsilon}(\textbf{p},s';m,j)+\frac{1}{\sqrt{2\omega_{\textbf{p}}}}(\mathrm{e}^{(\ln\frac{|\textbf{p}|+\omega_{\textbf{p}}}{m})\sum_{l=1}^{3}\frac{p^{l}}{|\textbf{p}|}    \mathcal{J}^{(j)}_{l}})_{s(-s')}\\                                          &(-1)^{3j+s'}\mathrm{e}^{ip.x}b_{\epsilon'}^{*}(\textbf{p},s';m,j))\bigg)\ .
    \end{split}
\end{equation}

We also obtain
\begin{equation}
    \begin{split}
    &^{C}\widetilde\Psi^{[0,j]\epsilon}_{s}(x)  \\
    &=(2\pi)^{-\frac{3}{2}}\sum_{s'}\int \d^{3}\textbf{p}\frac{1}{\sqrt{2\omega_{\textbf{p}}}}(\mathrm{e}^{(\ln\frac{|\textbf{p}|+\omega_{\textbf{p}}}{m})\sum_{l=1}^{3}\frac{p^{l}}{|\textbf{p}|}    \mathcal{J}^{(j)}_{l}})_{ss'}\\    &(\mathrm{e}^{-ip.x}b_{\epsilon}(\textbf{p},s';m,j)+(-1)^{3j-s'}\mathrm{e}^{ip.x}b_{\epsilon'}^{*}(\textbf{p},-s';m,j)) .
    \end{split}
\end{equation}

and

\begin{equation}
    \begin{split}
    &^{H}\widetilde\Psi^{[0,j]\epsilon}_{s}(x)=(2\pi)^{-\frac{3}{2}}\sum_{s'}\int \d^{3}\textbf{p}\bigg(\frac{1}{\sqrt{2\omega_{\textbf{p}}}}\big(\mathrm{e}^{-i\arccos\frac{p^{3}}{|\textbf{p}|}(-\frac{p^{2}}{|\textbf{p}|}\mathcal{J}^{(j)}_{1}+ \frac{p^{1}}{|\textbf{p}|}\mathcal{J}^{(j)}_{2})}\big)_{ss'}\\                              &\big(\frac{|\textbf{p}|+\omega_{\textbf{p}}}{m}\big)^{s'}\mathrm{e}^{-ip.x}b_{\epsilon}(\textbf{p},s';m,j)
    +\frac{1}{\sqrt{2\omega_{\textbf{p}}}}\big(\mathrm{e}^{-i\arccos\frac{p^{3}}{|\textbf{p}|}(-\frac{p^{2}}{|\textbf{p}|}\mathcal{J}^{(j)}_{1}+ \frac{p^{1}}{|\textbf{p}|}\mathcal{J}^{(j)}_{2})}\big)_{s(-s')}\\                   &\big(\frac{|\textbf{p}|+\omega_{\textbf{p}}}{m}\big)^{-s'}(-1)^{3j+s'}\mathrm{e}^{-ip.x}b^{*}_{\epsilon}(\textbf{p},s';m,j)\bigg)\ .
    \end{split}
\end{equation}

and also

\begin{equation}
    \begin{split}
    &^{H}\widetilde\Psi^{[0,j]\epsilon}_{s}(x)  \\
    &=(2\pi)^{-\frac{3}{2}}\sum_{s'}\int\d^{3}\textbf{p}\frac{1}{\sqrt{2\omega_{\textbf{p}}}}\big(\mathrm{e}^{-i\arccos\frac{p^{3}}{|\textbf{p}|}(-\frac{p^{2}}{|\textbf{p}|}\mathcal{J}^{(j)}_{1}+ \frac{p^{1}}{|\textbf{p}|}\mathcal{J}^{(j)}_{2})}\big)_{ss'}\\
     &\big(\frac{|\textbf{p}|+\omega_{\textbf{p}}}{m}\big)^{s'}(\mathrm{e}^{-ip.x}b_{\epsilon}(\textbf{p},s';m,j)
     +(-1)^{3j-s'}\mathrm{e}^{ip.x}b_{\epsilon'}^{*}(\textbf{p},-s';m,j)).
    \end{split}
\end{equation}

\section{Free causal fields for a massless particle of any finite helicity }\mbox{}

\setcounter{equation}{0}

In this chapter we introduce the construction of free causal fields for massless particles of helicity $j$ by still following the formalism of S.Weinberg in
\cite{WeinbergII1964},\cite{WeinbergIII1964} and \cite[section 5.9]{Weinberg1995}. Note that the construction of free fields for photons and gravitons is not included in this approach. See \cite{Weinberg1965}. Photons and gravitons are properly associated with potentials instead of fields. The approach that we now follow will be adapted to massless fermions as neutrinos and antineutrinos in the Standard model.

 Let $\gF^{[j]}_{s}$ (resp.$\gF^{[j]}_{a}$) be the bosonic (resp. fermionic)Fock space for massless bosons(resp.massless fermions) of helicity $j$.
 We have
 \begin{equation}\label{4.1}
    \gF^{[j]}_{s}
 = \bigotimes \left(\oplus_{n=0}^\infty \otimes_s^n
 L^2(\R^3)\right)\ .
\end{equation}
 where $\otimes_s^n$ denotes the symmetric $n$-th tensor product
and $\otimes_s^0 L^2(\Sigma_j)$ =$ \C$.

and
\begin{equation}\label{4.2}
   \gF^{[j]}_{a}
 = \bigotimes\left(\oplus_{n=0}^\infty \otimes_a^n
 L^2(\R^3)\right)\ .
\end{equation}
where $\otimes_a^n$ denotes the antisymmetric $n$-th tensor
product and $\otimes_a^0 L^2(\Sigma_j)$ =$ \C$.


The unitary irreducible representations  $\widetilde{U}^{[j]}$ of $\mathcal{P}$ induce two unitary representations of $\mathcal{P}$ in $\gF^{[j]}_{s}$ and $\gF^{[j]}_{a}$. Each representation is $\Gamma(\widetilde{U}^{[j]})$.

$a_{\epsilon}(\textbf{p},j)$  (resp. $a^{*}_{\epsilon}(\textbf{p},j)$ ) is the annihilation  (resp.creation)  operator for a massless boson of helicity $j$ if $\epsilon=+$ and for a antiparticle of helicity $j$ if $\epsilon=-$ .


Similarly, $b_{\epsilon}(\textbf{p},j)$ (resp. $b^{*}_{\epsilon}(\textbf{p},j)$ ) is the annihilation (resp.creation)  operator for a massless fermion of helicity $j$ if $\epsilon=+$ and  for a antiparticle of helicity $j$ if $\epsilon=-$.


The operators $a_{\epsilon}(\textbf{p},j)$ and $a^{*}_{\epsilon}(\textbf{p},j)$ fulfil the usual commutation relations (CCR), whereas $b_{\epsilon}(\textbf{p},j)$ and $b^{*}_{\epsilon}(\textbf{p},j)$ fulfil the canonical anticommutation relation (CAR). See \cite{Weinberg1995}. Futhermore, the $a's$ commute with the $b's$.

In addition, in the case where several fermions are involved we follow the convention described in \cite[sections 4.1 and 4.2]{Weinberg1995}. This means that we will assume that fermionic annihilation and creation operators of different species of particles anticommute for both massive and massless fermions.

Therefore, the following canonical anticommutation and commutation
relations hold for a couple of massless particles with helicities $j$ and $j'\neq j$ together with a massive particle with $m>0$ and spin $\widetilde j$.
\begin{equation}\label{4.3}
\begin{split}
 &\{ b_{\epsilon}(\textbf{p},j), b^*_{\epsilon'}(\textbf{p}',j)\} =
 \delta_{\epsilon \epsilon'} \delta(\textbf{p}-\textbf{p}') \ ,\\
 &[ a_{\epsilon}(\textbf{p},j), a^*_{\epsilon'}(\textbf{p}',j)] =
 \delta_{\epsilon \epsilon'} \delta(\textbf{p} - \textbf{p}') \ .
\end{split}
\end{equation}
\begin{equation}\label{4.4}
\begin{split}
 &\{ b_{ \epsilon}^{\sharp}(\textbf{p},j), b_{ \epsilon'}^{\sharp}(\textbf{p}',j')\}
 =0\ ,\\
 &\{ b_{ \epsilon}^{\sharp}(\xi;m,\widetilde j), b_{ \epsilon'}^{\sharp}(\textbf{p},j')\}
 =0\ .
\end{split}
\end{equation}
\begin{equation}\label{4.5}
\begin{split}
& [ a_{\epsilon}^{\sharp}(\textbf{p},j), a_{\epsilon'}^{\sharp}(\textbf{p}',j') ] = 0\ ,\\
& [ b_{ \epsilon}^{\sharp}(\textbf{p},j), a_{\epsilon'}^{\sharp}(\textbf{p}',j')] = 0\ ,\\
&[ b_{ \epsilon}^{\sharp}(\xi;m,\widetilde j), a_{\epsilon'}^{\sharp}(\textbf{p}',j)] = 0\ .
 \end{split}
\end{equation}
where $a^{\sharp}(resp.b^{\sharp})$ is $a(resp.b)$ or $a^{*}(resp.b^{*})$.

We now introduce
\begin{equation}\label{4.6}
\begin{split}
   & a_{ \epsilon}(j)(\vp) = \int_{\R^{3}} a_{
  \epsilon}(\textbf{p},j) \overline{\vp(\textbf{p})} \d^{3}\textbf{p}\ ,
   \quad a^*_{ \epsilon}(j)(\vp) = \int_{\R^{3}} a^*_{
  \epsilon}(\textbf{p},j) {\vp(\textbf{p})} \d^{3} \textbf{p}\ , \\
   & b_{ \epsilon}(j)(\vp) = \int_{\R^{3}} b_{
  \epsilon}(\textbf{p},j) \overline{\vp(\textbf{p})} \d^{3} \textbf{p}\ ,
  \quad  b^*_{ \epsilon}(j)(\vp) = \int_{\R^{3}} b^*_{
  \epsilon}(\textbf{p},j) {\vp(\textbf{p})} \d^{3} \textbf{p}\ .
\end{split}
\end{equation}

Moreover,
  for $\vp\in L^2(\R^{3})$, the operators $b_{ \epsilon}(j)$ and $b^*_{ \epsilon}(j)$ are bounded operators on $\gF^{j}_{a}$ satisfying
\begin{equation}\label{4.7}
  \| b^\sharp_{\epsilon}(j)(\vp)\| = \|\vp\|_{L^2}\ .
\end{equation}

From now on we only consider the helicity formalism because it is very useful in Physics.

Furthermore we restrict ourselves to the case of a massless fermion of helicity $j$ and we suppose that the massless fermions we consider are not their own antiparticles.

In that case S.Weinberg (see \cite{WeinbergII1964,WeinbergIII1964},\cite{Weinberg1965} and \cite[section 5.9]{Weinberg1995}) has shown that, if we construct a causal field for a massless particle of helicity $j$ by mimicking the construction for a massive particle of spin $j$, the associated causal field can be constructed only with the annihilation for the massless particle of helicity $j$ and the creation operator for the antiparticle with helicity $-j$. Moreover only the representations $(J_{1}, J_{2})$ of $SL(2,\C)$ such that $j= J_{2}- J_{1}$ are involved in the construction.

It follows that, if a massless fermion of helicity $j$ is not its own antiparticle, the helicity of the antiparticle is $-j$.

The massless fermion of helicity $j$ is associated to the unitary irreducible representation $\widetilde{U}^{[j]}$ and its antiparticle to the unitary irreducible representation $\widetilde{U}^{[-j]}$.

Let
\begin{equation}\label{4.8}
    \widetilde{U}^{[|j|]}=\widetilde{U}^{[j]}\oplus\widetilde{U}^{[-j]}
\end{equation}

Let $(J_{1},J_{2})$ be two spins. For every $M_{1}\in (-J_{1},-J_{1} + 1,\dots,J_{1}- 1,J_{1})$ and for every $M_{2}\in (-J_{2},-J_{2} + 1,\dots,J_{2}- 1,J_{2})$ we look for causal free fields, denoted by $\Big(\Phi^{[J_{1},J_{2}]\epsilon}_{M_{1}M_{2}}(x)\Big)_{M_{1}M_{2}}$, involving particles and antiparticles and satisfying the two fundamental conditions:

(a)The relativistic covariance law:
\begin{equation}\label{4.9}
    \begin{split}
       & \Gamma(\widetilde{U}^{[|j|]}(A,a)) (\Phi^{[J_{1},J_{2}]\epsilon}_{M_{1}M_{2}})(x) \Gamma(\widetilde{U}^{[|j|]}(A,a))^{-1} \\
       & =\sum_{M'_{1}M'_{2}}D^{[J_{1},J_{2}]}_{M_{1}M_{2}M'_{1}M'_{2}}(A^{-1})(\Phi^{[J_{1},J_{2}]\epsilon}_{M'_{1}M'_{2}})(\Lambda(A)x+ a)\ ,
    \end{split}
\end{equation}
where $x\in \R^4$.

and

(b)The microscopic causality
\begin{equation}\label{4.10}
\{\Phi^{[J_{1},J_{2}]\epsilon}_{M_{1}M_{2}}(x),\Phi^{[J_{1},J_{2}]\epsilon}_{M'_{1}M'_{2}}(y)\}=
        \{\Phi^{[J_{1},J_{2}]\epsilon}_{M_{1}M_{2}}(x),\Phi^{[J_{1},J_{2}]\epsilon,\dag}_{M'_{1}M'_{2}}(y)\}=0 \ .
\end{equation}
for x-y space-like.

As in \cite[3.47]{WeinbergIII1964} and \cite[section~5.9]{Weinberg1995} we set
\begin{equation}\label{4.11}
    \begin{split}
       & (\Phi^{[J_{1},J_{2}]\epsilon}_{ M_{1}M_{2}})(x) \\
       & =(\frac{1}{2\pi})^{\frac{3}{2}}\int \d^{3}\textbf{p}\big(\alpha (u_{ M_{1}M_{2}}^{[J_{1},J_{2}]})(\textbf{p},j)\mathrm{e}^{-ip.x}b_{\epsilon}(\textbf{p},j)\\
       &\quad\quad\quad\quad\quad +\beta(v_{ M_{1}M_{2}}^{[J_{1},J_{2}]})(\textbf{p},-j)\mathrm{e}^{ip.x}b_{\epsilon'}^{*}(\textbf{p},-j)\big)\ .
    \end{split}
\end{equation}
where $\epsilon\neq\epsilon'$.

We now study the transformation rules of the annihilation and creation operators by $\Gamma(\widetilde{U}^{[|j|]})$. By \cite{WeinbergIII1964} and \cite[section~5.9]{Weinberg1995} we easily get
\begin{equation}\label{4.12}
    \begin{split}
       &\Gamma(\widetilde{U}^{[j]}(A,a)\oplus I) b_{\epsilon}(\textbf{p}\ ,j)\Gamma(\widetilde{U}^{[j]}(A,a)\oplus I)^{-1}  \\
       &=\Big(\frac{|\textbf{p}_{\Lambda(A)p}|}{|\textbf{p}|}\Big)^{\frac{1}{2}}\mathrm{e}^{-ia.\Lambda(A)p}L^{-j}\big((A_{\Lambda(A)p}^{2})^{-1} A  A^{2}_{p}\big)b_{\epsilon}(\textbf{p}_{\Lambda(A)p}\ ,j)\ .
    \end{split}
\end{equation}

\begin{equation}\label{4.13}
    \begin{split}
       &\Gamma(\widetilde{U}^{[j]}(A,a)\oplus I) b^{*}_{\epsilon}(\textbf{p}\ ,j)\Gamma(\widetilde{U}^{[j]}(A,a)\oplus I)^{-1}  \\
       &=\Big(\frac{|\textbf{p}_{\Lambda(A)p}|}{|\textbf{p}|}\Big)^{\frac{1}{2}}\mathrm{e}^{ia.\Lambda(A)p}L^{j}\big((A_{\Lambda(A)p}^{2})^{-1} A  A^{2}_{p}\big)b^{*}_{\epsilon}(\textbf{p}_{\Lambda(A)p}\ ,j)\ .
    \end{split}
\end{equation}

\begin{equation}\label{4.14}
    \begin{split}
       &\Gamma(I\oplus\widetilde{U}^{[-j]}(A,a)) b^{*}_{\epsilon'}(\textbf{p}\ ,-j)\Gamma(I\oplus\widetilde{U}^{[-j]}(A,a))^{-1}  \\
        &=\Big(\frac{|\textbf{p}_{\Lambda(A)p}|}{|\textbf{p}|}\Big)^{\frac{1}{2}}\mathrm{e}^{ia.\Lambda(A)p}L^{-j}\big((A_{\Lambda(A)p}^{2})^{-1} A A^{2}_{p}\big)b^*_{\epsilon'}(\textbf{p}_{\Lambda(A)p}\ ,-j)\ .
    \end{split}
\end{equation}

and

\begin{equation}\label{4.15}
    \begin{split}
       &\Gamma(I\oplus\widetilde{U}^{[-j]}(A,a)) b_{\epsilon'}(\textbf{p}\ ,-j)\Gamma(I\oplus\widetilde{U}^{[-j]}(A,a))^{-1}  \\
       &=\Big(\frac{|\textbf{p}_{\Lambda(A)p}|}{|\textbf{p}|}\Big)^{\frac{1}{2}}\mathrm{e}^{-ia.\Lambda(A)p}L^{j}\big((A_{\Lambda(A)p}^{2})^{-1} A A^{2}_{p}\big)b_{\epsilon'}(\textbf{p}_{\Lambda(A)p}\ ,-j)\ .
    \end{split}
\end{equation}

From now on we omit the superscript $[J_{1},J_{2}]$. We shall introduce it again later on. By \eqref{4.9}),\eqref{4.11} and \eqref{4.12} we obtain, for $A\in SL(2,\C)$,
\begin{equation}\label{4.16}
    \begin{split}
       &\sum_{M'_{1}M'_{2}}\big(\frac{|\textbf{p}|}{|\textbf{p}_{\Lambda(A)p}|}\big)^{\frac{1}{2}}D_{M_{1}M_{2}M'_{1}M'_{2}}(A)u_{M'_{1}M'_{2}}(\textbf{p}\ ,j)  \\
       &=L^{j}\big((A_{\Lambda(A)p}^{2})^{-1} A  A^{2}_{p}\big)u_{M_{1}M_{2}}(\textbf{p}_{\Lambda(A)p}\ ,j)  \ .
    \end{split}
\end{equation}

and by \eqref{4.9}, \eqref{4.11} and \eqref{4.14} we get

\begin{equation}\label{4.17}
    \begin{split}
       &\sum_{M'_{1}M'_{2}}\big(\frac{|\textbf{p}|}{|\textbf{p}_{\Lambda(A)p}|}\big)^{\frac{1}{2}}D_{M_{1}M_{2}M'_{1}M'_{2}}(A)v_{M'_{1}M'_{2}}(\textbf{p}\ ,-j)  \\
       &=L^{j}\big((A_{\Lambda(A)p}^{2})^{-1} A  A^{2}_{p}\big)v_{M_{1}M_{2}}(\textbf{p}_{\Lambda(A)p}\ ,-j)  \ .
    \end{split}
\end{equation}

Setting $p$=$k_{0}$ we then get

\begin{equation}\label{4.18}
    \begin{split}
     &  u_{M_{1}M_{2}}(\textbf{p} ,j) =(|\textbf{p}|)^{-\frac{1}{2}}\sum_{M'_{1}M'_{2}}D_{M_{1}M_{2}M'_{1}M'_{2}}(A^{2}_{p})u_{M'_{1}M'_{2}}(\textbf{k}_{0},j) \ , \\
     & v_{M_{1}M_{2}}(\textbf{p} ,-j) =(|\textbf{p}|)^{-\frac{1}{2}}\sum_{M'_{1}M'_{2}}D_{M_{1}M_{2}M'_{1}M'_{2}}(A^{2}_{p})v_{M'_{1}M'_{2}}(\textbf{k}_{0} ,-j)\ .
    \end{split}
\end{equation}

Recall that $A^{2}_{p}$ is given by \eqref{2.44} .

\subsection{Computation of $u_{M_{1}M_{2}}(\textbf{k}_{0},j)$ and $v_{M_{1}M_{2}}(\textbf{k}_{0},-j)$ }\mbox{}

 Let $A_{\varphi}$ be the following rotation
 \begin{equation}\label{4.19}
    A_{\varphi}=\begin{pmatrix}
                  \mathrm{e}^{-i\frac{\varphi}{2}} & 0 \\
                  0 & \mathrm{e}^{i\frac{\varphi}{2}} \\
                \end{pmatrix}
 \end{equation}

 We have

 \begin{equation}\label{4.20}
    \begin{split}
    &\Lambda(A_{\varphi})k_{0}= k_{0}\ ,\\
    &(A_{\Lambda(A_{\varphi})k_{0}}^{2})^{-1} A_{\varphi}  A^{2}_{k_{0}}= A_{\varphi}\ ,  \\
    & A_{\varphi}= \mathrm{e}^{-i\varphi\frac{\sigma_{3}}{2}}\ .
    \end{split}
 \end{equation}

Combining this with \eqref{3.28},\eqref{3.29},\eqref{3.30},\eqref{3.35}, \eqref{3.38}, \eqref{4.16} and \eqref{4.17} we easily get

\begin{equation}\label{4.21}
    \begin{split}
     & \mathrm{e}^{-ij\varphi}u_{M_{1}M_{2}}(\textbf{k}_{0}, j)= \mathrm{e}^{-i\varphi(M_{1}+ M_{2})}u_{M_{1}M_{2}}(\textbf{k}_{0}, j)\ , \\
     & \mathrm{e}^{-ij\varphi}v_{M_{1}M_{2}}(\textbf{k}_{0}, -j)= \mathrm{e}^{-i\varphi(M_{1}+ M_{2})}v_{M_{1}M_{2}}(\textbf{k}_{0}, -j)\ .
    \end{split}
\end{equation}

This proves that $u_{M_{1}M_{2}}(\textbf{k}_{0}, j)$ and $v_{M_{1}M_{2}}(\textbf{k}_{0}, -j)$ are different from zero if and only if
\begin{equation}\label{4.22}
    M_{1} + M_{2} = j
\end{equation}

Let $A_{z}$ be the following transformation
\begin{equation}\label{4.23}
    A_{z} = \begin{pmatrix}
              1 & z \\
              0 & 1 \\
            \end{pmatrix}
\end{equation}

We have

\begin{equation}\label{4.24}
    \begin{split}
     & \Lambda(A_{z})k_{0}= k_{0}\ , \\
     & (A_{\Lambda(A_{z})k_{0}}^{2})^{-1} A_{z} A^{2}_{k_{0}}= A_{z} \ .
    \end{split}
\end{equation}

We get, for $z=\lambda + i\mu$,

\begin{equation}\label{4.25}
    \Lambda(A_{z}) = \begin{pmatrix}
                       1+ \frac{|z|^{2}}{2} & \lambda & -\mu &- \frac{|z|^{2}}{2} \\
                       \lambda & 1 & 0 & -\lambda \\
                       -\mu & 0 & 1 & \mu \\
                       \frac{|z|^{2}}{2} & \lambda & -\mu & 1 - \frac{|z|^{2}}{2} \\
                     \end{pmatrix}
\end{equation}

By \eqref{3.29} $A_{z}$ is the transformation

\begin{equation}\label{4.26}
    \mathrm{e}^{-i\big(\lambda(\mathrm{M}_{10}+ \mathrm{M}_{13}) -\mu(\mathrm{M}_{20}+ \mathrm{M}_{23})\big) }
\end{equation}

Here $\mathrm{M}_{10}$, $ \mathrm{M}_{13}$, $\mathrm{M}_{20}$ and $ \mathrm{M}_{23}$ are given in
\eqref{3.29}.

This yields
\begin{equation}\label{4.27}
   A_{z}= \mathrm{e}^{i\big(\lambda(\frac{\sigma_{2}}{2}-i\frac{\sigma_{1}}{2})+\mu(\frac{\sigma_{1}}{2}+i\frac{\sigma_{2}}{2})\big) }
\end{equation}

It follows from \eqref{2.37},\eqref{4.16},\eqref{4.17} and \eqref{4.24} that
\begin{equation}\label{4.28}
    \begin{split}
    & u_{M_{1}M_{2}}(\textbf{k}_{0}, j)=\sum_{M'_{1}M'_{2}}D^{[J_{1},J_{2}]}_{M_{1}M_{2}M'_{1}M'_{2}}(A_{z})u_{M'_{1}M'_{2}}(\textbf{k}_{0},j) \ , \\
    & v_{M_{1}M_{2}}(\textbf{k}_{0}, -j)=\sum_{M'_{1}M'_{2}}D^{[J_{1},J_{2}]}_{M_{1}M_{2}M'_{1}M'_{2}}(A_{z})v_{M'_{1}M'_{2}}(\textbf{k}_{0},-j) \ .
    \end{split}
\end{equation}

By \eqref{3.32} we have in the representation associated with $D^{[J_{1},J_{ 2}]}$
\begin{equation}\label{4.29}
    \begin{split}
    & \mathrm{M}_{10}+ \mathrm{M}_{13}= -i(\mathcal{A}_{1}-\mathcal{B}_{1})-(\mathcal{A}_{2}+\mathcal{B}_{2})\ , \\
    & \mathrm{M}_{20}+ \mathrm{M}_{23}=-i(\mathcal{A}_{2}-\mathcal{B}_{2})+(\mathcal{A}_{1}+\mathcal{B}_{1})\ .
    \end{split}
\end{equation}

By \eqref{4.26}, \eqref{4.28} and \eqref{4.29} we have
\begin{equation}\label{4.30}
    \sum_{M'_{1}M'_{2}}\big(-i(\mathcal{A}_{1}-\mathcal{B}_{1})-(\mathcal{A}_{2}+\mathcal{B}_{2})\big)_{M_{1}M_{2}M'_{1}M'_{2}}u_{M'_{1}M'_{2}}(\textbf{k}_{0},j)=0
\end{equation}
\begin{equation}\label{4.31}
    \sum_{M'_{1}M'_{2}}\big(-i(\mathcal{A}_{2}-\mathcal{B}_{2})+(\mathcal{A}_{1}+\mathcal{B}_{1})\big)_{M_{1}M_{2}M'_{1}M'_{2}}u_{M'_{1}M'_{2}}(\textbf{k}_{0},j)=0
\end{equation}

By \eqref{3.35} and \eqref{3.36} we get from \eqref{4.30} and \eqref{4.31}

\begin{equation}\label{4.32}
    \begin{split}
    &\sum_{M'_{1}}\big((\mathcal{J}^{(1)}_{2}+ i\mathcal{J}_{1}^{(1)})_{M_{1}M'_{1}}u_{M'_{1}M_{2}}(\textbf{k}_{0},j)+  \\
    &\sum_{M'_{2}}(\mathcal{J}_{2}^{(2)}-i\mathcal{J}_{1}^{(2)})_{M_{2}M'_{2}}u_{M_{1}M'_{2}}(\textbf{k}_{0},j)\big)=0\ .
    \end{split}
\end{equation}

\begin{equation}\label{4.33}
    \begin{split}
     &\sum_{M'_{1}}\big((-\mathcal{J}^{(1)}_{1}+ i\mathcal{J}_{2}^{(1)})_{M_{1}M'_{1}}u_{M'_{1}M_{2}}(\textbf{k}_{0},j)+  \\
     & \sum_{M'_{2}}(-\mathcal{J}_{1}^{(2)}-i\mathcal{J}_{2}^{(2)})_{M_{2}M'_{2}}u_{M_{1}M'_{2}}(\textbf{k}_{0},j)\big)=0\ .
    \end{split}
\end{equation}

It follows from \eqref{4.32} and \eqref{4.33} that
\begin{equation}\label{4.34}
    \sum_{M'_{1}}(\mathcal{J}^{(1)}_{1}- i\mathcal{J}_{2}^{(1)})_{M_{1}M'_{1}}u_{M'_{1}M_{2}}(\textbf{k}_{0},j)=0\ .
\end{equation}
\begin{equation}\label{4.35}
    \sum_{M'_{2}}(\mathcal{J}_{1}^{(2)}+i\mathcal{J}_{2}^{(2)})_{M_{2}M'_{2}}u_{M_{1}M'_{2}}(\textbf{k}_{0},j)=0\ .
\end{equation}

In view of \eqref{3.35},\eqref{4.34} and \eqref{4.35}  $u_{M_{1}M_{2}}(\textbf{k}_{0},j)$ is equal to zero unless
\begin{equation}\label{4.36}
    M_{1}=-J_{1}, M_{2}=J_{2}\ .
\end{equation}

By \eqref{4.28} the same is true for $v_{M'_{1}M'_{2}}(\textbf{k}_{0},-j)$ and by \eqref{4.22} we must have
\begin{equation}\label{4.37}
    j= J_{2}-J_{1}
\end{equation}

We finally set by applying the normalization used in Physics
\begin{equation}\label{4.38}
   u_{M_{1}M_{2}}(\textbf{k}_{0},j)= v_{M'_{1}M'_{2}}(\textbf{k}_{0},-j)= \delta_{M_{1},-J_{1}}\delta_{M_{2},J_{2}}2^{J_{1}+ J_{2}-1/2}
\end{equation}

This, together with \eqref{4.18}, yields
\begin{equation}\label{4.39}
    u_{M_{1}M_{2}}(\textbf{p} ,j)=v_{M_{1}M_{2}}(\textbf{p} ,-j) =(2|\textbf{p}|)^{-\frac{1}{2}}D^{[J_{1},J_{2}]}_{M_{1}M_{2}-J_{1}J_{2}}(A^{2}_{p}) \ .
\end{equation}

In view of \eqref{2.43} and \eqref{2.44} we obtain in the representation $D^{[J_{1},J_{2}]}(\textbf{.})$
\begin{equation}\label{4.40}
   D^{[J_{1},J_{2}]}( A^{2}_{p}) =D^{[J_{1},J_{2}]}(B_{\frac{\textbf{p}}{|\textbf{p}|}}\mathrm{e}^{-i\ln|\textbf{p}|(\widetilde{K_{3}})})
\end{equation}                                                                                                                                      where $B_{\frac{\textbf{p}}{|\textbf{p}|}}$ is given by \eqref{3.42} .

This, together with \eqref{3.35},\eqref{3.37} and \eqref{3.38}, yields

\begin{equation}\label{4.41}
     D^{[J_{1},J_{2}]}_{M_{1}M_{2}(-J_{1})J_{2}}(A^{2}_{p})=|\textbf{p}|^{J_{1}+J_{2}}({D}^{J_{1}}_{M_{1}(-J_{1})}(B_{\frac{\textbf{p}}{|\textbf{p}|}})D^{J_{2}}_{M_{2}J_{2}}(B_{\frac{\textbf{p}}{|\textbf{p}|}}))\ .
\end{equation}

Combining this with \eqref{3.46} and \eqref{4.39} we then get
\begin{equation}\label{4.42}
    \begin{split}
     &u^{[J_{1},J_{2}]}_{M_{1}M_{2}}(\textbf{p} ,j)=v^{[J_{1},J_{2}]}_{M_{1}M_{2}}(\textbf{p} ,-j)=(2|\textbf{p}|)^{J_{1}+J_{2}-1/2}\\  &\big(\mathrm{e}^{-i\arccos\frac{p^{3}}{|\textbf{p}|}(-\frac{p^{2}}{|\textbf{p}|}\mathcal{J}^{(1)}_{1}+\frac{p^{1}}{|\textbf{p}|}\mathcal{J}^{(1)}_{2})}\big)_{M_{1}(-J_{1})}\big(\mathrm{e}^{-i\arccos\frac{p^{3}}{|\textbf{p}|}(-\frac{p^{2}}{|\textbf{p}|}\mathcal{J}^{(2)}_{1}+ \frac{p^{1}}{|\textbf{p}|}\mathcal{J}^{(2)}_{2})}\big)_{M_{2}J_{2}} \ .
    \end{split}
\end{equation}

In \cite{WeinbergII1964} two particular cases are considered. For a left-handed particle with helicity $j<0$ one can choose $J_{2}=0$ and $J_{1}=-j=|j|$ and we have
\begin{equation}\label{4.43}
    u^{[-j,0]}_{s}(\textbf{p}, j)=(2|\textbf{p}|)^{|j|-1/2}\big(\mathrm{e}^{-i\arccos\frac{p^{3}}{|\textbf{p}|}(-\frac{p^{2}}{|\textbf{p}|}\mathcal{J}^{(-j)}_{1}+ \frac{p^{1}}{|\textbf{p}|}\mathcal{J}^{(-j)}_{2})}\big)_{sj}\ .
\end{equation}
where $s=(-|j|,-|j|+1,\cdots,|j|-1,|j|)$.

For a right-handed particle with helicity $j>0$ one can choose $J_{1}=0$ and $J_{2}=j$. We then get
\begin{equation}\label{4.44}
     u^{[0,j]}_{s}(\textbf{p}, j)=(2|\textbf{p}|)^{j-1/2}\big(\mathrm{e}^{-i\arccos\frac{p^{3}}{|\textbf{p}|}(-\frac{p^{2}}{|\textbf{p}|}\mathcal{J}^{(j)}_{1}+ \frac{p^{1}}{|\textbf{p}|}\mathcal{J}^{(j)}_{2})}\big)_{sj}\ .
\end{equation}
where $s=(-j,-j+1,\cdots,j-1,j)$.

This gives for a neutrino
\begin{equation}\label{4.45}
    u^{[-1/2,0]}_{s}(\textbf{p}, -1/2)=\big(\mathrm{e}^{-i\arccos\frac{p^{3}}{|\textbf{p}|}(-\frac{p^{2}}{|\textbf{p}|}\mathcal{J}^{(1/2)}_{1}+ \frac{p^{1}}{|\textbf{p}|}\mathcal{J}^{(1/2)}_{2})}\big)_{s(-1/2)}\ .
\end{equation}
where $s=-1/2,1/2.$
and for an antineutrino
\begin{equation}\label{4.46}
    u^{[0,1/2]}_{s}(\textbf{p}, 1/2)=\big(\mathrm{e}^{-i\arccos\frac{p^{3}}{|\textbf{p}|}(-\frac{p^{2}}{|\textbf{p}|}\mathcal{J}^{(1/2)}_{1}+ \frac{p^{1}}{|\textbf{p}|}\mathcal{J}^{(1/2)}_{2})}\big)_{s(1/2)}\ .
\end{equation}
where $s=-1/2,1/2.$

In order to satisfy the microscopic condition \eqref{4.10} with $u^{[J_{1},J_{2}]}_{M_{1}M_{2}}(\textbf{p} ,j)$ and $v^{[J_{1},J_{2}]}_{M_{1}M_{2}}(\textbf{p} ,-j)$ given by \eqref{4.42} S.Weinberg has shown in \cite[section~5.9]{Weinberg1995} that we must have $|\alpha|=|\beta|$ and that we can choose $\alpha=\beta$.

Thus, up to an over-all scale of the fields, we finally get
\begin{equation}\label{4.47}
    \begin{split}
       & (\Phi^{[J_{1},J_{2}]\epsilon}_{ M_{1}M_{2}})(x) =(\frac{1}{2\pi})^{\frac{3}{2}}\int \d^{3}\textbf{p} (2|\textbf{p}|)^{J_{1}+J_{2}-1/2}  \\
     &\big(\mathrm{e}^{-i\arccos\frac{p^{3}}{|\textbf{p}|}(-\frac{p^{2}}{|\textbf{p}|}\mathcal{J}^{(1)}_{1}+ \frac{p^{1}}{|\textbf{p}|}\mathcal{J}^{(1)}_{2})}\big)_{M_{1}(-J_{1})}
     \big(\mathrm{e}^{-i\arccos\frac{p^{3}}{|\textbf{p}|}(-\frac{p^{2}}{|\textbf{p}|}\mathcal{J}^{(2)}_{1}+ \frac{p^{1}}{|\textbf{p}|}\mathcal{J}^{(2)}_{2})}\big)_{M_{2}J_{2}}\\                          &\big(\mathrm{e}^{-ip.x}b_{\epsilon}(\textbf{p},j)+\mathrm{e}^{ip.x}b_{\epsilon'}^{*}(\textbf{p},-j)\big)\ .
    \end{split}
\end{equation}
where $J_{2}-J_{1}=j$.



For a left-handed particle of helicity $j<0$ we get

\begin{equation}\label{4.48}
    \begin{split}
       & (\Phi^{[-j,0]+}_{s})(x) =(\frac{1}{2\pi})^{\frac{3}{2}}\int \d^{3}\textbf{p} (2|\textbf{p}|)^{-j-1/2}
     \big(\mathrm{e}^{-i\arccos\frac{p^{3}}{|\textbf{p}|}(-\frac{p^{2}}{|\textbf{p}|}\mathcal{J}^{(-j)}_{1}+ \frac{p^{1}}{|\textbf{p}|}\mathcal{J}^{(-j)}_{2})}\big)_{sj}\\                          &\big(\mathrm{e}^{-ip.x}b_{+}(\textbf{p},j)+\mathrm{e}^{ip.x}b_{-}^{*}(\textbf{p},-j)\big)\ .
    \end{split}
\end{equation}
where $s=(-|j|,-|j|+1,\cdots,|j|-1,|j|)$

For a right- handed particle of helicity $j>0$ we obtain

\begin{equation}\label{4.49}
    \begin{split}
       & (\Phi^{[0,j]+}_{s})(x) =(\frac{1}{2\pi})^{\frac{3}{2}}\int \d^{3}\textbf{p} (2|\textbf{p}|)^{j-1/2}
       \big(\mathrm{e}^{-i\arccos\frac{p^{3}}{|\textbf{p}|}(-\frac{p^{2}}{|\textbf{p}|}\mathcal{J}^{(j)}_{1}+ \frac{p^{1}}{|\textbf{p}|}\mathcal{J}^{(j)}_{2})}\big)_{sj}\\                          &\big(\mathrm{e}^{-ip.x}b_{+}(\textbf{p},j)+\mathrm{e}^{ip.x}b_{-}^{*}(\textbf{p},-j)\big)\ .
    \end{split}
\end{equation}
where $s=(-j,-j+1,\cdots,j-1,j)$

For a neutrino we get

\begin{equation}\label{4.50}
    \begin{split}
     &(\Phi^{[-1/2,0]+}_{s})(x) =(\frac{1}{2\pi})^{\frac{3}{2}}\int \d^{3}\textbf{p}
     \big(\mathrm{e}^{-i\arccos\frac{p^{3}}{|\textbf{p}|}(-\frac{p^{2}}{|\textbf{p}|}\frac{\sigma_{1}}{2}+ \frac{p^{1}}{|\textbf{p}|}\frac{\sigma_{2}}{2})}\big)_{s(-1/2)}\\                          &\big(\mathrm{e}^{-ip.x}b_{+}(\textbf{p},-1/2)+\mathrm{e}^{ip.x}b_{-}^{*}(\textbf{p},1/2)\big)\ .
    \end{split}
\end{equation}

and for an antineutrino we obtain

\begin{equation}\label{4.51}
   \begin{split}
    &(\Phi^{[0,1/2]+}_{s})(x) =(\frac{1}{2\pi})^{\frac{3}{2}}\int \d^{3}\textbf{p}
     \big(\mathrm{e}^{-i\arccos\frac{p^{3}}{|\textbf{p}|}(-\frac{p^{2}}{|\textbf{p}|}\frac{\sigma_{1}}{2}+ \frac{p^{1}}{|\textbf{p}|}\frac{\sigma_{2}}{2})}\big)_{s(1/2)}\\                          &\big(\mathrm{e}^{-ip.x}b_{+}(\textbf{p},1/2)+\mathrm{e}^{ip.x}b_{-}^{*}(\textbf{p},-1/2)\big)\ .
    \end{split}
\end{equation}
Here $s=(1/2,-1/2)$.

\section{Definition of the model}

\setcounter{equation}{0}




 We consider a model which is a generalization of the weak decay of the nucleus  $^{60}_{27}Co$ into the nucleus $^{60}_{28}Ni^{*}$ , $e^{-}$ and $\overline{\nu}_{e}$.
\begin{equation}\label{5.1}
    ^{60}_{27}Co\rightarrow ^{60}_{28}Ni^{*} +  e^{-} + \overline{\nu}_{e}
\end{equation}
 Spin($^{60}_{27}Co)=5$ and Spin($^{60}_{28}Ni^{*})=4$. In this decay parity is not conserved.

Our model involves four particles : two bosons of mass $m_{1}>0$ and spin $j_{1}$ and of mass $m_{2}>0$ and spin $j_{2}$ respectively, a fermion of mass $m_{3}>0$ and spin $j_{3}$ and a massless fermion of helicity $-j_{4}$ which is the antiparticle of a massless fermion of helicity $j_{4}<0$ as it follows from the conservation of the leptonic number.

Set \,$\xi_{i}=(\textbf{p}_{i},s_{i})$ for each $i=1,2,3$,\ i.e., for the  massive bosons and fermion.We have, for each $i=1,2,3$, $\int \d \xi_{i}=\sum_{s_{i}}\int \d^{3}\textbf{p}_{i}$.

For the massless fermion we set $\xi_{4}=(\textbf{p}_{4},j_{4})$ and $\widetilde{\xi_{4}}=(\textbf{p}_{4},-j_{4})$ with $\int \d \xi_{4}=\int\d^{3}\textbf{p}_{4}$.

The Fock space of the system is

\begin{equation}\label{5.2}
    \gF=\gF^{[m_{1},j_{1}]}_{s}\otimes\gF^{[m_{2},j_{2}]}_{s}\otimes\gF^{[m_{3},j_{3}]}_{a}\otimes\gF^{[-j_{4}]}_{a}
\end{equation}

$\Omega$ shall denote the vacuum in $\gF$.

The free Hamiltonian $H_{0}$ is given by
\begin{multline}\label{5.3}
    H_{0}=\sum_{i=1}^{2}\int w^{i}(\xi_{i})a_{+}^{*}(\xi_{i};m_{i},j_{i})a_{+}(\xi_{i};m_{i},j_{i}) \d \xi_{i}\\ + \int w^{3}(\xi_{3})b_{+}^{*}(\xi_{3};m_{3},j_{3})b_{+}(\xi_{3};m_{3},j_{3}) \d \xi_{3} \\
    +\int w^{4}(\xi_{4})b_{-}^{*}(\widetilde{\xi}_{4})b_{-}(\widetilde{\xi}_{4}) \d \xi_{4}\quad\quad\quad\quad\quad\quad\quad\quad\quad
\end{multline}

The free relativistic energies of the massive bosons and fermion and of the massless fermion are given by

\begin{equation}\label{5.4}
  w^{i}(\xi_{i})=(|\textbf{p}_{i}|^{2}+m_{i}^{2})^{1/2}, i=1,2,3
\end{equation}
\begin{equation}\label{5.5}
   w^{4}(\xi_{4})=|\textbf{p}_{4}|
\end{equation}

From now on we suppose that
\begin{equation}\label{5.6}
\begin{split}
 & m_{1}>m_{2}> m_{3} \\
 &  m_{1}>m_{2}+ m_{3}\ .
\end{split}
\end{equation}

$H_{0}$ is a self-adjoint operator in $\gF$.

In the interaction representation the formal interaction,denoted by $H_{I}(t)$, is given by
\begin{equation}\label{5.7}
   H_{I}(t)=\int\d^{3}\textbf{x}\mathcal{H}(t,\textbf{x})
\end{equation}
The formal S-matrix, as defined in \cite[chapter 3]{Weinberg1995}, will be Poincar\'{e}-invariant if
\begin{equation}\label{5.8}
   \Gamma(U(A,a))\mathcal{H}(x)\Gamma(U(A,a))^{-1}=\mathcal{H}(\Lambda(A)x + a)
\end{equation}
\begin{equation}\label{5.9}
   [\mathcal{H}(x),\mathcal{H}(y)]=0, (x-y)^{2}\leq0
\end{equation}
See \cite[(3.5.12) and (3.5.14)]{Weinberg1995}.

The general form of of $\mathcal{H}(x)$ in terms of the causal free fields is given in \cite[(5.1.9) and (5.1.10)]{Weinberg1995}.

.The proofs of the Poincar\'{e} invariance of the S-matrix and of the causality property \eqref{5.9} are  formal ones. Later on Poincar\'{e} invariance will be broken because of the regularization of the kernels which appear in \cite[(4.4.1) and (4.4.2)]{Weinberg1995}




By \eqref{3.63} and \eqref{3.65} we get for each $i=1,2,3$

\begin{multline}\label{5.10}
  (^{C}u^{[J^{i}_{1},J^{i}_{2}]}_{M^{i}_{1}M^{i}_{2}})(\xi_{i};m_{i},j_{i})\\
        =(2\pi)^{-3/2}\frac{1}{\sqrt{2\omega_{\textbf{p}_{i}}}}\sum_{M^{i'}_{1}M^{i'}_{2}}\big(\mathrm{e}^{-(\ln\frac{|\textbf{p}_{i}|+\omega_{\textbf{p}_{i}}}{m})\sum_{l=1}^{3}\frac{p^{l}_{i}}{|\textbf{p}_{i}|} \mathcal{J}^{(i,1)}_{l}})_{M^{i}_{1}M^{i'}_{1}}\\(\mathrm{e}^{\ln\frac{|\textbf{p}_{i}|+\omega_{\textbf{p}_{i}}}{m})\sum_{l=1}^{3}\frac{p^{l}_{i}}{|\textbf{p}_{i}|}\mathcal{J}^{(i,2)}_{l}})_{M^{i}_{2}M^{i'}_{2}}.
       (J^{i}_{1}J^{i}_{2}j_{i}s_{i}|J^{i}_{1}M^{i}_{1}J^{i}_{2}M^{i'}_{2})\big).
\end{multline}

\begin{multline}\label{5.11}
   (^{H}u^{[J^{i}_{1},J^{i}_{2}]}_{M^{i}_{1}M^{i}_{2}})(\xi_{i};m_{i},j_{i})=(2\pi)^{-3/2}\\\frac{1}{\sqrt{2\omega_{\textbf{p}_{i}}}}\Big(\sum_{M^{i'}_{1}M^{i'}_{2}} (\frac{|\textbf{p}_{i}|+\omega_{\textbf{p}_{i}}}{m})^{M^{i'}_{2}-M^{i'}_{1}}\big(\mathrm{e}^{-i\arccos\frac{p^{3}_{i}}{|\textbf{p}_{i}|}(-\frac{p^{2}_{i}}{|\textbf{p}_{i}|}\mathcal{J}^{(i,1)}_{1}+ \frac{p^{1}_{i}}{|\textbf{p}_{i}|}\mathcal{J}^{(i,1)}_{2})}\big)_{M^{i}_{1}M^{i'}_{1}}\\   \big(\mathrm{e}^{-i\arccos\frac{p^{3}_{i}}{|\textbf{p}_{i}|}(-\frac{p^{2}_{i}}{|\textbf{p}_{i}|}\mathcal{J}^{(i,2)}_{1}+ \frac{p^{1}_{i}}{|\textbf{p}_{i}|}\mathcal{J}^{(i,2)}_{2})}\big)_{M^{i}_{2}M^{i'}_{2}}
   (J^{i}_{1}J^{i}_{2}j_{i}s_{i}|J^{i}_{1}M^{i'}_{1}J^{i}_{2}M^{i'}_{2})\Big).
\end{multline}

Here $J^{i}_{\textbf{.}}$ and $M^{i}_{\textbf{.}}$ are associated to the spin of the particle $i$. $\mathcal{J}^{(i,\textbf{.})}_{\textbf{.}}$ are the generators of the rotations in the representation $D^{J^{i}_{\textbf{.}}}(\textbf{.})$.

 For the massless fermion we only consider the helicity formalism and, by \eqref{4.42}, we set

\begin{multline}\label{5.12}
   u^{[J^{4}_{1},J^{4}_{2}]}_{M^{4}_{1}M^{4}_{2}}(\xi_{4})=(2|\textbf{p}_{4}|)^{J^{4}_{1}+J^{4}_{2}-1/2}  \big(\mathrm{e}^{-i\arccos\frac{p^{3}_{4}}{|\textbf{p}_{4}|}(-\frac{p^{2}_{4}}{|\textbf{p}_{4}|}\mathcal{J}^{(4,1)}_{1}+\frac{p^{1}_{4}}{|\textbf{p}_{4}|}\mathcal{J}^{(4,1)}_{2})}\big)_{M^{4}_{1}(-J^{4}_{1})}\\\big(\mathrm{e}^{-i\arccos\frac{p^{3}_{2}}{|\textbf{p}_{2}|}(-\frac{p^{2}_{2}}{|\textbf{p}_{2}|}\mathcal{J}^{(4,2)}_{1}+ \frac{p^{1}_{4}}{|\textbf{p}_{4}|}\mathcal{J}^{(4,2)}_{2})}\big)_{M^{4}_{2}J^{4}_{2}} \ .
\end{multline}

where $J^{4}_{\textbf{.}}$ and $M^{4}_{\textbf{.}}$ are associated to the spin of the massless fermion.

$\mathcal{J}^{(4,\textbf{.})}_{\textbf{.}}$
are the generators of the translations in the representation $D^{J^{4}_{\textbf{.}}}(\textbf{.})$.

By \eqref{3.70} we now set for the massive bosons, $i=1,2$,

\begin{equation}\label{5.13}
    ^{\sharp}_{1}\Phi^{[J^{i}_{1},J^{i}_{2}]}_{M^{i}_{1}M^{i}_{2}}(x)=(2\pi)^{-\frac{3}{2}}\int \d\xi_{i}(^{\sharp}u^{[J^{i}_{1},J^{i}_{2}]}_{M^{i}_{1}M^{i}_{2}})(\xi_{i};m_{i},j_{i})\mathrm{e}^{-ip_{3}.x}a_{+}(\xi_{i};m_{i},j_{i}).
\end{equation}

and, by \eqref{3.76}, for the massive fermion

\begin{equation}\label{5.14}
    ^{\sharp}_{1}\Psi^{[J^{3}_{1},J^{3}_{2}]}_{M^{3}_{1}M^{3}_{2}}(x)=(2\pi)^{-\frac{3}{2}}\int \d\xi_{3}(^{\sharp}u^{[J^{3}_{1},J^{3}_{2}]}_{M^{3}_{1}M^{3}_{2}})(\xi_{3};m_{3},j_{3})\mathrm{e}^{-ip_{3}.x}b_{+}(\xi_{3};m_{3},j_{3}).
\end{equation}

Finally, by \eqref{4.47}, for the massless fermion we let
\begin{equation}\label{5.15}
    (_{2}\Psi^{[J^{4}_{1},J^{4}_{2}]}_{ M^{4}_{1}M^{4}_{2}})(x) =(\frac{1}{2\pi})^{\frac{3}{2}}\int \d\xi_{4}u^{[J^{4}_{1},J^{4}_{2}]}_{M^{4}_{1}M^{4}_{2}}(\xi_{4})\mathrm{e}^{ip_{4}.x}b^{*}_{-}(\widetilde{\xi_{4}}).
\end{equation}

Let us now write down the formal interaction,denoted by $V_{I}$, of the three particles and antiparticles in the Schr\"{o}dinger representation.We have
\begin{equation}\label{5.16}
    V_{I}=\big(V^{(1)}_{I}+ V^{(2)}_{I}+\widetilde{V}^{(1)}_{I}+\widetilde{V}^{(2)}_{I}\big)
\end{equation}

$V^{(1)}_{I}$ is given by
\begin{equation}\label{5.17}
    \begin{split}
     V^{(1)}_{I}&=\int\d^{3}\textbf{x}\sum_{M^{1}_{1}M^{1}_{2}M^{2}_{1}M^{2}_{2}M^{3}_{1}M^{3}_{2}M^{4}_{1}M^{4}_{2}}   \big(g^{(1)}_{M^{1}_{1}M^{1}_{2}M^{2}_{1}M^{2}_{2}M^{3}_{1}M^{3}_{2}M^{4}_{1}M^{4}_{2}}\\
    &\big(_{2}\Psi^{[J^{4}_{1},J^{4}_{2}]}_{M^{4}_{1}M^{4}_{2}}\big)(0,\textbf{x})\big(^{\sharp}_{1}\Psi^{[J^{3}_{1},J^{3}_{2}]}_{M^{3}_{1}M^{3}_{2}}\big)^{\textbf{*}}(0,\textbf{x})
    \big(^{\sharp}_{1}\Phi^{[J^{2}_{1},J^{2}_{2}]}_{M^{2}_{1}M^{2}_{2}}\big)^{\textbf{*}}(0,\textbf{x})\big(^{\sharp}_{1}\Phi^{[J^{1}_{1},J^{1}_{2}]}_{M^{1}_{1}M^{1}_{2}}\big)(0,\textbf{x})\big).
    \end{split}
\end{equation}
$V^{(2)}_{I}$ is given by
\begin{equation}\label{5.18}
    \begin{split}
     V^{(2)}_{I}&=\int\d^{3}\textbf{x}\sum_{M^{1}_{1}M^{1}_{2}M^{2}_{1}M^{2}_{2}M^{3}_{1}M^{3}_{2}M^{4}_{1}M^{4}_{2}}   \big(\overline{g^{(1)}_{M^{1}_{1}M^{1}_{2}M^{2}_{1}M^{2}_{2}M^{3}_{1}M^{3}_{2}M^{4}_{1}M^{4}_{2}}}\\
    &\big(^{\sharp}_{1}\Phi^{[J^{1}_{1},J^{1}_{2}]}_{M^{1}_{1}M^{1}_{2}}\big)^{\textbf{*}}(0,\textbf{x})
    \big(^{\sharp}_{1}\Phi^{[J^{2}_{1},J^{2}_{2}]}_{M^{2}_{1}M^{2}_{2}}\big)(0,\textbf{x})\big(^{\sharp}_{1}\Psi^{[J^{3}_{1},J^{3}_{2}]}_{M^{3}_{1}M^{3}_{2}}\big)(0,\textbf{x})\big)
    \big(_{2}\Psi^{[J^{4}_{1},J^{4}_{2}]}_{M^{4}_{1}M^{4}_{2}}\big)^{\textbf{*}}(0,\textbf{x})\big).
    \end{split}
\end{equation}

and we have
\begin{equation}\label{5.19}
    \begin{split}
     \widetilde{V}^{(1)}_{I}&=\int\d^{3}\textbf{x}\sum_{M^{1}_{1}M^{1}_{2}M^{2}_{1}M^{2}_{2}M^{3}_{1}M^{3}_{2}M^{4}_{1}M^{4}_{2}}   \big(g^{(2)}_{M^{1}_{1}M^{1}_{2}M^{2}_{1}M^{2}_{2}M^{3}_{1}M^{3}_{2}M^{4}_{1}M^{4}_{2}}\\
    &\big(_{2}\Psi^{[J^{4}_{1},J^{4}_{2}]}_{M^{4}_{1}M^{4}_{2}}\big)(0,\textbf{x})\big(^{\sharp}_{1}\Psi^{[J^{3}_{1},J^{3}_{2}]}_{M^{3}_{1}M^{3}_{2}}\big)^{\textbf{*}}(0,\textbf{x})
    \big(^{\sharp}_{1}\Phi^{[J^{2}_{1},J^{2}_{2}]}_{M^{2}_{1}M^{2}_{2}}\big)^{\textbf{*}}(0,\textbf{x})\big(^{\sharp}_{1}\Phi^{[J^{1}_{1},J^{1}_{2}]}_{M^{1}_{1}M^{1}_{2}}\big)^{\textbf{*}}(0,\textbf{x})\big).
    \end{split}
\end{equation}

and

\begin{equation}\label{5.20}
    \begin{split}
     \widetilde{V}^{(2)}_{I}&=\int\d^{3}\textbf{x}\sum_{M^{1}_{1}M^{1}_{2}M^{2}_{1}M^{2}_{2}M^{3}_{1}M^{3}_{2}M^{4}_{1}M^{4}_{2}}   \big(\overline{g^{(2)}_{M^{1}_{1}M^{1}_{2}M^{2}_{1}M^{2}_{2}M^{3}_{1}M^{3}_{2}M^{4}_{1}M^{4}_{2}}}\\
    &\big(^{\sharp}_{1}\Phi^{[J^{1}_{1},J^{1}_{2}]}_{M^{1}_{1}M^{1}_{2}}\big)(0,\textbf{x})
    \big(^{\sharp}_{1}\Phi^{[J^{2}_{1},J^{2}_{2}]}_{M^{2}_{1}M^{2}_{2}}\big)(0,\textbf{x})\big(^{\sharp}_{1}\Psi^{[J^{3}_{1},J^{3}_{2}]}_{M^{3}_{1}M^{3}_{2}}\big)(0,\textbf{x})\big)
    \big(_{2}\Psi^{[J^{4}_{1},J^{4}_{2}]}_{M^{4}_{1}M^{4}_{2}}\big)^{\textbf{*}}(0,\textbf{x})\big).
    \end{split}
\end{equation}

$ \widetilde{V}^{(1)}_{I}$ and $ \widetilde{V}^{(2)}_{I}$ are responsible for the fact that the bare vacuum will not be an eigenvector of the total Hamiltonian asz expected in Physics.

$V_{I}$ is formally self adjoint.

By \cite[5.1.10]{Weinberg1995} the constants $g^{(i)}_{\textbf{.}}$,$i=1,2$, have to satisfy the following condition for $i=1,2$, and for every $A\in SL(2,\C)$
\begin{multline}\label{5.21}
    g^{(i)}_{M^{1}_{1}M^{1}_{2}M^{2}_{1}M^{2}_{2}M^{3}_{1}M^{3}_{2}M^{4}_{1}M^{4}_{2}}=\\
    \sum_{M^{1'}_{1}M^{1'}_{2}M^{2'}_{1}M^{2'}_{2}M^{3'}_{1}M^{3'}_{2}M^{4'}_{1}M^{4'}_{2}}D^{[J^{1}_{1},J^{1}_{2}]}_{M^{1'}_{1}M^{1'}_{2}M^{1}_{1}M^{1}_{2}}(A^{-1})
    D^{[J^{2}_{1},J^{2}_{2}]}_{M^{2'}_{1}M^{2'}_{2}M^{2}_{1}M^{2}_{2}}(A^{-1})\\    D^{[J^{3}_{1},J^{3}_{2}]}_{M^{3'}_{1}M^{3'}_{2}M^{3}_{1}M^{3}_{2}}(A^{-1})D^{[J^{4}_{1},J^{4}_{2}]}_{M^{4'}_{1}M^{4'}_{2}M^{4}_{1}M^{4}_{2}}(A^{-1})g^{(i)}_{M^{1'}_{1}M^{1'}_{2}M^{2'}_{1}M^{2'}_{2}M^{3'}_{1}M^{3'}_{2}M^{4'}_{1}M^{4'}_{2}}.
\end{multline}

The coefficients $g^{(i)}_{\textbf{.}}$,$i=1,2$, are associated with the coupling of the spins $J^{1}_{1}$,$J^{2}_{1}$, $J^{3}_{1}$ and $J^{4}_{1}$  and with the coupling of the spins $J^{1}_{2}$,$J^{2}_{2}$, $J^{3}_{2}$ and $J^{3}_{1}$ to make scalars.See \cite[section5]{Weinberg1995},\cite{WeinbergI1964} and \cite{Weinberg1969}.

After integrating with respect to $\textbf{x}$ we obtain
\begin{multline}\label{5.22}
  V^{(1)}_{I}=\\(2\pi)^{-3}\sum_{M^{1}_{1}M^{1}_{2}M^{2}_{1}M^{2}_{2}M^{3}_{1}M^{3}_{2}M^{4}_{1}M^{4}_{2}}g^{(1)}_{M^{1}_{1}M^{1}_{2}M^{2}_{1}M^{2}_{2}M^{3}_{1}M^{3}_{2}M^{4}_{1}M^{4}_{2}}\int\d\xi_{1}\d\xi_{2}\d\xi_{3}\d\xi_{4}\\\delta^{3}(\textbf{p}_{1}- \textbf{p}_{2}-\textbf{p}_{3}-\textbf{p}_{4}) \\ \big(u^{[J^{4}_{1},J^{4}_{2}]}_{M^{4}_{1}M^{4}_{2}}(\xi_{4})b^{*}_{-}(\widetilde{\xi_{4}})\overline{^{\sharp}u^{[J^{3}_{1},J^{3}_{2}]}_{M^{3}_{1}M^{3}_{2}})(\xi_{3};m_{3},j_{3})}b^{*}_{+}(\xi_{3};m_{3},j_{3})\big)\times\\                (\overline{^{\sharp}u^{[J^{2}_{1},J^{2}_{2}]}_{M^{2}_{1}M^{2}_{2}})(\xi_{2};m_{2},j_{2})}a^{*}_{+}(\xi_{2};m_{2},j_{2})^{\sharp}u^{[J^{1}_{1},J^{1}_{2}]}_{M^{1}_{1}M^{1}_{2}})(\xi_{1};m_{1},j_{1})a_{+}(\xi_{1};m_{1},j_{1})\big)\ .
\end{multline}
and

\begin{multline}\label{5.23}
   V^{(2)}_{I}=\\ (2\pi)^{-3}\sum_{M^{1}_{1}M^{1}_{2}M^{2}_{1}M^{2}_{2}M^{3}_{1}M^{3}_{2}M^{4}_{1}M^{4}_{2}}\overline{g^{(1)}_{M^{1}_{1}M^{1}_{2}M^{2}_{1}M^{2}_{2}M^{3}_{1}M^{3}_{2}M^{4}_{1}M^{4}_{2}}}\int\d\xi_{1}\d\xi_{2}\d\xi_{3}\d\xi_{4}\\\delta^{3}(\textbf{p}_{1}- \textbf{p}_{2}-\textbf{p}_{3}-\textbf{p}_{4}) \\   \big(\overline{^{\sharp}u^{[J^{1}_{1},J^{1}_{2}]}_{M^{1}_{1}M^{1}_{2}}(\xi_{1};m_{1},j_{1})}a^{*}_{+}(\xi_{1};m_{1},j_{1})^{\sharp}u^{[J^{2}_{1},J^{2}_{2}]}_{M^{2}_{1}M^{2}_{2}}(\xi_{2};m_{2},j_{2})a_{+}(\xi_{2};m_{2},j_{2})\big)\times\\\big(^{\sharp}u^{[J^{3}_{1},J^{3}_{2}]}_{M^{3}_{1}M^{3}_{2}}(\xi_{3};m_{3},j_{3}) b_{+}(\xi_{3}; m_{3},j_{3})\overline{^{\sharp}u^{[J^{4}_{1},J^{4}_{2}]}_{M^{4}_{1}M^{}_{2}}}(\xi_{4})b_{-}(\widetilde{\xi_{4}})\big)\ .
\end{multline}

together with
\begin{multline}\label{5.24}
  \widetilde{V}^{(1)}_{I}=\\ (2\pi)^{-3}\sum_{M^{1}_{1}M^{1}_{2}M^{2}_{1}M^{2}_{2}M^{3}_{1}M^{3}_{2}M^{4}_{1}M^{4}_{2}}g^{(2)}_{M^{1}_{1}M^{1}_{2}M^{2}_{1}M^{2}_{2}M^{3}_{1}M^{3}_{2}M^{4}_{1}M^{4}_{2}}\int\d\xi_{1}\d\xi_{2}\d\xi_{3}\d\xi_{4}\\\delta^{3}(\textbf{p}_{1}+ \textbf{p}_{2}+\textbf{p}_{3}+\textbf{p}_{4})  \\
  \big(u^{[J^{4}_{1},J^{4}_{2}]}_{M^{4}_{1}M^{4}_{2}}(\xi_{4})b^{*}_{-}(\widetilde{\xi_{4}})\overline{^{\sharp}u^{[J^{3}_{1},J^{3}_{2}]}_{M^{3}_{1}M^{3}_{2}})(\xi_{3};m_{3},j_{3})}b^{*}_{+}(\xi_{3};m_{3},j_{3})\big)\times\\                (\overline{^{\sharp}u^{[J^{2}_{1},J^{2}_{2}]}_{M^{2}_{1}M^{2}_{2}})(\xi_{2};m_{2},j_{2})}a^{*}_{+}(\xi_{2};m_{2},j_{2})\overline{^{\sharp}u^{[J^{1}_{1},J^{1}_{2}]}_{M^{1}_{1}M^{1}_{2}})(\xi_{1};m_{1},j_{1})}a^{*}_{+}(\xi_{1};m_{1},j_{1})\big)\ .
\end{multline}

and

\begin{multline}\label{5.25}
  \widetilde{V}^{(2)}_{I}= (2\pi)^{-3}\sum_{M^{1}_{1}M^{1}_{2}M^{2}_{1}M^{2}_{2}M^{3}_{1}M^{3}_{2}M^{4}_{1}M^{4}_{2}}\overline{g^{(2)}_{M^{1}_{1}M^{1}_{2}M^{2}_{1}M^{2}_{2}M^{3}_{1}M^{3}_{2}M^{4}_{1}M^{4}_{2}}}\int\d\xi_{1}\d\xi_{2}\d\xi_{3}\d\xi_{4}\\\delta^{3}(\textbf{p}_{1}+ \textbf{p}_{2}+\textbf{p}_{3}+\textbf{p}_{4}) \\
  \big(^{\sharp}u^{[J^{1}_{1},J^{1}_{2}]}_{M^{1}_{1}M^{1}_{2}}(\xi_{1};m_{1},j_{1})a_{+}(\xi_{1};m_{1},j_{1})^{\sharp}u^{[J^{2}_{1},J^{2}_{2}]}_{M^{2}_{1}M^{2}_{2}}(\xi_{2};m_{2},j_{2})a_{+}(\xi_{2};m_{2},j_{2})\big)\times\\\big(^{\sharp}u^{[J^{3}_{1},J^{3}_{2}]}_{M^{3}_{1}M^{3}_{2}}(\xi_{3};m_{3},j_{3}) b_{+}(\xi_{3}; m_{3},j_{3})\overline{^{\sharp}u^{[J^{4}_{1},J^{4}_{2}]}_{M^{4}_{1}M^{}_{2}}}(\xi_{4})b_{-}(\widetilde{\xi_{4}})\big)\ .
\end{multline}

In the Fock space $\gF$ the interaction $V_{I}$ is a highly singular operator due to the $\delta$-distributions that occur in the  $(V^{(\textbf{*})}_{I})'s$ and the $(\widetilde{V}^{(\textbf{*})}_{I})'s$ and because of the ultraviolet behavior of the functions $u^{[J^{(\textbf{.})}_{1},J^{(\textbf{.})}_{2}]}(\textbf{.})$ involved.

In order to get well defined operators in $\gF$ we have to substitute smoother kernels $F^{(\alpha)}(\xi_{1},\xi_{2})$, $G^{(\alpha)}(\xi_{3})$ and $\widetilde{G}^{(\alpha)}(\xi_{4})$, where $\alpha=1,2$, for the $\delta$-distributions.

 We then obtain a new operator denoted by $H_{I}$ and defined as follows in the Schr\"{o}dinger representation.

\begin{equation}\label{5.26}
    H_{I}=H^{(1)}_{I}+ (H^{(1)}_{I})^{\textbf{*}}+H^{(2)}_{I}+(H^{(2)}_{I})^{\textbf{*}}
\end{equation}

\begin{remark}\label{Rem5.1}
For the fermionic part of the interaction one could consider  kernels $G^{(\alpha)}(\xi_{3},\xi_{4})$ which are not products of $G^{(\alpha)}(\xi_{3})$ and $ \widetilde{G}^{(\alpha)}(\xi_{4})$ . Nevertheless this kernel should satisfy implicit conditions or should be very regular. It is better to consider kernels which are products of $G^{(\alpha)}(\xi_{3})$ and $ \widetilde{G}^{(\alpha)}(\xi_{4})$ because the conditions that each element of the product will have to satisfy will be more explicit and general.
\end{remark}

We have

\begin{multline}\label{5.27}
  H^{(1)}_{I}=\\(2\pi)^{-3}\sum_{M^{1}_{1}M^{1}_{2}M^{2}_{1}M^{2}_{2}M^{3}_{1}M^{3}_{2}M^{4}_{1}M^{4}_{2}}g^{(1)}_{M^{1}_{1}M^{1}_{2}M^{2}_{1}M^{2}_{2}M^{3}_{1}M^{3}_{2}M^{4}_{1}M^{4}_{2}}\int\d\xi_{1}\d\xi_{2}\d\xi_{3}\d\xi_{4}\\F^{(1)}(\xi_{1},\xi_{2}) G^{(1)}(\xi_{3})\widetilde{G}^{(1)}(\xi_{4})\\ \big(u^{[J^{4}_{1},J^{4}_{2}]}_{M^{4}_{1}M^{4}_{2}}(\xi_{4})b^{*}_{-}(\widetilde{\xi_{4}})\overline{^{\sharp}u^{[J^{3}_{1},J^{3}_{2}]}_{M^{3}_{1}M^{3}_{2}})(\xi_{3};m_{3},j_{3})}b^{*}_{+}(\xi_{3};m_{3},j_{3})\big)\times\\                (\overline{^{\sharp}u^{[J^{2}_{1},J^{2}_{2}]}_{M^{2}_{1}M^{2}_{2}})(\xi_{2};m_{2},j_{2})}a^{*}_{+}(\xi_{2};m_{2},j_{2})^{\sharp}u^{[J^{1}_{1},J^{1}_{2}]}_{M^{1}_{1}M^{1}_{2}})(\xi_{1};m_{1},j_{1})a_{+}(\xi_{1};m_{1},j_{1})\big)\ .
\end{multline}

\begin{multline}\label{5.28}
   (H^{(1)}_{I})^{\textbf{*}}=\\ (2\pi)^{-3}\sum_{M^{1}_{1}M^{1}_{2}M^{2}_{1}M^{2}_{2}M^{3}_{1}M^{3}_{2}M^{4}_{1}M^{4}_{2}}\overline{g^{(1)}_{M^{1}_{1}M^{1}_{2}M^{2}_{1}M^{2}_{2}M^{3}_{1}M^{3}_{2}M^{4}_{1}M^{4}_{2}}}\int\d\xi_{1}\d\xi_{2}\d\xi_{3}\d\xi_{4}\\\overline{F^{(1)}(\xi_{1},\xi_{2}) G^{(1)}(\xi_{3})\widetilde{G}^{(1)}(\xi_{4})}\\    \big(\overline{^{\sharp}u^{[J^{1}_{1},J^{1}_{2}]}_{M^{1}_{1}M^{1}_{2}}(\xi_{1};m_{1},j_{1})}a^{*}_{+}(\xi_{1};m_{1},j_{1})^{\sharp}u^{[J^{2}_{1},J^{2}_{2}]}_{M^{2}_{1}M^{2}_{2}}(\xi_{2};m_{2},j_{2})a_{+}(\xi_{2};m_{2},j_{2})\big)\times\\\big(^{\sharp}u^{[J^{3}_{1},J^{3}_{2}]}_{M^{3}_{1}M^{3}_{2}}(\xi_{3};m_{3},j_{3}) b_{+}(\xi_{3}; m_{3},j_{3})\overline{^{\sharp}u^{[J^{4}_{1},J^{4}_{2}]}_{M^{4}_{1}M^{}_{2}}}(\xi_{4})b_{-}(\widetilde{\xi_{4}})\big)\ .
\end{multline}

\begin{multline}\label{5.29}
  H^{(2)}_{I}=\\ (2\pi)^{-3}\sum_{M^{1}_{1}M^{1}_{2}M^{2}_{1}M^{2}_{2}M^{3}_{1}M^{3}_{2}M^{4}_{1}M^{4}_{2}}g^{(2)}_{M^{1}_{1}M^{1}_{2}M^{2}_{1}M^{2}_{2}M^{3}_{1}M^{3}_{2}M^{4}_{1}M^{4}_{2}}\int\d\xi_{1}\d\xi_{2}\d\xi_{3}\d\xi_{4}\\F^{(2)}(\xi_{1},\xi_{2}) G^{(2)}(\xi_{3})\widetilde{G}^{(2)}(\xi_{4})\\
  \big(u^{[J^{4}_{1},J^{4}_{2}]}_{M^{4}_{1}M^{4}_{2}}(\xi_{4})b^{*}_{-}(\widetilde{\xi_{4}})\overline{^{\sharp}u^{[J^{3}_{1},J^{3}_{2}]}_{M^{3}_{1}M^{3}_{2}})(\xi_{3};m_{3},j_{3})}b^{*}_{+}(\xi_{3};m_{3},j_{3})\big)\times\\                (\overline{^{\sharp}u^{[J^{2}_{1},J^{2}_{2}]}_{M^{2}_{1}M^{2}_{2}})(\xi_{2};m_{2},j_{2})}a^{*}_{+}(\xi_{2};m_{2},j_{2})\overline{^{\sharp}u^{[J^{1}_{1},J^{1}_{2}]}_{M^{1}_{1}M^{1}_{2}})(\xi_{1};m_{1},j_{1})}a^{*}_{+}(\xi_{1};m_{1},j_{1})\big)\ .
\end{multline}

\begin{multline}\label{5.30}
  (H^{(2)}_{I})^{\textbf{*}}= (2\pi)^{-3}\sum_{M^{1}_{1}M^{1}_{2}M^{2}_{1}M^{2}_{2}M^{3}_{1}M^{3}_{2}M^{4}_{1}M^{4}_{2}}\overline{g^{(2)}_{M^{1}_{1}M^{1}_{2}M^{2}_{1}M^{2}_{2}M^{3}_{1}M^{3}_{2}M^{4}_{1}M^{4}_{2}}}\int\d\xi_{1}\d\xi_{2}\d\xi_{3}\d\xi_{4}\\\overline{F^{(2)}(\xi_{1},\xi_{2}) G^{(2)}(\xi_{3})\widetilde{G}^{(2)}(\xi_{4})}\\
  \big(^{\sharp}u^{[J^{1}_{1},J^{1}_{2}]}_{M^{1}_{1}M^{1}_{2}}(\xi_{1};m_{1},j_{1})a_{+}(\xi_{1};m_{1},j_{1})^{\sharp}u^{[J^{2}_{1},J^{2}_{2}]}_{M^{2}_{1}M^{2}_{2}}(\xi_{2};m_{2},j_{2})a_{+}(\xi_{2};m_{2},j_{2})\big)\times\\\big(^{\sharp}u^{[J^{3}_{1},J^{3}_{2}]}_{M^{3}_{1}M^{3}_{2}}(\xi_{3};m_{3},j_{3}) b_{+}(\xi_{3}; m_{3},j_{3})\overline{^{\sharp}u^{[J^{4}_{1},J^{4}_{2}]}_{M^{4}_{1}M^{}_{2}}}(\xi_{4})b_{-}(\widetilde{\xi_{4}})\big)\ .
\end{multline}

The total Hamiltonian is then
\begin{equation}\label{5.31}
    H=H_{0}+ H_{I}
\end{equation}

We now give the conditions that the kernels $F^{\alpha}(.,.),G^{(\alpha)}(.),\widetilde{G}^{(\alpha)}(.)$ and the couplings constants $g^{(\alpha)}_{.}$ have to satisfy in order to associate with the formal operator $H$ a well defined self-adjoint operator in $\gF$.

\section{A self-adjoint Hamiltonian}

\setcounter{equation}{0}

Let $\mathfrak{D}$ denote the set of smooth vectors in $\gF$ for which only a finite number of components are different from zero and each component is smooth with a compact support. See \cite{BarbarouxGuillot2009} for a careful definition.
$H_{0}$ is essentially self-adjoint on $\mathfrak{D}$. The spectrum of $H_{0}$ is $[0,\infty)$ and $0$ is  a simple eigenvalue with $\Omega$ as eigenvector.

The set of thresholds of $H_{0}$, denoted by $\mathcal{T}$, is given by
\begin{equation}\label{6.1}
    \mathcal{T}=\{ p\,m_1 + q\,m_2 + r\,m_{3} ; (p,\,q,\,r)
 \in\N^3 \mbox{ and } p+q+r\geq 1  \}\ ,
\end{equation}

For each causal field corresponding to the massive particles we can choose either the canonical formalism or the helicity one. Nevertheless, from the physical point of view, the helicity formalism is very important and from now on we restrict ourselves to this formalism for each particle. For any other choice of formalisms  our results will the same because we can apply the same proof. Only constants and smallness conditions on the couplings constants would vary. We omit the details.

Thus, from now on, we omit the superscript $H$ in the formulae.

We now need to estimate the functions $(u^{[J^{i}_{1},J^{i}_{2}]}_{M^{i}_{1}M^{i}_{2}})(\xi_{i}; m_{i},j_{i})$ ,where $i=1,2,3$, and $u^{[J^{4}_{1},J^{4}_{2}]}_{M^{4}_{1}M^{4}_{2}}(\xi_{4})$.

By \eqref{5.11} and  one easily shows that there exist two constants $C^{i}$ for $i=1,2,3$ such that

\begin{equation}\label{6.2}
    |(u^{[J^{i}_{1},J^{i}_{2}]}_{M^{i}_{1}M^{i}_{2}})(\xi_{i}; m_{i},j_{i})|\leq C^{i}\big(1+ |\textbf{p}_{i}|\big)^{J^{i}_{1}+ J^{i}_{2}-1/2}
\end{equation}

Remark that $ C^{i}$ depends on $J^{i}_{1}, J^{i}_{2}$ and $j_{i}$.

By \eqref{5.12} we obtain

\begin{equation}\label{6.3}
    |u^{[J^{4}_{1},J^{4}_{2}]}_{M^{4}_{1}M^{4}_{2}}(\xi_{4})|\leq (2|\textbf{p}_{4}|)^{J^{4}_{1}+J^{4}_{2}-1/2}
\end{equation}

The estimate \eqref{6.3} is verified in the case of neutrinos and antineutrinos in the Standard Model.

From now on the kernels $F^{(\alpha)}(\xi_{1},\xi_{2})$,$G^{(\alpha)}(\xi_{3})$ and $\widetilde{G}^{(\alpha)}(\xi_{4})$ are supposed to satisfy the following hypothesis

\begin{hypothesis}\label{hypothesis:6.1}
For $\alpha=1,2$ we assume
\begin{equation*}
    \begin{split}
        &(i) \prod_{\beta=1,2,}\big(1 +|\textbf{p}_{\beta}|\big)^{J^{\beta}_{1}+ J^{\beta}_{2}-1/2}F^{(\alpha)}(.,.)\in L^{2}(\Sigma_{j_{1}}\times\Sigma_{j_{2}})  \\
        &(ii)\  \big(1 +|\textbf{p}_{3}|\big)^{J^{3}_{1}+ J^{3}_{2}-1/2}G^{(\alpha)}(.)\in L^{2}(\Sigma_{j_{3}})\\                                        &(iii)\ |\textbf{p}_{4}|^{J^{4}_{1}+J^{4}_{2}-1/2}\widetilde{G}^{(\alpha)}(.)\in L^{2}(\R^{3})
    \end{split}
\end{equation*}
\end{hypothesis}
\begin{remark}
Hypothesis 6.1 is an ultraviolet regularization of the model such that the Poincaré invariance is broken.
\end{remark}
Set
\begin{equation}\label{6.4}
   _{1}F^{(1)}(\xi_{1},\xi_{2})=\overline{u^{[J^{2}_{1},J^{2}_{2}]}_{M^{2}_{1}M^{2}_{2}}(\xi_{2};m_{2},j_{2})}u^{[J^{1}_{1},J^{1}_{2}]}_{M^{1}_{1}M^{1}_{2}})(\xi_{1};m_{1},j_{1})F^{(1)}(\xi_{1},\xi_{2}).  \end{equation}
\begin{equation}\label{6.5}
    _{2}F^{(1)}(\xi_{1},\xi_{2})=u^{[J^{2}_{1},J^{2}_{2}]}_{M^{2}_{1}M^{2}_{2}}(\xi_{2};m_{2},j_{2})\overline{u^{[J^{1}_{1},J^{1}_{2}]}_{M^{1}_{1}M^{1}_{2}})(\xi_{1};m_{1},j_{1})F^{(1)}(\xi_{1},\xi_{2})}.  \end{equation}
\begin{equation}\label{6.6}
    _{1}F^{(2)}(\xi_{1},\xi_{2})=\overline{u^{[J^{2}_{1},J^{2}_{2}]}_{M^{2}_{1}M^{2}_{2}}(\xi_{2};m_{2},j_{2})}\overline{u^{[J^{1}_{1},J^{1}_{2}]}_{M^{1}_{1}M^{1}_{2}})(\xi_{1};m_{1},j_{1})}F^{(2)}(\xi_{1},\xi_{2}). \end{equation}
\begin{equation}\label{6.7}
    _{2}F^{(2)}(\xi_{1},\xi_{2})=u^{[J^{2}_{1},J^{2}_{2}]}_{M^{2}_{1}M^{2}_{2}}(\xi_{2};m_{2},j_{2})u^{[J^{1}_{1},J^{1}_{2}]}_{M^{1}_{1}M^{1}_{2}})(\xi_{1};m_{1},j_{1})\overline{F^{(2)}(\xi_{1},\xi_{2})}. \end{equation}

For every $\Psi\in\mathfrak{D}$ we have

\begin{multline}\label{6.8}
    \|\int\d\xi_{1}\d\xi_{2}(_{1}F^{(1)}(\xi_{1},\xi_{2}))a^{*}_{+}(\xi_{2};m_{2},j_{2})a_{+}(\xi_{1};m_{1},j_{1})\Psi\| \\
    \leq C^{1}C^{2}\|\big( \prod_{\beta=1,2}(1 +|\textbf{p}_{\beta}|)^{J^{\beta}_{1}+ J^{\beta}_{2}-1/2}\big) F^{(1)}(.,.)\|_{L^{2}(\Sigma_{j_{1}}\times\Sigma_{j_{2}})}\\\times\big((\frac{1}{m_{1}}+\frac{1}{m_{2}})\|H_{0}\Psi\| +\frac{1}{2} \|\Psi\|\ \big)\quad\quad\quad\quad\quad\quad\quad\quad
    \end{multline}

\begin{multline}\label{6.9}
    \|\int\d\xi_{1}\d\xi_{2}(_{2}F^{(1)}(\xi_{1},\xi_{2}))a^{*}_{+}(\xi_{1};m_{1},j_{1})a_{+}(\xi_{2};m_{2},j_{2})\Psi\| \\
    \leq C^{1}C^{2}\|\big( \prod_{\beta=1,2}(1 +|\textbf{p}_{\beta}|)^{J^{\beta}_{1}+ J^{\beta}_{2}-1/2}\big) F^{(1)}(.,.)\|_{L^{2}(\Sigma_{j_{1}}\times\Sigma_{j_{2}})}\\\times\big((\frac{1}{m_{1}}+\frac{1}{m_{2}})\|H_{0}\Psi\| +\frac{1}{2} \|\Psi\| \big)\quad\quad\quad\quad\quad\quad\quad\quad
    \end{multline}

\begin{multline}\label{6.10}
    \|\int\d\xi_{1}\d\xi_{2}(_{1}F^{(2)}(\xi_{1},\xi_{2}))a^{*}_{+}(\xi_{2};m_{2},j_{2})a^{*}_{+}(\xi_{1};m_{1},j_{1})\Psi\| \\
    \leq C^{1}C^{2}\|\big( \prod_{\beta=1,2}(1 +|\textbf{p}_{\beta}|)^{J^{\beta}_{1}+ J^{\beta}_{2}-1/2}\big) F^{(2)}(.,.)\|_{L^{2}(\Sigma_{j_{1}}\times\Sigma_{j_{2}})}\\\times\big((\frac{1}{m_{1}}+\frac{1}{m_{2}})\|H_{0}\Psi\| + \|\Psi\| \big)\quad\quad\quad\quad\quad\quad\quad\quad
    \end{multline}

\begin{multline}\label{6.11}
    \|\int\d\xi_{1}\d\xi_{2}(_{2}F^{(2)}(\xi_{1},\xi_{2}))a_{+}(\xi_{1};m_{1},j_{1})a_{+}(\xi_{2};m_{2},j_{2})\Psi\| \\
    \leq C^{1}C^{2}\|\big( \prod_{\beta=1,2}(1 +|\textbf{p}_{\beta}|)^{J^{\beta}_{1}+ J^{\beta}_{2}-1/2}\big) F^{(2)}(.,.)\|_{L^{2}(\Sigma_{j_{1}}\times\Sigma_{j_{2}})}\\\times(\frac{1}{m_{1}}+\frac{1}{m_{2}})\|H_{0}\Psi\|\quad\quad\quad\quad\quad\quad\quad\quad
    \end{multline}

The estimates \eqref{6.8}-\eqref{6.11} are examples of $N_{\tau}$ estimates.The proof is similar to the one of \cite[Proposition 3.7]{BarbarouxDGuillot2004} and details are omitted.

Set

\begin{equation}\label{6.12}
    W(\textbf{p}_{1},\textbf{p}_{2},\textbf{p}_{3},\textbf{p}_{4})=\Big(|\textbf{p}_{4}|^{J^{4}_{1}+J^{4}-1/2}  \prod_{\beta=1,2,3}\big(1 +|\textbf{p}_{\beta}|\big)^{J^{\beta}_{1}+ J^{\beta}_{2}-1/2}\Big)
\end{equation}

\begin{equation}\label{6.13}
    C_{0}=(2\pi)^{-3}C^{1}C^{2}C^{3}\big(\prod_{\beta=1}^{3}(1 + 2J^{\beta}_{1})^{2}(1 + 2J^{\beta}_{2})^{2}\big)2^{J^{4}_{1}+J^{4}_{2}-1/2}(1 + 2J^{4}_{1})(1 + 2J^{4}_{2})                                                                                                         \end{equation}

\begin{equation}\label{6.14}
    g=\sup_{\alpha}\sup_{M^{1}_{1}M^{1}_{2}M^{2}_{1}M^{2}_{2}M^{3}_{1}M^{3}_{2}M^{4}_{1}M^{4}_{2}}|g^{(\alpha)}_{M^{1}_{1}M^{1}_{2}M^{2}_{1}M^{2}_{2}M^{3}_{1}M^{3}_{2}M^{4}_{1}M^{4}_{2}}|
\end{equation}

and
\begin{equation}\label{6.15}
    b=\frac{m_{1}m_{2}}{2(m_{1}+ m_{2})}
\end{equation}

By \eqref{3.11},\eqref{4.7},\eqref{5.26}-\eqref{5.30} and \eqref{6.8}-\eqref{6.11} we finally get for every $\Psi\in\mathfrak{D}$
\begin{multline}\label{6.16}
\|H_{I}\Psi\|\leq2gC_{0}(\frac{1}{m_{1}} + \frac{1}{m_{2}})\times\\\sum_{\alpha=1,2}\| W(\textbf{p}_{1},\textbf{p}_{2},\textbf{p}_{3},\textbf{p}_{4})F^{(\alpha)}(.,.)G^{(\alpha)}(.)\widetilde{G}^{(\alpha)}(.)\|_{ L^{2}(\Sigma_{j_{1}}\times\Sigma_{j_{2}}\times\Sigma_{j_{3}}\times\R^{3})}\\
\times(\|H_{0}\Psi\| +b \|\Psi\|)\ .\quad\quad\quad\quad\quad\quad\quad\quad\quad
    \end{multline}

We then have the following theorem

\begin{theorem}\label{thm6.2}
Let $g_{1}>0$ be such that
\begin{equation}\label{6.17}
\begin{split}
  &2C_{0}g_{1}(\frac{1}{m_{1}} + \frac{1}{m_{2}})\sum_{\alpha=1,2}\\                                     &\bigg(\int\big( W(\textbf{p}_{1},\textbf{p}_{2},\textbf{p}_{3},\textbf{p}_{4})\big)^{2}\\ &|F^{(\alpha)}(\xi_{1},\xi_{2})|^{2}|G^{(\alpha)}(\xi_{3})|^{2}|\widetilde{G}^{(\alpha)}(\xi_{4})|^{2}\d\xi_{1}\d\xi_{2}\d\xi_{3}\d\xi_{4}\bigg)^{1/2}<1\ .
\end{split}
\end{equation}
Then, for every $g$ satisfying $g\leq g_{1}$, $H$ is a self-adjoint operator in $\gF$ with domain $\mathcal{D}(H_{0})$ and $\mathfrak{D}$ is a core for $H$.
\end{theorem}

By \eqref{6.16} the proof of the theorem follows from the Kato-Rellich theorem.

We now set
\begin{equation}\label{6.18}
  \begin{split}                                                                                                                                                                          &K(F,G,\widetilde{G})=\sum_{\alpha=1,2}\| W(\textbf{p}_{1},\textbf{p}_{2},\textbf{p}_{3},\textbf{p}_{4})F^{(\alpha)}(.,.)G^{(\alpha)}(.)\widetilde{G}^{(\alpha)}(.)\|_{ L^{2}(\Sigma_{j_{1}}\times\Sigma_{j_{2}}\times\Sigma_{j_{3}}\times\R^{3})}.\\
  &C=\frac{C_{0}}{b}\\
  &B=C_{0}.
  \end{split}
\end{equation}

By\eqref{6.16} we get for every $\Psi\in\mathfrak{D}$

\begin{equation}\label{6.19}
  \|H_{I}\Psi\|\leq K(F,G,\widetilde{G})(C\|H_{0}\Psi\| +B \|\Psi\|)
\end{equation}

\section{Main results}

\setcounter{equation}{0}

We now wish to give statements about the existence of a ground state for the Hamiltonian $H$ together with the location of its spectrum and of its absolutely continuous spectrum. This is our first main result.

As in \cite{BarbarouxGuillot2009} and \cite{AschbacherGuillot2011} our second main result is the proof that the spectrum of $H$ is absolutely continuous in any interval  $(\inf\sigma(H),\inf\sigma(H)+m_{1}-\delta]$ for  $\delta<m_{1}$ and for $g$ sufficiently small whose smallness depends on $\delta$.

We shall now make the following additional assumptions on the kernels $\widetilde{G}^{(\alpha)}(\xi_{4})$ .
\begin{hypothesis}\label{hypothesis:7.1}There  exist constants $K(\widetilde{G})$ and $\widetilde{K}(\widetilde{G})$ such that for $\alpha=1,2,$ $i,l=1,2,3$,\,and $\sigma\geq 0$,
\begin{equation*}
    {(i)}\quad\big(|\textbf{p}_{4}|^{J^{4}_{1} + J^{4}_{2}-3/2} \big) \widetilde{G}^{(\alpha)}(\xi_{4})                                                               \in L^{2}(\R^{3})\ .
\end{equation*}

 \begin{equation*}
    {(ii)}\quad\Bigg(\int_{|\textbf{p}_{4}|\leq\sigma}\Big(|\textbf{p}_{4}|^{2(J^{4}_{1} + J^{4}_{2})-1}|\widetilde{G}^{(\alpha)}(\xi_{4})|^{2}\d\xi_{4}\Bigg)^{1/2}
    \leq K(\widetilde{G})\sigma\ .
 \end{equation*}

\begin{equation*}
    {(iii-a)}\quad\big(|\textbf{p}_{4}|^{J^{4}_{1} + J^{4}_{2}-1/2} \big) \big((\textbf{p}_{4}\cdot\nabla_{\textbf{p}_{4}}\big) \widetilde{G}^{(\alpha)}\big)(\xi_{4}) \in L^{2}(\R^{3})\ .
\end{equation*}

\begin{equation*}
    {(iii-b)}\Bigg(\int_{|\textbf{p}_{4}|\leq\sigma}|\textbf{p}_{4}|^{2(J^{4}_{1} + J^{4}_{2})-1}|\big((\textbf{p}_{4}\cdot\nabla_{\textbf{p}_{4}})\widetilde{G}^{(\alpha)}\big)(\xi_{4})|^{2}\d\xi_{4}\Bigg)^{1/2}\leq \widetilde{K}(\widetilde{G})\sigma\ .
\end{equation*}
\begin{equation*}
    {(iii-c)}\int_{\R^{3}}|\textbf{p}_{4}|^{2(J^{4}_{1} + J^{4}_{2})-1}(p_{2}^{i})^{2}(p_{2}^{l})^{2}\left|\frac{\partial^{2} \widetilde{G}^{(\alpha)}}{\partial p_{4}^{i} \partial p_{4}^{l}}(\xi_{4})\right|^{2}\d\xi_{4}<\infty\ .
\end{equation*}
\end{hypothesis}

The first main result is concerned with the existence of an unique ground state for $H$ and with the location of the spectrum of $H$ and of its absolutely continuous spectrum.

\begin{theorem}\label{theorem 7.2}
Assume that the kernels $F^{(\alpha)}$, $G^{(\alpha)}$ and $\widetilde{G}^{(\alpha)}$, $\alpha=1,2$, satisfy Hypothesis~\ref{6.1}, Hypothesis~ \ref{7.1}(i) and \ref{7.1}(ii).
Then there exists $g_{2} \in (0,g_{1}]$ such that $H$ has a ground state for $g\leq g_{2}$.
Furthermore, setting
\begin{equation*}
    E=\inf\sigma(H)
\end{equation*}
we have
\begin{equation*}
    \sigma(H)=\sigma_{\mathrm{ess}}(H)= [E,\infty)
\end{equation*}
with $E\leq 0$ .
\end{theorem}

$\sigma_{\mathrm{ess}}(H)$ is the essential spectrum of $H$.

In order to prove theorem 7.2 we first need to get an important result about the spectrum of the Hamiltonians with infrared cutoff.

Let us first define the cutoff operators which are the
Hamiltonians with infrared cutoff with respect to the momentum of
the massless fermion.

For that purpose, let $\chi_0(.)\in C^\infty(\R, [0,1])$ with $\chi_0=1$ on
$(-\infty, 1]$ and $\chi_0=0$ on $[2,\infty]$. For $\sigma>0$ and $\textbf{p}\in\R^3$, we set
\begin{equation}\label{7.1}
\begin{split}
 &  \chi_\sigma(\textbf{p}) = \chi_0(|\textbf{p}|/\sigma)\ , \\
 &  \tilde\chi^\sigma(\textbf{p}) = 1 - \chi_\sigma(\textbf{p})\ .
\end{split}
\end{equation}

The operator $H_{I, \sigma}$ is the interaction given by $(5.26)$-$(5.30)$
 associated with the kernel
$\tilde{\chi}^\sigma(\textbf{p}_4)\widetilde{G}^{(\alpha)}(\xi_4)$ instead of $\widetilde{G}^{(\alpha)}(\xi_4)$.

We then set
\begin{equation}\label{7.2}
  H_\sigma = H_0 + g H_{I,\sigma}\ .
\end{equation}

We now introduce
\begin{equation}\label{7.3}
    \begin{split}
      &\Sigma_{4,\sigma} =  \R^{3} \cap \{|\textbf{p}_4|<\sigma\}\ , \quad \Sigma_4^{\sigma} =\R^{3} \cap  \{|\textbf{p}_4|\geq\sigma\}  \\
      &\gF_{4,\sigma} = \gF_a(L^2(\Sigma_{4,\sigma})) \ ,   \quad\quad\quad  \gF_{4}^{\sigma} = \gF_a(L^2(\Sigma_{4}^{\sigma}))\ .
    \end{split}
 \end{equation}

 $\gF_{4,\sigma} \otimes \gF_{}^{\sigma}$ is the Fock space for the massless fermion.

 Now, we set
\begin{eqnarray}\label{7.4}
 & & \gF^{\sigma}   = \gF^{[m_{1},j_{1}]}_{s}\otimes\gF^{[m_{2},j_{2}]}_{s}\otimes\gF^{[m_{3},j_{3}]}_{a} \otimes \gF_{4}^{\sigma}\ \mbox{and}\quad
 \gF_{\sigma} = \gF_{4,\sigma}\ .
\end{eqnarray}
and we have
\begin{equation}\label{7.5}
 \gF \simeq \gF^{\sigma} \otimes \gF_{\sigma}\ .
\end{equation}

We further set

\begin{equation}\label{7.6}
\begin{split}
 & H_0^{i} =\int w^{i}(\xi_i)a^*_{+}(\xi_i)a_{+}(\xi_i)\d\xi_i\ ,\ \mbox{i=1,2,} \\
 & H_0^{3} =\int w^{3}(\xi_3)b^*_{+}(\xi_3)b_{+}(\xi_3)\d\xi_3\ , \\
 & H_0^{4} =\int w^{4}(\xi_4)b^*_{-}(\widetilde{\xi_4})b_{-}(\widetilde{\xi_4})\d\xi_4\ .
\end{split}
\end{equation}

and
\begin{equation}\label{7.7}
\begin{split}
 & H_0^{4,\sigma} =\int_{|\textbf{p}_{4}|>\sigma} w^{4}(\xi_4)b^*(\widetilde{\xi_4})b(\widetilde{\xi_4})\d\xi_4\ , \\
 & H_{0,\sigma}^{4} =
 \int_{|\textbf{p}_{4}|\leq\sigma} w^{4}(\xi_4)b^*(\widetilde{\xi_4})b(\widetilde{\xi_4})\d\xi_4\ .
\end{split}
\end{equation}

Then, on $\gF^\sigma\otimes\gF_\sigma$, we have
\begin{equation}\label{7.8}
  H_0^{4} = H_0^{4,\sigma} \otimes \1_\sigma +
  \1^\sigma\otimes H_{0,\sigma}^{4}\ ,
\end{equation}
where $\1^{\sigma}$ (resp. $\1_\sigma$) is the identity operator
on $\gF^{\sigma}$ (resp. $\gF_\sigma$).

Using the definitions
\begin{equation}\label{7.9}
 H^{\sigma} = H_\sigma|_{\gF^{\,\sigma}}\quad\mbox{and}\quad
 H_0^{\sigma} = H_0|_{\gF^\sigma}\ ,
\end{equation}
we get
\begin{equation}\label{7.10}
  H^{\sigma} = H_0^{1} + H_0^{2}+ H_0^{3}+ H_0^{4,\sigma} +g H_{I,\sigma}\quad\mbox{on}\ \gF^{\,\sigma}\ ,
\end{equation}
and
\begin{equation}\label{7.11}
  H_\sigma = H^{\sigma}\otimes\1_\sigma + \1^{\sigma}\otimes H_{0,\sigma}^{4}
  \quad\mbox{on}\ \gF^{\,\sigma}\otimes\gF_\sigma\ .
\end{equation}

Now, for $\delta\in\R$ with $  0< \delta< m_3$, we define the sequence $(\sigma_n)_{n\geq 0}$ by
\begin{equation}\label{7.12}
\begin{split}
 & \sigma_0 = 2m_3 +1 \ ,\\
 & \sigma_1 = m_{3}-\frac{\delta}{2}\ , \\
 & \sigma_{n+1} = \gamma \sigma_n \mbox{ for } n\geq 1\ ,
\end{split}
\end{equation}
where
\begin{equation}\label{7.13}
 \gamma = 1 - \frac{\delta}{2m_3 - \delta}\ .
\end{equation}
For $n\geq 0$, we then define the cutoff operators on $\gF^n=\gF^{\sigma_n}$ by
\begin{equation}\label{7.14}
H^n = H^{\sigma_n},\quad H_0^n = H_0^{\sigma_n},
\end{equation}
and we denote, for $n\geq 0$,
\begin{equation}\label{7.15}
 E^n = \inf\sigma(H^n) .
\end{equation}

We also define the cutoff operators on $\gF$ by

\begin{equation}\label{7.16}
H_{n} = H_{\sigma_n},\quad H_{0,n} = H_{0,\sigma_n},
\end{equation}
and we denote, for $n\geq 0$,
\begin{equation}\label{7.17}
 E_{n} = \inf\sigma(H_{n}) .
\end{equation}
Note that
\begin{equation}\label{7.18}
   E^n = E_{n}
\end{equation}

One easily shows that, for $g\leq g_{1}$,
\begin{equation}\label{7.19}
   |E^n| = |E_{n}|\leq \frac{gK(F,G,\widetilde{G})B}{1-g_{1}K(F,G,\widetilde{G})C}
\end{equation}
See \cite{BarbarouxGuillot2009,AschbacherGuillot2011} for a proof.

We now set

\begin{equation}\label{7.22}
  \widetilde{K}(F,G)=\sum_{\alpha=1,2} \|\big( \prod_{\beta=1,2,3}(1 +|\textbf{p}_{\beta}|)^{J^{\beta}_{1}+ J^{\beta}_{2}-1/2}\big) F^{(\alpha)}(.,.)G^{(\alpha)}(.)\|_{L^{2}(\Sigma_{j_{1}}\times\Sigma_{j_{2}}\times\Sigma_{j_{3}})}\ .
\end{equation}

\begin{equation}\label{7.21}
  \widetilde{C}=\frac{C}{1-g_{1}K(F,G,\widetilde{G})C}
\end{equation}

\begin{equation}\label{7.22}
  \widetilde{B}=\frac{B}{(1-g_{1}K(F,G,\widetilde{G})C)^{2}}
\end{equation}

\begin{multline}\label{7.23}
\widetilde{D}(F,G,\widetilde{G})=\max\left\{ \frac{4(2m_3+1)\gamma}{2 m_3 - \delta},\ 2\right\}\times\\
\widetilde{K}(F,G)K(\widetilde{G})\big( 2m_{3}\widetilde{C}+\widetilde{B})\big)
\end{multline}

 Let $g_1^{(\delta)}$ be such that
\begin{equation}\label{7.24}
 0 < g_1^{(\delta)} < \min\left\{1,\ g_1,\
 \frac{\gamma-\gamma^2}{3\widetilde{D}(F,G,\widetilde{G})}\right\}\ .
\end{equation}
and let
\begin{equation}\label{7.25}
    g_{3}=\frac{1}{2K(F,G,\widetilde{G})(2C+ B}
\end{equation}

Setting
\begin{equation}\label{7.26}
    g_2^{(\delta)}=\inf\{g_{3},g_1^{\delta)}\}
\end{equation}

and applying the same method as the one used for proving proposition 4.1 in \cite{AschbacherGuillot2011} we finally get the following result

\begin{proposition}\label{proposition7.3}
 Suppose that the kernels $F^{(\alpha)}(.,.)$, $G^{(\alpha)}(.)$ and $\widetilde{G}^{(\alpha)}(.)$,\\ $\alpha=1,2$, satisfy
 Hypothesis~\ref{hypothesis:6.1} and \ref{hypothesis:7.1}(ii) .
 Then, for $g\leq g_2^{(\delta)}$, $E^n$ is a simple eigenvalue of $H^n$ for $n\geq 1$, and $H^n$ does not have spectrum in the interval
$(E^n,\, E^n+(1 - 3 g\frac{\widetilde{D}(F,G,\widetilde{G})}{\gamma})\sigma_n)$.
\end{proposition}

\subsubsection{\textbf{ Proof of theorem 7.2}}\mbox{ }

\begin{proof}.We adapt the proof of theorem 3.3 in \cite{BarbarouxGuillot2009}. By Proposition 7.3  $H^n$ has
an unique ground state, denoted by $\phi^n$, in $\gF^n$ such that
\begin{equation}\label{7.27}
 H^n \phi^n = E^n \phi^n,\quad \phi^n\in\cD(H^n),\quad
 \|\phi^n\|=1,\quad n\geq 1\ .
\end{equation}
Therefore $H_n$ has an unique normalized ground state in $\gF$,
given by $\tilde\phi_n = \phi^n \otimes \Omega_n$, where
$\Omega_n$ is the vacuum state in $\gF_n$,
\begin{equation}\label{7.28}
 H_n \tilde\phi_n = E^n \tilde\phi_n,\quad \tilde\phi_n\in\cD(H_n),\quad
 \|\tilde\phi_n\|=1,\quad n\geq 1\ .
\end{equation}

Let $H_{I,n}$ be the interaction $ H_{I,\sigma_{n}}$. It follows from the pull-through formula that
\begin{equation}\label{7.29}
    (H_{0}+ H_{I,n})b_{-}(\widetilde{\xi_{4}})\tilde\phi_n=E_{n}b_{-}(\widetilde{\xi_{4}})\tilde\phi_n-w^{4}(\xi_4) b_{-}(\widetilde{\xi_{4}})\tilde\phi_n-\big(V_{n}^{1}(\xi_{4})+ V_{n}^{2}(\xi_{4})\big)\tilde\phi_n
\end{equation}
where

\begin{multline}\label{7.30}
  V_{n}^{1}(\xi_{4}) = \\
  (2\pi)^{-3}\sum_{M^{1}_{1}M^{1}_{2}M^{2}_{1}M^{2}_{2}M^{3}_{1}M^{3}_{2}M^{4}_{1}M^{4}_{2}}g^{(1)}_{M^{1}_{1}M^{1}_{2}M^{2}_{1}M^{2}_{2}M^{3}_{1}M^{3}_{2}M^{4}_{1}M^{4}_{2}}\int\d\xi_{1}\d\xi_{2}\d\xi_{3}\\  F^{(1)}(\xi_{1},\xi_{2}) G^{(1)}(\xi_{3})\tilde{\chi}^{\sigma_{n}}(\textbf{p}_4)\widetilde{G}^{(1)}(\xi_{4})\\
  \big(u^{[J^{4}_{1},J^{4}_{2}]}_{M^{4}_{1}M^{4}_{2}}(\xi_{4})\overline{u^{[J^{3}_{1},J^{3}_{2}]}_{M^{3}_{1}M^{3}_{2}})(\xi_{3};m_{3},j_{3})}b^{*}_{+}(\xi_{3};m_{3},j_{3})\big)\\               (\overline{u^{[J^{2}_{1},J^{2}_{2}]}_{M^{2}_{1}M^{2}_{2}})(\xi_{2};m_{2},j_{2})}a^{*}_{+}(\xi_{2};m_{2},j_{2})u^{[J^{1}_{1},J^{1}_{2}]}_{M^{1}_{1}M^{1}_{2}})(\xi_{1};m_{1},j_{1})a_{+}(\xi_{1};m_{1},j_{1})\big)\ .
\end{multline}

and

\begin{multline}\label{7.31}
   V_{n}^{2}(\xi_{4}) = \\
   (2\pi)^{-3}\sum_{M^{1}_{1}M^{1}_{2}M^{2}_{1}M^{2}_{2}M^{3}_{1}M^{3}_{2}M^{4}_{1}M^{4}_{2}}g^{(2)}_{M^{1}_{1}M^{1}_{2}M^{2}_{1}M^{2}_{2}M^{3}_{1}M^{3}_{2}M^{4}_{1}M^{4}_{2}}\int\d\xi_{1}\d\xi_{2}\d\xi_{3}\\  F^{(2)}(\xi_{1},\xi_{2}) G^{(2)}(\xi_{3})\tilde{\chi}^{\sigma_{n}}(\textbf{p}_4)\widetilde{G}^{(2)}(\xi_{4})\\
  \big(u^{[J^{4}_{1},J^{4}_{2}]}_{M^{4}_{1}M^{4}_{2}}(\xi_{4})\overline{u^{[J^{3}_{1},J^{3}_{2}]}_{M^{3}_{1}M^{3}_{2}})(\xi_{3};m_{3},j_{3})}b^{*}_{+}(\xi_{3};m_{3},j_{3})\big)\\               (\overline{u^{[J^{2}_{1},J^{2}_{2}]}_{M^{2}_{1}M^{2}_{2}})(\xi_{2};m_{2},j_{2})}a^{*}_{+}(\xi_{2};m_{2},j_{2})\overline{u^{[J^{1}_{1},J^{1}_{2}]}_{M^{1}_{1}M^{1}_{2}})(\xi_{1};m_{1},j_{1})}a^{*}_{+}(\xi_{1};m_{1},j_{1})\big)\ .
\end{multline}

We obtain
\begin{multline}\label{7.32}
    \|\big( V_{n}^{1}(\xi_{4})+V_{n}^{2}(\xi_{4})\big)\tilde\phi_n\|\leq\frac{g}{2} |\textbf{p}_{4}|^{J^{4}_{1}+ J^{4}_{2} -1/2}\\                                                    \big(\sum_{\alpha=1,2}\widetilde{K}(F,G)\big(C\|H_{0}\tilde\phi_n\| + 3B \big).
\end{multline}

It follows from \eqref{6.19} that, for every $g\leq g_{1}$,

\begin{equation}\label{7.33}
    \|H_{I,n}\tilde\phi_n\|\leq g K(F,G,\widetilde{G})(C\|H_{0}\tilde\phi_n\| + B)
\end{equation}
This yields
\begin{equation}\label{7.34}
    \|H_{0}\tilde\phi_n\|\leq |E_{n}| +  g K(F,G,\widetilde{G})(C\|H_{0}\tilde\phi_n\| + B)
\end{equation}

By \eqref{7.19} and \eqref{7.34} we get
\begin{equation}\label{7.35}
    \|H_{0}\tilde\phi_n\|\leq gK(F,G,\widetilde{G})\big(\frac{1}{(1-g_1K(F,G,\widetilde{G})C)^2} + \frac{1}{1-g_1K(F,G,\widetilde{G})C}\big)
\end{equation}
uniformly with respect to $n$.

We set
\begin{equation}\label{7.36}
  M = gK(F,G,\widetilde{G})\big((\frac{1}{(1-g_1K(F,G,\widetilde{G})C)^2} + \frac{1}{1-g_1K(F,G,\widetilde{G})C}\big)
\end{equation}

We then obtain

\begin{multline}\label{7.37}
    \|b_{-}(\widetilde{\xi_{4}})\tilde\phi_n\|\leq gC_{0}|\textbf{p}_{4}|^{J^{4}_{1}+ J^{4}_{2} -3/2}\\
   \big( \sum_{\alpha=1,2}|\widetilde{G}^{(\alpha)}(\xi_{4})\big)\widetilde{K}(F,G)
    \big((\frac{1}{m_{1}}+\frac{1}{m_{2}})M + 3/2 \big) \qquad \qquad \qquad \qquad \qquad \qquad \qquad \qquad
\end{multline}


Thus by Hypothesis~\eqref{6.1} and \eqref{7.1}(i) and from \eqref{7.37} there exists a constant $\mathcal{O}(F,G,\widetilde{G})>0$ such that
\begin{equation}\label{7.38}
    \int \|b_{-}(\widetilde{\xi_{4}})\tilde\phi_n\|^{2}\d\xi_{4}\leq g^{2}\mathcal{O}(F,G,\widetilde{G})
\end{equation}
uniformly with respect to n.

Since $\|\tilde\phi_n\|=1$, there exists a subsequence
$(n_k)_{k\geq 1}$, converging to $\infty$ such that
$(\tilde\phi_{n_k})_{k\geq 1}$ converges weakly to a state
$\tilde\phi\in\gF$. By adapting the proof of theorem 4.1 in \cite{BarbarouxDGuillot2004,Amouretal2007} it follows from \eqref{7.38} that
there exists $g_1$ such that $0<g_2\leq g_2^{(\delta)}$ and $\tilde\phi\neq0$ for any $g\leq g_2$. Thus $\tilde\phi$ is a ground state of $H$.

\end{proof}
Our second main result is devoted to the study of spectrum above the energy of a ground state.

Let $p$ be the operator in $L^{2}(\R^{3})$ associated to the position of the neutrinos and antineutrinos:
\begin{equation}\label{7.39}
  p=i\nabla_{\textbf{p}_{2}}\ ,
\end{equation}
and set
\begin{equation}\label{7.40}
  \langle p \rangle=( 1 + |p|^{2})^{1/2}
\end{equation}

The second quantized version $\d\Gamma(\langle p \rangle )$ is a self-adjoint operator in $\gF_{a}(L^{2}(\R^{3})$. We then define the position operator $P$ for the neutrinos and antineutrinos in $\gF$ by
\begin{equation}\label{7.41}
    P=\1\otimes\1\otimes\d\Gamma(\langle p \rangle )\otimes\1\otimes\1\otimes\1 + \1\otimes\1\otimes\1\otimes\d\Gamma(\langle p \rangle )\otimes\1\otimes\1\ .
\end{equation}
We then have our second main result devoted to the spectrum above the energy of the ground state and below the first threshold.
\begin{theorem}\label{theorem:7.4}
Suppose that the kernels $F^{(\alpha)}(.,.)$, $G^{(\alpha)}(.)$ and $\widetilde{G}^{(\alpha)}(.)$, $\alpha=1,2$, satisfy Hypothesis~\ref{6.1} and \ref{7.1}.
For any $\delta>0$ satisfying
$0<\delta<m_{3}$ there exists $g_{\delta}>0$ for $0<g\leq g_{\delta}$:
\begin{itemize}
\item[(i)] The spectrum of $H$ in $(E,E + m_{3}-\delta]$ is absolutely continuous.

\item[(ii)] For $s>1/2$,\, $\varphi$ and $\psi \in \gF$ the limits
\begin{equation*}
    \lim_{\epsilon\to 0}\big(\varphi,\, \langle P \rangle^{-s}(H-\lambda\pm i\epsilon)^{-1}\langle P \rangle^{-s} \psi\big)
\end{equation*}
exist uniformly for $\lambda$ in every compact subset of $(E,E + m_{3}-\delta]$.

\item[(iii)] For $s\in (1/2,\, 1)$ the map
\begin{equation*}
    \lambda\rightarrow\langle P \rangle^{-s}(H-\lambda\pm i0)^{-1}\langle P \rangle^{-s}
\end{equation*}
is locally H\"{o}lder continuous of degree $s-1/2$ in $(E,E + m_{1}-\delta]$.

\item[(iv)] For $s\in (1/2,\, 1)$ and $f\in C_{0}^{\infty}\big((E,E + m_{1}-\delta)\big)$ we have
\begin{equation*}
    \|(P + 1)^{-s} \mathrm{e}^{-itH}f(H)(P + 1)^{-s}\|= \mathcal{O}\big(t^{-(s-1/2)}\big)\, .
\end{equation*}
\end{itemize}
\end{theorem}

\subsubsection{\textbf{Proof of theorem 7.4}}\mbox{ }

\begin{proof}The following proposition will be fundamental for the proof of theorem~\ref{theorem:7.4}
A straightforward but lengthy computation shows the following fundamental estimates

\begin{proposition}\label{proposition:7.5}
There exists $C(J^{4}_{1},J^{4}_{2})>0$ such that we have
\begin{equation*}
    |p_{4}^{i}|\left|\frac{\partial u^{[J^{4}_{1},J^{4}_{2}]}_{M^{4}_{1}M^{4}_{2}}}{\partial p_{4}^{i} }(\xi_{4})\right|\leq C(J^{4}_{1},J^{4}_{2})|\textbf{p}_{4}|^{J^{4}_{1} + J^{4}_{2}-1/2}
\end{equation*}
\begin{equation*}
   |p_{4}^{i}||p_{4}^{l}|\left|\frac{\partial^{2} u^{[J^{4}_{1},J^{4}_{2}]}_{M^{4}_{1}M^{4}_{2}}}{\partial p_{4}^{i}  \partial p_{4}^{l}}(\xi_{4})\right|\leq C(J^{4}_{1},J^{4}_{2})|\textbf{p}_{4}|^{J^{4}_{1} + J^{4}_{2}-1/2}
\end{equation*}
for $i,l=1,2,3$.
\end{proposition}

In the proof of proposition~\ref{proposition:7.5} we explicitly use the norm of the operators $\mathcal{J}^{(2,\textbf{.})}_{\textbf{.}}$ associated with the $l^{2}$-norm of $\C^{(2J^{2}+ 1)}$.

We now introduce a strict Mourre inequality.

Let us set
\begin{equation}\label{7.42}
 \tau = 1 - \frac{\delta}{2(2m_3 - \delta)}\ .
\end{equation}

We now introduce $\chi^{(\tau)}\in C^\infty(\R,\, [0,1])$ be such that
\begin{equation}\label{7.43}
 \chi^{(\tau)}(\lambda) = \left\{
  \begin{array}{ll}
   1 & \mbox{ for } \lambda\in(-\infty,\, \tau]\, , \\
   0 & \mbox{ for } \lambda\in[1,\, \infty) \, .
  \end{array}
  \right.
\end{equation}

and we set, for all $\textbf{p}_{4}\in\R^3$ and $n\geq 1$,
\begin{equation}\label{7.44}
 \chi_n^{(\tau)} (\textbf{p}_{4}) =
 \chi^{(\tau)}\left(\frac{|\textbf{p}_{4}|}{\sigma_n}\right),
\end{equation}
\begin{equation}\label{7.45}
 a_n^{(\tau)} = \chi_n^{(\tau)}(\textbf{p}_{4}) \frac12
 \left(\textbf{p}_{4}\cdot i\nabla_{\textbf{p}_{4}}
 + i\nabla_{\textbf{p}_{4}}\cdot \textbf{p}_{4}\right)
 \chi_n^{(\tau)}(\textbf{p}_{4})\ ,
\end{equation}
and
\begin{equation}\label{7.46}
 A^{(\tau)}_{n} = \1\otimes\1\otimes\1\otimes\d \Gamma(a_n^{(\tau)})\, ,
\end{equation}
The operators $a_n^{(\tau)}$ and $A^{(\tau)}_{n}$ are
self-adjoint and we also have
\begin{equation}\label{7.47}
 a_n^{(\tau)} = \frac12
 \left(\chi_n^{(\tau)}(\textbf{p}_{4})^{2} \textbf{p}_{4}\cdot
 i\nabla_{\textbf{p}_{4}} +
 i\nabla_{\textbf{p}_{4}} \cdot\textbf{p}_{4} \, \chi_n^{(\tau)}(\textbf{p}_{4})^{2}\right) \ .
\end{equation}
Let now $N$ be the smallest integer such that
\begin{equation}\label{7.48}
 N\gamma \geq 1\ .
\end{equation}

Let us define
\begin{equation}\label{7.49}
 \epsilon_\gamma = \min\left\{\frac{1}{2N} \Big(1-\frac{3 g\widetilde{D}_{\delta}(F,G,\widetilde{G})
 }{\gamma} -\gamma\Big),\, \frac{\tau-\gamma}{4}\right\}\ ,
\end{equation}
and choose $f\in C_0^\infty(\R)$ such that $0\leq f \leq 1$ and
\begin{equation}\label{7.50}
f(\lambda)=
 \left\{
 \begin{array}{ll}
  1 & \mbox{ if } \lambda\in [(\gamma-\epsilon_\gamma)^2,\, \gamma
  + \epsilon_\gamma]\, , \\
  0 & \mbox{ if } \lambda > \gamma+ 2\epsilon_\gamma\, , \\
  0 & \mbox{ if } \lambda < (\gamma - 2\epsilon_\gamma)^2\, .
 \end{array}
 \right.
\end{equation}

We now define, for $n\geq 1$,
\begin{equation}\label{7.51}
 f_n(\lambda) = f\left(\frac{\lambda}{\sigma_n}\right)\ ,
\end{equation}

Let $P^n$ denote the ground state
projection of $H^n$ and let $H_{0, n}^{4}$ denote $H_{0, \sigma_{n}}^{4}$.

It follows from Proposition 7.3
that, for $n\geq 1$ and $g\leq g_\delta^{(2)}$,
\begin{equation}\label{7.52}
 f_n(H_n - E_n) = P^n\otimes f_n(H_{0, n}^{4})\ .
\end{equation}

For $E=\inf\sigma(H)$ and any interval $\Delta$, let
$E_\Delta(H-E)$ be the spectral projection for the operator
$(H-E)$ onto $\Delta$. Consider, for $n\geq 1$,
\begin{equation}\label{7.53}
 \Delta_n = [(\gamma - \epsilon_\gamma)^2\sigma_n,\,
 (\gamma+\epsilon_\gamma)\sigma_n]\, .
\end{equation}

Now, by adapting the proof of theorem~\ref{5.1} (Mourre inequality) in \cite{AschbacherGuillot2011} and by applying proposition~\ref{7.5} together with Hypothesis~\ref{6.1}and Hypothesis~\ref{7.1}  we prove the existence of a constant $\widetilde{C}_{\delta}(F,G,\widetilde{G})>0$ such that for every $g\leq\inf\big(g_{2},g^{(2)}_{\delta}\big)$ we have

\begin{equation}\label{7.54}
 f_n(H -E) [H,\, iA^{(\tau)}_{n}] f_n(H-E)
 \geq \frac{\gamma^2}{N^2} \sigma_n f_n(H-E)^2 - g\sigma_n
 \tilde{C}_\delta(F,G,\widetilde{G})\ .
\end{equation}

Multiplying both sides of \eqref{7.54} with $E_{\Delta_n}(H-E)$ we obtain

\begin{equation}\label{7.55}
 E_{\Delta_n} (H-E) [H,\, iA^{(\tau)}_{n}] E_{\Delta_n} (H-E)
 \geq \left( \frac{\gamma^2}{N^2} - g\tilde{C}_\delta(F,G,\widetilde{G})\right)
 \sigma_n E_{\Delta_n}(H-E)\, .
\end{equation}


Choosing a constant $g_\delta^{(3)}$ such that
\begin{equation}\label{7.56}
 g_\delta^{(3)} < \min\Big\{ g_2,\, g_\delta^{(2)},
 \frac{\gamma^2}{N^2} \frac{1}{\tilde{C}_\delta(F,G,\widetilde{G})}\Big\}\, ,
\end{equation}
we finally get the following strict Mourre inequality for every $g\leq g_\delta^{(3)}$ and for $n\geq1$
\begin{equation}\label{7.57}
 E_{\Delta_n}(H-E) \, [H,\, iA^{(\tau)}_{n}]\, E_{\Delta_n}(H-E)
 \geq C_\delta(F,G,\widetilde{G}) \frac{\gamma^2}{N^2} \,\sigma_n\, E_{\Delta_n}(H-E)\, .
\end{equation}
where
\begin{equation}\label{7.58}
    C_\delta(F,G,\widetilde{G}) =
(1 - N^2
 \tilde{C}_\delta(F,G,\widetilde{G}) g_\delta^{(3)}/\gamma^2)>0 \ .
\end{equation}

After proving a strict Mourre inequality it remains to prove that H is of class $C^{2}\big(A^{(\tau)}_{n}\big)$ in order to apply the commutator theory. See \cite{Mourre1981,Amreinetal1996,Sahbani1997,GeorgescuGerard1999,GoleniaJecko2007,Gerard2008}.

In fact, according to \cite{Sahbani1997}, it suffices to prove that H is locally of class $C^{2}\big(A^{(\tau)}_{n}\big)$ in $(-\infty,m_{3}-\delta/2)$.

This means that, for any $\varphi\in C_{0}^{\infty}(-\infty,m_{3}-\delta/2)$, $\varphi(H)$ is of class $C^{2}\big(A^{(\tau)}_{n}\big)$, i.e., $t\rightarrow \textrm{e}^{-iA^{(\tau)}_{n}t}\varphi(H)\textrm{e}^{iA^{(\tau)}_{n}t}\psi$ is twice continuously differentiable for all $\varphi\in C_{0}^{\infty}(-\infty,m_{3}-\delta/2)$ and $\psi\in \gF$.

Set

\begin{equation}\label{7.59}
    A_{n,t}^{(\tau)}=\frac{\textrm{e}^{-iA^{(\tau)}_{n}t}-1}{t}
\end{equation}

By using the proof given in \cite{BarbarouxGuillot2009} H is locally of class $C^{2}\big(A^{(\tau)}_{n}\big)$ in $(-\infty,m_{3}-\delta/2)$ if
we show that

\begin{equation}\label{7.60}
 \sup_{0<|t| \leq 1} \|[A_{n,t}^{(\tau)},\, [A_{n,t}^{(\tau)},\, H]] (H+i)^{-1} \| <\infty
\end{equation}

The operator $a_n^{(\tau)}$ is associated to the following $C^{\infty}$- vector field in $\R^{3}$:
\begin{equation}\label{7.61}
    V_{n}^{(\tau)}(\textbf{p}_{4})= \chi_n^{(\tau)} (\textbf{p}_{4})^{2}\textbf{p}_{4}
\end{equation}

Let $\phi_{n,t}^{(\tau)}(.):\R^{3}\longmapsto \R^{3}$ be the corresponding flow generated by $ V_{n}^{(\tau)}(\textbf{p}_{4})$:
\begin{equation}\label{7.62}
    \begin{split}
      \frac{\textrm{d}}{\textrm{d}t}\phi_{n,t}^{(\tau)}(\textbf{p}_{4}) & = V_{n}^{(\tau)}\big(\phi_{n,t}(\textbf{p}_{4})\big) \\
       \phi_{n,0}^{(\tau)}(\textbf{p}_{4}) & = \textbf{p}_{4}
    \end{split}
\end{equation}

We have
\begin{equation}\label{7.63}
    \mathrm{e}^{-|t|}|\textbf{p}_{4}|\leq |\phi_{n,t}^{(\tau)}(\textbf{p}_{4})|\leq \mathrm{e}^{|t|}|\textbf{p}_{4}|
\end{equation}

$\phi_{n,t}^{(\tau)}(\textbf{p}_{4})$ induces a one-parameter group of unitary operators $U_{n}^{(\tau)}(t)$
  in $L^{2}(\R^{3})$ defined by

  \begin{equation}\label{7.64}
    \left(U_{n}^{(\tau)}(t)f\right)(\textbf{p}_{4})=\left(\det\nabla\phi_{n,t}^{(\tau)}(\textbf{p}_{4})\right)^{\frac{1}{2}}f(\phi_{n,t}^{(\tau)}(\textbf{p}_{4})).
  \end{equation}
 $a_n^{(\tau)}$ is the generator of $U_{n}^{(\tau)}(t)$, i.e.,

\begin{equation}\label{7.65}
    U_{n}^{(\tau)}(t)=\textrm{e}^{-ia_n^{(\tau)}t}
\end{equation}

We have,for every $\psi\in\mathcal{D}(H)$
\begin{equation}\label{7.66}
    [A_{n,t}^{(\tau)},\, [A_{n,t}^{(\tau)},\, H]]\psi=\frac{1}{t^{2}}\textrm{e}^{2iA_n^{(\tau)}t}\left(\textrm{e}^{-2iA_n^{(\tau)}t}H\textrm{e}^{2iA_n^{(\tau)}t}-2\textrm{e}^{-iA_n^{(\tau)}t}H\textrm{e}^{iA_n^{(\tau)}t} + H\right)\psi
\end{equation}

In particular we get
\begin{equation}\label{7.67}
   [A_{n,t}^{(\tau)},\, [A_{n,t}^{(\tau)},\, H_{0}]]\psi=\frac{1}{t^{2}}\textrm{e}^{2iA_n^{(\tau)}t}\left(\mathrm{d}\Gamma(|\phi_{n,2t}^{(\tau)}(\textbf{p}_{4})|-2|\phi_{n,t}^{(\tau)}(\textbf{p}_{4})|+ |\textbf{p}_{4}|)\right)\psi\ .
\end{equation}

We note that

\begin{equation}\label{7.68}
    \frac{1}{t^{2}}\left||\phi_{n,2t}^{(\tau)}(\textbf{p}_{4})|-2|\phi_{n,t}^{(\tau)}(\textbf{p}_{4})|+ |\textbf{p}_{4}|\right|\leq  \sup_{|s|\leq 2|t|} \left| \frac{\partial^2}{\partial s^2}
 |\phi_{n,s}(\textbf{p}_{4})|\, \right|\ ,
\end{equation}

Moreover we get

\begin{equation}\label{7.69}
  \begin{split}
   \frac{\partial^2}{\partial s^2}|\phi_{n,s}(\textbf{p}_{4})| = -\frac{1}{ |\phi_{n,s}(\textbf{p}_{4})|^{3}}(\phi_{n,t}^{(\tau)}(\textbf{p}_{4},V_{n}^{(\tau)}(\phi_{n,t}(\textbf{p}_{4})))\\
       + \frac{1}{ |\phi_{n,s}(\textbf{p}_{4})|} \|V_{n}^{(\tau)}(\phi_{n,t}(\textbf{p}_{4}))\|^{2}\\
       +  \frac{1}{ |\phi_{n,s}(\textbf{p}_{4})|}\Big((\phi_{n,s}(\textbf{p}_{4}),V_{n}^{(\tau)}(\phi_{n,t}(\textbf{p}_{4}))\chi_n^{(\tau)} (\textbf{p}_{4})^{2})\\
      +(\frac{\textbf{p}_{4}}{|\textbf{p}_{4}|},V_{n}^{(\tau)}(\phi_{n,t}(\textbf{p}_{4})))(\textbf{p}_{4},\phi_{n,t}(\textbf{p}_{4}))\frac{\textrm{d}}{\textrm{d}|\textbf{p}_{4}|}\chi_n^{(\tau)} (\textbf{p}_{4})^{2}\Big) \ .
  \end{split}
\end{equation}
where $(.,.)$ is the scalar product in $\R^3$ and $\|.\|$ the corresponding norm.

By \eqref{7.63} there exists a constant $c_{n}>0$ such that
\begin{equation}\label{7.70}
    \left| \frac{\partial^2}{\partial s^2}
 |\phi_{n,s}(\textbf{p}_{4})|\, \right|\leq c_{n} |\phi_{n,t}^{(\tau)}(\textbf{p}_{4})|\leq c_{n}|\textbf{p}_{4}|
\end{equation}

This yields

\begin{equation}\label{7.71}
   \sup_{0\leq|t|\leq 1} \|[A_{n,t}^{(\tau)},\, [A_{n,t}^{(\tau)},\, H_{0}]]\left(H_{0}+ 1\right)^{-1}\|\leq c_{n} \mathrm{e}^{2}
\end{equation}

Let

\begin{equation}\label{7.72}
    \mathcal{G}^{(\alpha)}(\textbf{p}_{4})=u^{[J^{4}_{1},J^{4}_{2}]}_{M^{4}_{1}M^{4}_{2}}(\xi_{4})\widetilde{G}^{(\alpha)}(\xi_{4})
\end{equation}

and

\begin{equation}\label{7.73}
    \mathcal{G}^{(\alpha)}_{t}(\textbf{p}_{4})= \left(\textrm{e}^{-ia_n^{(\tau)}t}\mathcal{G}^{(\alpha)}\right)(\textbf{p}_{4})
\end{equation}

It follows from \eqref{5.26} and \eqref{5.27}--\eqref{5.30} that we can write

\begin{equation}\label{7.74}
    H_{I}= \sum_{\alpha=1,2} H_{I}(F^{(\alpha)},G^{(\alpha)},\mathcal{G}^{(\alpha)})
\end{equation}

We then have, for every $\psi\in\mathcal{D}(H)$

\begin{multline}\label{7.75}
     [A_{n,t}^{(\tau)},\, [A_{n,t}^{(\tau)},\, H_{I}]]\psi =\sum_{\alpha=1,2}\frac{1}{t^{2}}\textrm{e}^{2iA_n^{(\tau)}t}\\   \left(H_{I}(F^{(\alpha)},G^{(\alpha)},\mathcal{G}^{(\alpha)}_{2t})-2H_{I}(F^{(\alpha)},G^{(\alpha)},\mathcal{G}^{(\alpha)}_{t}) +H_{I}(F^{(\alpha)},G^{(\alpha)},\mathcal{G}^{(\alpha)})\right)\psi
\end{multline}

By \eqref{6.16} and \eqref{7.20} we get

\begin{multline}\label{7.76}
    \|[A_{n,t}^{(\tau)},\, [A_{n,t}^{(\tau)},\, H_{I}]]\psi\|\leq g\widetilde{K}(F,G)\\
    \left(\frac{1}{t^{2}}\sum_{\alpha=1,2}\|\mathcal{G}^{(\alpha)}_{2t}(.)-2\mathcal{G}^{(\alpha)}_{t}(.) + \mathcal{G}^{(\alpha)}(.)\|_{L^{2}(\R^{3})}\right)^{1/2}\left(\|H_{0}\psi\| + b\|\psi\|\right).
\end{multline}

Note that, for $0\leq|t|\leq 1$,

\begin{multline}\label{7.77}
   \left(\frac{1}{t^{2}}\sum_{\alpha=1,2}\|\mathcal{G}^{(\alpha)}_{2t}(.)-2\mathcal{G}^{(\alpha)}_{t}(.) + \mathcal{G}^{(\alpha)}(.)\|_{L^{2}(\R^{3})}\right)^{1/2}\\
   \leq \sup_{0<|s|\leq 2}\Big( \sum_{\alpha=1,2}\left\| \frac{\partial^2}{\partial s^2}\mathcal{G}^{(\alpha)}_{s}(.)\right\|^{2} _{L^2(\R^{3})}\Big)^{1/2} .
\end{multline}

with

\begin{equation}\label{7.78}
\begin{split}
 & \left( \frac{\partial^2}{\partial s^2} (\mathrm{e}^{- i a^{(\tau)}_{n}s}\mathcal{G}^{(\alpha)})\right)(\textbf{p}_{4})\\
 & =
 \frac14 \left( \mathrm{e}^{- ia^{(\tau)}_{n}s}((\mathrm{div}  V_{n}^{(\tau)}(\textbf{p}_{4}))^2\mathcal{G}^{(\alpha)})\right)(\textbf{p}_{4}) \\
 & + \frac12 \left( \mathrm{e}^{- ia^{(\tau)}_{n}s}((\mathrm{div} V_{n}^{(\tau)}(\textbf{p}_{4})) V_{n}^{(\tau)}(\textbf{p}_{4})\cdot\nabla_{\textbf{p}_{4}}\mathcal{G}^{(\alpha)})\right)(\textbf{p}_{4}) \\
 & +
 \frac12 \left(
 \mathrm{e}^{- ia^{(\tau)}_{n}s}(\sum_{i,j=1}^3
 ( V_{n}^{(\tau),i}(\textbf{p}_{4})(\frac{\partial^2}
 {\partial p^{i}_{4} \partial p^{j}_{4}}  V_{n}^{(\tau),j}(\textbf{p}_{4})))\mathcal{G}^{(\alpha)})\right)(\textbf{p}_{4})\\
 & +
 \frac12 \left(
 \mathrm{e}^{- ia^{(\tau)}_{n}s}(\sum_{i,j=1}^3
  V_{n}^{(\tau),i}(\textbf{p}_{4}) \frac{\partial  V_{n}^{(\tau),j}}{\partial p^{i}_{4}}(\textbf{p}_{4})
 \frac{\partial}{\partial p^{j}_{4}}\mathcal{G}^{(\alpha)})\right)(\textbf{p}_{4})\\
 & + \frac12 \left(\mathrm{e}^{-a^{(\tau)}_{n}s}
 (\sum_{i,j=1}^3  V_{n}^{(\tau),i}(\textbf{p}_{4})  V_{n}^{(\tau),j}(\textbf{p}_{4}) \frac{\partial^2}
 {\partial p^{i}_{4} \partial p^{j}_{4}}
 \mathcal{G}^{(\alpha)})\right)(\textbf{p}_{4})\ .
\end{split}
\end{equation}

Combining the properties of the $C^{\infty}$ field $V_{n}^{(\tau)}(\textbf{p}_{4})$ with Hypothesis~\ref{6.1} and Hypothesis~\ref{7.1} together with Proposition~\ref{7.5} and by mimicking the proof of theorem 5.1 in \cite{BarbarouxGuillot2009} we finally prove \eqref{7.60}. It follows H is locally of class $C^{2}\big(A^{(\tau)}_{n}\big)$ in $(-\infty,m_{3}-\delta/2)$.

By applying the commutator theory (see \cite{Mourre1981,Amreinetal1996,Sahbani1997,GeorgescuGerard1999,GoleniaJecko2007,Gerard2008}), we then get the following Limiting Absorption Principle

\begin{theorem}\label{theorem:7.6}
Suppose that the kernels $F^{(\alpha)}(.,.)$,$G^{(\alpha)}(.)$ and $\widetilde{G}^{(\alpha)}(.)$, $\alpha=1,2,$ satisfy
Hypothesis~\ref{hypothesis:6.1} and Hypothesis~\ref{hypothesis:7.1}. Then, for any $\delta>0$ satisfying
$0<\delta<m_3$, there exists $g_\delta>0$ such that, for $0<g \leq
g_\delta$, for $s>1/2$, $\varphi$, $\psi\in\gF$ and for $n\geq 1$,
the limits
\begin{equation*}
 \lim_{\epsilon\to 0} (\varphi,\, \langle A_{n}^{(\tau)}\rangle^{-s}
 (H-\lambda\pm i \epsilon) \langle A_{n}^{(\tau)}\rangle^{-s} \psi)
\end{equation*}
exist uniformly for $\lambda\in\Delta_n$.
Moreover, for $1/2<s<1$, the map
\begin{equation*}
  \lambda \mapsto \langle A_{n}^{(\tau)}\rangle^{-s} (H-\lambda\pm i0)^{-1} \langle A_{n}^{(\tau)}\rangle^{-s}
\end{equation*}
is H\"older continuous of degree $s-1/2$ in $\Delta_n$.
\end{theorem}

Here $g_\delta=g_\delta^{(3)}$.

Note that there exists a constant $d_{n}>0$ such that
\begin{equation}\label{7.79}
    |a_{n}^{(\tau)}|^{2} \leq d_{n} \langle b \rangle^{2}
\end{equation}

and

\begin{equation}\label{7.80}
    \left(A_{n}^{(\tau)}\right)^{2} \leq d_{n} P^{2}
\end{equation}

Now,by adapting the proof of theorem~\ref{3.3} in \cite{AschbacherGuillot2011}, we deduce theorem~\ref{7.4} from theorem~\ref{7.6} and from the following lemma

\begin{lemma}\label{lemma: 7.7}
Suppose that $s\in(1/2,\, 1)$ and that for some $n$, $f\in
C_0^\infty(\Delta_n)$. Then,
\begin{equation*}
 \left\|
  \langle A_{n}^{(\tau)} \rangle^{-s} \mathrm{e}^{-i t H} f(H) \langle A_{n}^{(\tau)}\rangle^{-s}
 \right\|
 =
 \mathcal{O}\left(t^{-(s-\frac12)}\right)\, .
\end{equation*}
\end{lemma}

We omit the details.
\end{proof}

\section*{Acknowledgements}
J.-C.~G. acknowledges W.~Aschbacher,J.-M~ Barbaroux and J.~Faupin for helpful discussions.

\end{document}